\documentclass[aps notitlepage, twocolumn,superscriptaddress,floatfix,letterpaper,nofootinbib]{revtex4-1}

\pdfoutput=1

\usepackage[all]{xy}

\usepackage{euscript}

\newcommand{\w}{{\rm w}}
\newcommand{\Tor}{\text{Tor}}

\renewcommand{\mod}{\;\text{mod}\;}
\newcommand{\gSq}{\mathbb{Sq}}
\newcommand{\Sq}{\text{Sq}}
\newcommand{\Bs}{\beta}
\newcommand{\RZ}{{\mathbb{R}/\mathbb{Z}}}
\newcommand\se[1]{\overset{\scriptscriptstyle #1}{=}}

\newcommand\hcup[1]{\underset{{\scriptscriptstyle #1}}{\smile}}
\newcommand{\hcupB}[2]{\overset{\scriptscriptstyle #2}{\underset{{\scriptscriptstyle #1}}{\smile}}}

\newcommand{\sC}{\mathscr{C}}
\newcommand{\sD}{\mathscr{D}}

\newcommand{\A}{{\mathcal{A}}}

\newcommand{\SO}{{\rm SO}}
\newcommand{\Spin}{{\rm Spin}}
\newcommand{\U}{{\rm U}}
\newcommand{\SU}{{\rm SU}}
\newcommand{\CP}{{\mathbb{CP}}}

\def\RP{\mathbb{RP}}
\def\TP{\mathrm{TP}}
\def\PD{\mathrm{PD}}
\def\CS{\mathrm{CS}}
\def\U{\mathrm{U}}
\def\rF{\mathrm{F}}
\newcommand{\Table}[1]{Table \ref{#1}} 
\def\B{\mathrm{B}}

\newcommand{\Sec}[1]{Sec.~\ref{#1}} 

\DeclareFontFamily{U}{mathx}{\hyphenchar\font45}
\DeclareFontShape{U}{mathx}{m}{n}{<-> mathx10}{}
\DeclareSymbolFont{mathx}{U}{mathx}{m}{n}
\DeclareMathAccent{\widebar}{0}{mathx}{"73}

\newcommand\ointint{\begingroup \displaystyle \unitlength 1pt
\int\mkern-7mu\begin{picture}(0,3)\put(0,3){\oval(10,8)}\end{picture}
\mkern-8mu\int\endgroup}

\newcommand{\Fig}[1]{Fig.~\ref{#1}}

\usepackage{tikz}
\usetikzlibrary{calc}

\newcommand{\nn}{\nonumber}
\usepackage{upgreek}

\makeatletter
                
\def\@fnsymbol#1{{\ifcase#1\or \fmarki\or \fmarkii\or \fmarkiii\or \fmarkiv\or \fmarkv\or \fmarkvi\or \fmarkvii\or \fmarkviii\or \fmarkix \else\@ctrerr\fi}}
\makeatother

\newcommand{\fmarki}{w$_1$}
\newcommand{\fmarkii}{w$_2$}
\newcommand{\fmarkiii}{w$_3$}

\def\@fnsymbol#1{{\ifcase#1\or \fmarki\or \fmarkii\or \fmarkiii\or \fmarkiv\or \fmarkv\or \fmarkvi\or \fmarkvii\or \fmarkviii\or \fmarkix \else\@ctrerr\fi}}
\makeatother

\newcommand{\jw}[1]{\color{black} #1 \color{black}}

\begin{document}

\begin{titlepage}

\title{3+1d Boundaries with Gravitational Anomaly of 4+1d Invertible
Topological Order\\ for Branch-Independent Bosonic Systems }
 
\author{Zheyan Wan}
\email{wanzheyan@mail.tsinghua.edu.cn}
\affiliation{Yau Mathematical Sciences Center, Tsinghua University, Beijing 100084, China}

\author{Juven Wang}
\email{jw@cmsa.fas.harvard.edu}
\affiliation{Center of Mathematical Sciences and Applications, Harvard University, MA 02138, USA}
\affiliation{School of Natural Sciences, Institute for Advanced Study, 
Einstein Drive, Princeton, NJ 08540, USA}

\author{Xiao-Gang Wen}
\email{xgwen@mit.edu}
\affiliation{Department of Physics, Massachusetts Institute of
Technology, Cambridge, Massachusetts 02139, USA}

\begin{abstract} 

We study bosonic systems on a spacetime lattice (with imaginary time) defined by
path integrals of commuting fields.  We introduce a concept of
\emph{branch-independent bosonic} (BIB) \emph{systems}, whose path integral is independent of
the branch structure of the spacetime simplicial complex, even for a spacetime
with boundaries.  In contrast, a \emph{generic lattice bosonic} (GLB) \emph{system}'s path
integral may depend on the branch structure.  \jw{We find the invertible
topological order characterized by the Stiefel-Whitney cocycle (such as 4+1d w$_2$w$_3$) to be
nontrivial for \emph{branch-independent bosonic systems}, but 
this topological order and a trivial gapped tensor product state belong to the same
phase (via a smooth deformation without any phase transition) for \emph{generic
lattice bosonic systems}.}  This implies that the invertible topological orders
in generic lattice bosonic systems on a spacetime lattice are not classified by the oriented
cobordism.  {The branch dependence on a lattice may be related to the orthonormal
frame of smooth manifolds and the framing anomaly of continuum field theories.}
 \jw{In general, the branch structure on a 
discretized lattice may be related to a frame structure on a smooth manifold that trivializes any Stiefel-Whitney classes.}
We construct branch-independent bosonic systems to realize the w$_2$w$_3$
topological order, and its 3+1d gapped or gapless boundaries.  One of
the gapped boundaries is a 3+1d $\mathbb{Z}_2$ gauge theory with (1) fermionic
$\mathbb{Z}_2$ gauge charge particle which trivializing w$_2$ and (2)
``fermionic'' $\mathbb{Z}_2$ gauge flux line trivializing w$_3$. In
particular, if the flux loop's worldsheet is unorientable, then
the orientation-reversal 1d worldline must correspond to a fermion worldline that
\emph{does not carry the $\mathbb{Z}_2$ gauge charge}.  We also explain why
Spin and Spin$^c$ structures trivialize the w$_2$w$_3$ nonperturbative global
pure gravitational anomaly to zero (which helps to construct the anomalous 3+1d
gapped $\mathbb{Z}_2$ and gapless all-fermion U(1) gauge theories), but the
Spin$^h$ and Spin$\times_{\Z_2}$Spin$(n \geq 3)$ structures modify the
w$_2$w$_3$ into a nonperturbative global mixed gauge-gravitational anomaly,
which helps to constrain Grand Unifications (e.g., $n=10,18$) or construct new
models.

\end{abstract}


\maketitle

\end{titlepage}

{\small \setcounter{tocdepth}{1} \tableofcontents }

\section{Introduction}

Gapped quantum states of matter (or more precisely, gapped quantum liquids
\cite{ZW1490,SM1403}) with no symmetry can be divided
into two classes \cite{W9039}:\\
(1) States with no topological order. All those states belong to
the trivial phase represented  by tensor product states, 
with no entanglement or short-range entanglement.\\ 
(2) States with topological order \cite{W8951,WN9077}
(\ie gapped states with long-range entanglement \cite{CGW1038}).\\
In the presence of global symmetry that is not spontaneously broken,
the above two classes can be further divided into some subclasses:
\\
(1a) States with no topological order and no symmetry-protected topological (SPT)
order, or synonymously, symmetry-protected trivial (SPT) order, with no entanglement or short-range entanglement.\\
(1b) States with no topological order but with nontrivial SPT order \cite{GW0931,Kitaev-2011-talk,
CLW1141,CGL1314,LV1219,VS1306,GW1441,Senthil1405.4015,KTT1429,FH160406527,KT170108264,WG170310937,BM210911039},
with short-range entanglement.\\
(2a) States with both topological order and
symmetry. Those states are said to have symmetry enriched topological (SET)
orders \cite{W0213,W0303,Kitaev2006, WVc0608129,KLW0834,KW0906,LV1334,EH1306,BZ14104540,LW160205946,B210910913,AK210910911},
with long-range entanglement. \\
In this work, we aim to study those topological states of matter and their
boundaries for bosonic systems.  We realize that even  bosonic systems without
any symmetries can have many different types, such as  bosonic systems on
a spacetime lattice with imaginary time, bosonic systems in continuum spacetime
with real or imaginary time, and bosonic systems defined via lattice
Hamiltonian with real continuous time.  Those different bosonic theories
require different mathematical descriptions.  In this work, we will only study
bosonic systems on a spacetime lattice with imaginary time, that satisfy the
reflection positivity.  In fact, we will study a simpler problem -- the so
called invertible topological states of matter in the bulk, and their
boundaries, with or without symmetry.

Stacking two topological states, $\sC_1$ and $\sC_2$, give use a third
topological state $\sC_3 = \sC_1\boxtimes \sC_2$.  The stacking operation
$\boxtimes$ makes the set of topological states into a monoid. (A monoid is
like a group except its elements may not have inverse.) If a topological state
$\sC$ has an inverse under the stacking operation $\boxtimes$, \ie there exists
a topological state $\sD$ such that $\sC\boxtimes \sD$ is a trivial product
state, then $\sC$ will be called an invertible topological state.  It turns out
that all SPT states are invertible.  Only a small fraction of topologically
ordered states and SET states are invertible.  For example, a fermionic integer
quantum Hall state is an invertible topologically ordered state (a fermionic
invertible SET with U(1) symmetry).  An anti-ferromagnetic spin-1 Haldane chain
is a SPT state protected by spin rotation SO(3) or time reversal $\Z_2^T$
symmetry, which is always invertible.  (In contrast, an anti-ferromagnetic
spin-2 chain, another Haldane phase, has trivial SPT order.)

Invertible topological states of bosonic systems are characterized by a simple
class of invertible topological invariants
\cite{W1447,HW1339,K1459,KW1458,WGW1489, Witten1508.04715,
FH160406527,1711.11587GPW}. In this article, we will derive the possible
boundary theory from those topological invariants, especially the 4+1d
invertible topological order characterized by the Stiefel-Whitney $\w_2\w_3$
topological invariant in 5d.  There are some earlier works in this direction
\cite{WangPotterSenthil1306.3238,
KS14098339,K1459,T14044385,T171002545,WW170506728,
WangWenWitten2018qoy1810.00844, Wang2106.16248, WangYou2111.10369GEQC,YouWang2202.13498,
CH211014644, FH211014654} which construct boundary theories of the $\w_2\w_3$
invertible topological order.  In this work, we will present a more complete
and systematic derivation.

\begin{figure}[t]
\begin{center}
\includegraphics[scale=0.7]{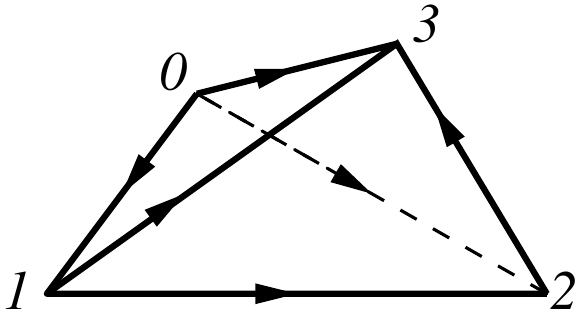} 
\end{center}
\caption{A branch structure of a simplicial complex is given by assigning
arrows to the links which do not form loops around every triangle.  The branch structure gives rise
to a local order of vertices for every simplex.  Stiefel-Whitney cocycle a
simplicial complex can be constructed after assigning a branch structure
(see Appendix \ref{app:SW}, \Fig{figure:tetrahedron} and \Fig{figure:flip}).  }
\label{branch}
\end{figure}

{\bf {Branch-Independent Bosonic System} vs {Generic Lattice Bosonic System}}:
Even bosonic systems on a spacetime lattice can have different types.  We find
that in order to discuss invertible topological order, we need to introduce a
concept of \emph{branch-independent bosonic system} (the L-type system studied
in \Rf{W1477} happens to be a branch-independent bosonic system, L for
Lagrangian formulated on the spacetime lattice), 
whose partition function computed from path integral is
independent of the branch structures (or branching structures) of spacetime
lattice (\ie spacetime simplicial complex), even for spacetime with boundaries.
Here a branch structure is a local ordering of the vertices for each simplex
(see Fig. \ref{branch} and Appendix \ref{cochain}).  \emph{Generic lattice
bosonic systems} do not have this requirement, and their path integrals may
depend on the branch structures of spacetime complex.  

Branch-independence is a property of lattice bosonic systems.  It appears that
such a lattice property is related to frame-independence, a property of
continuum field theory.  An orthonormal frame of a $d$-dimensional manifold is
a set of $d$ vector fields, which is orthonormal at every point respect to the
metrics tensor of the manifold.  We will abbreviate orthonormal frame as frame
(which is also called the vielbein).  After assigning a frame to a manifold, we
can define an SO$(d)$ connection to describe the curvature.  Continuum field
theory explicitly depends on the SO$(d)$ connection and the frame.  Thus the
partition function of the continuum field theory may depend on the frame of the
spacetime manifold.  If the partition function does depend on the frame, we say
that the theory has a \emph{framing anomaly} \cite{W8951}. 
Otherwise, the theory is 
\emph{free of framing anomaly}.

In \Rf{DW190608270}, 2+1d generic lattice bosonic systems are constructed to
realize topological orders with any integral chiral central charge and the
corresponding gravitational Chern-Simons term.  Since the central charge is not
0 mod 24, those models contain framing anomaly and are not frame-independent.
This implies that the partition function of generic lattice bosonic systems may
depend on the frames of the spacetime manifold in the continuum limit.  This
example and the above discussions suggest a close relation between the branch
structure on a lattice and the frame structure on a continuum manifold.\footnote{\jw{{\bf Branch structure on a discretized lattice 
vs Frame structure on a smooth manifold that trivializes any Stiefel-Whitney classes}:
A framing of a $d$-dimensional manifold of $M$ is a choice of trivialization of its tangent bundle, 
hence a choice of a section of the corresponding frame bundle.
If there exists a framing on the tangent bundle $TM$, 
then there exist $d$ linearly independent sections of the tangent bundle. 
If the $i$-th Stiefel-Whitney class $w_i( TM )\neq 0$, then there can not exist $d-i+1$ 
linearly independent sections of the tangent bundle. 
So $w_i( TM )=0$ for $1\leq i \leq d$.
Thus, \\
(1) The tangential frame structure on a smooth manifold trivializes any Stiefel-Whitney classes.\\
(2) We will later suggest that a branch structure on a discretized lattice or on a simplicial complex also trivializes Stiefel-Whitney classes.\\
(3) Thus, our work suggests a possible relation: a branch structure on a discretized lattice may be related to
the frame structure on a smooth manifold that trivializes any Stiefel-Whitney classes.}
}

We conjecture that the independence of the branch structure of spacetime
complex for lattice models implies the independence of frame structure of
spacetime manifold in the continuum limit.  Under such a conjecture, the
branch-independent bosonic systems on lattice only give rise to continuum
effective field theories that are free of framing anomaly.  As a result, a 2+1d
branch-independent bosonic system can only realize topological orders with
chiral central charge $c=0$ mod 24, where the 2+1d invertible topological
orders is generated by three copies of E$_8$ quantum Hall states (say E$_8^3$
topological order).  Indeed, using $\SO(\infty)$ non-linear $\si$-model,
\Rf{W1477} constructed a branch-independent bosonic model that realize the
E$_8^3$ topological order with $c=24$. For more details, see Appendix
\ref{pullback}.

In this work, we find that the 4+1d invertible topological order characterized
by the Stiefel-Whitney class $\w_2\w_3$ is nontrivial for branch-independent
bosonic systems.  On the other hand, The $\w_2\w_3$ topological order and
trivial tensor product states belong to the same phase for generic lattice bosonic
systems.  This implies that we cannot use the oriented cobordism
 \jw{(\ie manifolds with special orthogonal group SO structures)}
\cite{K1459,KW1458,WGW1489,FH160406527} 
to classify invertible topological
orders for the generic lattice bosonic systems (where the lattice means the
spacetime lattice).  

Can we use oriented cobordism to classify invertible topological orders for
branch-independent bosonic systems?  The oriented cobordism suggests a $\Z$
class of 2+1d invertible bosonic topological orders, generated by the E$_8$
topological order.\footnote{The 2+1d Abelian bosonic topological orders are
classified (in a many-to-one fashion) by symmetric integral matrices with even
diagonals\cite{WZ9290}, which are called $K$-matrices. Those topological orders
are realized by $K$-matrix quantum Hall wavefunctions
$\Psi(z_i^I)=\prod_{i<j;I,J} (z_i^I-z_j^J)^{K_{IJ}}\ee^{-\frac14 \sum
|z_i^I|^2}$ and are described by $K$-matrix U(1)-Chern-Simons theories
$\frac{K_{IJ}}{4\pi} a_I\dd a_J$.  The $K$-matrices with $|\text{det}(K)|=1$
classify invertible topological orders.  Here the E$_8$ topological order is an
invertible topological order described by a $K$-matrix given by the Cartan
matrix of E$_8$, denoted as $K_{E_8}$.  The 1+1d boundary carries the chiral
central charge $c=8$ \cite{W9505}. In contrast, the E$_8^3$ topological order
is described by a $K$-matrix $K= K_{E_8}\oplus K_{E_8}\oplus K_{E_8}$ and has
its boundary carrying the chiral central charge $c=24$.  It was suggested that
the $\Z$ class of the oriented cobordism is generated by the E$_8$ topological order,
see for example, Freed's work \cite{Freed1406.7278} or Freed-Hopkins
\cite{FH160406527}. See Section 7 of \cite{WW2019fxh1910.14668} for an
elaborated interpretation of the related cobordism invariants.}  However,
currently, we do not have a realization of E$_8$ topological order using
branch-independent bosonic systems.  We only know a branch-dependent
bosonic model that realizes the E$_8$ topological order \cite{DW190608270}, and
a branch-independent bosonic model that realizes the E$_8^3$ topological
order \cite{W1477}.  Thus it is not clear if oriented cobordism classifies
invertible topological orders for the branch-independent bosonic systems or
not.  
In this article, we will concentrate on branch-independent lattice  bosonic
systems.

The boundary theories of bulk invertible topological orders with no symmetry
belong to a special class: they are theories with gravitational anomalies.  In
fact, the gravitational anomalies in field theories are classified by
invertible topological orders in one higher dimension \cite{W1313,KW1458}. From
this point of view, we study various anomalous theories with a given
gravitational anomaly on the 3+1d boundary of 4+1d $\w_2\w_3$ topological
order.  We shall call this gravitational anomaly as the $\w_2\w_3$ anomaly,
which is a \emph{nonperturbative global gravitational anomaly}.\footnote{Perturbative local anomalies 
are detectable via infinitesimal gauge/diffeomorphism transformations continuously deformable from the identity, captured by perturbative Feynman diagrams \cite{AlvarezGaume1983igWitten1984}.
Nonperturbative global anomalies are detectable via large gauge/diffeomorphism transformations that cannot be continuously deformed from the identity \cite{Witten1985xe}.
}

For example, it is known that the 4+1d invertible topological order has a
gapped 3+1d boundary described by a $\Z_2$ gauge theory with the $\w_2\w_3$
anomaly, where the $\Z_2$ gauge charge is fermionic. However, the fermionic
$\Z_2$ gauge charge does not fully characterize the gravitational anomaly.  In
particular, there is also an anomaly-free 3+1d $\Z_2$ gauge theory with a
fermionic $\Z_2$ gauge charge, \ie there is a 3+1d lattice
bosonic system that can realize a $\Z_2$ gauge theory with a fermionic $\Z_2$
gauge charge \cite{LW0316}.  In this work, we show that the 2d worldsheet of the $\Z_2$ gauge
flux in the spacetime must carry a non-contractable 1d fermionic
orientation-reversal worldline if the 2d worldsheet is unorientable.  This 1d
fermionic orientation-reversal worldline with neutral gauge charge is however
distinct from the fermionic worldline of $\Z_2$ gauge charge.  This crucial
property, together with the fermionic $\Z_2$ gauge charge, characterizes the
gravitational anomaly. 

The above result about the fermion worldline on unorientable  worldsheet of
$\Z_2$ flux line was first obtained in Section 3.3 of
\Rf{WangWenWitten2018qoy1810.00844}. In this paper we give a different
derivation of the result using a path integral formulation on a spacetime
lattice. This result was also obtained recently in \Rf{CH211014644} using
Hamiltonian formulation on spatial lattice.  The ``fermionic'' nature of the
$\Z_2$ flux line can also be characterized by the statistical hopping algebra
for strings \cite{FH211014654}, a generalization of the statistical hopping
algebra for particles \cite{LW0316}.

\subsection{Notations and conventions}

We denote the ${\rm n'}$d for the spacetime dimensions to be ${\rm n'} =({\rm
n}+1)$ with an ${\rm n}$ dimensional space and a 1 dimensional time.
Typically, in this article, the dimension always refer to the spacetime
dimension altogether.  We may simply call the 0d $\Z_2$ gauge charge as $\Z_2$
charge, whose spacetime trajectory is a 1d worldline.  We may simply call the
1d $\Z_2$ gauge flux loop as $\Z_2$ flux, which can be a 1d loop which bounds a
2d surface enclosing gauge flux, whose spacetime trajectory is a 2d worldsheet.

In this work, we use a lot of formalisms of chain and cochain, as well as the associated
derivative cup product, Steenrod square, etc.  A brief coverage of those topic
can found in Appendix \ref{cochain}.  The $\Z_n$ values are chosen to be
$\{0,1,\cdots,n-1\}$.  In this article, we always use this set to extend $\Z_n$
values to $\Z$ values, and treat $\Z_n$-valued quantities as $\Z$-valued
quantities.  To help to express $\Z_n$-valued relation using $\Z$-valued
quantities, we denote $\se{n}$ to mean equal up to a multiple of $n$ {(thus it
is a mod $n$ relation: two sides of the equality are equal mod $n$)}, and use
$\se{\dd }$ to mean equal up to a coboundary $\dd f$ (\ie with the coboundary
operator $\dd$).

We denote the Lorentz group as $\SO$ (for bosonic systems) and $\Spin$ (for
fermionic systems graded by the fermion parity $\Z_2^f$).  In ${\rm n}+1$d
spacetime, the $\SO$ really means the $\SO({\rm n}+1)$ for the Euclidean
rotational symmetry group and the $\SO({\rm n},1)$ for the Lorentz rotational
$+$ boost symmetry group; the $\Spin$ really means the $\Spin({\rm n}+1)$ for
the Euclidean rotational symmetry group and the $\Spin({\rm n},1)$ for the
Lorentz rotational $+$ boost symmetry group. 
We denote  $\se{n,\dd}$ to mean equal up to a mod $n$ relation and also equal
up to a coboundary $\dd f$.  We will use the group $N\gext_{e_2,\al} G$ to
describe the extension of a group $G$ by an abelian group $N$ via $$ 1 \to N
\to N\gext_{e_2,\al} G \to G \to 1, $$ which is characterized by $e_2 \in
H^2(G;N)$ of the second cohomology class, where $N$ is an abelian group with a
$G$-action via $\al:G \to \text{Aut}[N]$.  {We will also use $G_1 \times_{N}
G_2 \equiv \frac{G_1 \times G_2}{N}$ to define as the product group of $G_1$
and $G_2$ modding out their common normal subgroup $N$.}
Other mathematical conventions and definitions (such as Stiefel-Whitney class)
can be found in Appendix \ref{sec:Conventions}.
We provide many appendices on 
the toolkits of co/chain, co/cycle, cohomology, characteristic class, and cobordism.

\emph{Dynamical gauge fields} are associated with the gauge connections of
gauge bundles that are summed over in the path integral (or partition
function).  \emph{Background gauge fields} are associated with the
non-dynamical gauge connections of gauge bundles that are fixed, not summed
over in the path integral.  We will distinguish their gauge transformations:
for dynamical fields as \emph{dynamical gauge transformations}, for background
fields as \emph{background gauge transformations}, 
see Appendix \ref{app:Gauge-Transformation}.

In this work, the \emph{anomalous gauge theory} merely means its partition
function alone is only \emph{non-invariant} under \emph{background gauge
transformations} (but still \emph{invariant} under \emph{dynamical gauge
transformations}) --- namely, the \emph{anomalous gauge theory} with only 't
Hooft anomaly of the global symmetry \cite{tHooft1979ratanomaly} can still be
\emph{well-defined} on the boundary of one-higher dimensional invertible
topological phase.  The cancellation of  \emph{background gauge
transformations}  between the bulk and boundary theories are known as the
anomaly inflow \cite{CH8527}. 

\section{Branch-Independent Bosonic System
and Generic lattice Bosonic System}

In this work, we are going to use cochains on a spacetime simplicial complex as
bosonic fields.  In order to construct the action $S$ in the path integral,
using local Lagrangian term on each simplex, it is important to give the
vertices of each simplex a local order.  A local scheme to order  the vertices
is given by a branch structure \cite{C0527,CGL1314,CGL1204}. A branch
structure is a choice of orientation of each link in the complex so that there
is no oriented loop on any triangle (see \Fig{branch}).  Relative to a base
branch structure, all other branch structure can be described by a
$\Z$-valued 1-cochain $s$ (see Appendix \ref{branchdep}).  After assigning
a branch structure to the spacetime complex, we can define cup product
$\hcupB{}{s}$ of cochains that depend on the  branch structure $s$.
For the base branch structure $s=0$, we abbreviate $\hcupB{}{s}$ by
$\hcupB{}{}$.  

We find that for two cocycles, $f$ and $g$, $f\hcupB{}{s} g -
f\hcupB{}{} g$ is coboundary, that depends on $s,f,g$. Let us use $\dd
\nu(s,f,g)$ to denote such a coboundary (for details, see Appendix
\ref{branchdep}):
\begin{align}
 f\hcupB{}{s} g +\dd \nu(s,f,g) = f\hcupB{}{} g.
\end{align}
Using derivative and cup product of the cochains, we can construct a local
action $S$.  So, in general, the action amplitude, $\ee^{-S}$, may depend on
the choices of the branch structures.  

\subsection{Branch-Independent Bosonic (BIB) System}

Now we are ready to define the \emph{branch-independent bosonic system}: A
branch-independent bosonic system is defined by a path integral on a branch
spacetime simplicial complex, such that the value of the path integral is
independent of the choices of branch structures, 
\jw{even when the spacetime has boundaries.}

Let us give an example of branch-independent bosonic system.  The bosonic
system has two fields: a $\Z_2$-valued 1-cochain field $a^{\Z_2}$ and a
$\Z_2$-valued 2-cochain field $b^{\Z_2}$, which give rise to the following
partition function
\begin{align}
\label{Zabww}
 Z=
 \sum_{a^{\Z_2}, b^{\Z_2}}
& \ee^{\ii \pi \int_{M^5} 
(\w_2^{s}+\dd a^{\Z_2})\hcupB{}{s}
(\w_3^{s}+\dd b^{\Z_2})} 
\nonumber\\
& \ee^{\ii \pi \int_{\prt M^5} 
\nu(s, \w_2^{s}+\dd a^{\Z_2},\w_3^{s}+\dd b^{\Z_2})} 
,
\end{align}
where $\sum_{a^{\Z_2}, b^{\Z_2}}$ sums over all the cochain fields.  The
summation $\sum_{a^{\Z_2}, b^{\Z_2}}$ in the path integral is known as a
summation of {\bf degrees of freedom}.  Here, the $\w_n^{s}$ is the $n^\text{th}$
Stiefel-Whitney cocycle computed from a simplicial complex $M^5$ with a branch
structure $s$, as described in \Rf{Goldstein-Turner} and in Appendix
\ref{app:SW}. The cocycle $\w_n^{s}$  is a representation of the Stiefel-Whitney class $w_n(TM)$ of
the tangent bundle $(TM)$ of the spacetime manifold $M$.\footnote{Here we treat $\w_n$ as the
Stiefel-Whitney cocycle. In contrast, the mathematical definition of
Stiefel-Whitney class as cohomology class and characteristic class $w_n$ is given in
Appendix \ref{sec:Conventions}. Because the cohomology class is also a cocycle, 
so we may also abuse the notation to write Stiefel-Whitney class $w_n$ in terms of cocycle $\w_n$.
\jw{A cohomology class is an equivalence class that has many representatives;
any elements in the equivalence class are representatives.
The $w_n$ is Stiefel-Whitney class, also it can be referred to as Stiefel-Whitney cocycle when we choose a representative $\w_n$; 
the $w_n$ is a Stiefel-Whitney number only when $n$ is the top spacetime dimension.\\
The Stiefel-Whitney product cocycle $\w_2 \w_3$ is 
a representative of the cup product of two Stiefel-Whitney classes ($w_2$ and $w_3$).
When the $\w_2 \w_3$ paired with the fundamental class, as $\int_{M^5} \w_2 \w_3$, is called the Stiefel-Whitney number.
}}
\jw{We omit the normalization factor here in the partition function \eq{Zabww}.}

Let $\w_n$ be the  Stiefel-Whitney cocycle for the base branch structure:
$\w_n \equiv \w_n^{s=0}$.  In general, $\w_n^{s}$ depends on the
branch structure $s$ on $M^5$.  However, $\w_n^{s}-\w_n$ is a
coboundary:
\begin{align}
 \w_n^{s} = \w_n +\dd v_{n-1}(s).
\end{align}

We can show that \eq{Zabww} is independent of branch structure $s$.
First, \jw{from the definition of $\nu$, 
with a bulk ${M^5}$ but without any boundary ${\prt M^5}$, we have
}
\jw{
\begin{align}
 Z &=\sum_{a^{\Z_2}, b^{\Z_2}}
 \ee^{\ii \pi \int_{M^5} 
(\w_2^{s}+\dd a^{\Z_2})\hcupB{}{}
(\w_3^{s}+\dd b^{\Z_2})} 
\nonumber\\
&=\sum_{a^{\Z_2}, b^{\Z_2}}
 \ee^{\ii \pi \int_{M^5} 
(\w_2^{s}+\dd a^{\Z_2})
(\w_3^{s}+\dd b^{\Z_2})} 
.
\end{align}}
In this article, we abbreviate the cup product for the base branch
structure, $f \smile g$, as $fg$.
From the relation between $\w_n^{s}$ and $\w_n$, we have
\begin{align}
 Z &=\sum_{a^{\Z_2}, b^{\Z_2}}
\ee^{\ii \pi \int_{M^5} 
\jw{(\w_2 +\dd v_1(s) +\dd a^{\Z_2})
(\w_3 +\dd v_2(s) +\dd b^{\Z_2})}} 
\nonumber\\
&= \sum_{a^{\Z_2}, b^{\Z_2}}
\ee^{\ii \pi \int_{M^5} 
(\w_2 +\dd a^{\Z_2})
(\w_3 +\dd b^{\Z_2})} 
.
\end{align}
We see that the partition function \eq{Zabww} is indeed independent of
branch structure.  

The  partition function for a generic bosonic system on a spacetime complex $M$
in a quantum liquid phase has a form
\begin{align}
\label{Ztop}
Z(M) &= \ee^{-S^\text{eff}}Z^\text{top}(M),
\end{align}
in a thermodynamic limit, where $S^\text{eff}$ $=\int_{M}$ $\text{energy-density}$ is
the non-universal volume part, and $Z^\text{top}(M)$ is the universal
topological partition function (\ie the topological invariant) that
characterizes the topological order.  (The topological partition function
$Z^\text{top}$ and its isolation are discussed in much more details in
\Rf{KW1458,W161201418,WW180109938}.) For a branch-independent bosonic system,
the topological invariant $Z^\text{top}(M)$ does not depend on the branch
structures on $M$, and possibly, nor the framing of the spacetime manifold in
continuum limit.  This leads to the conjecture that the topological invariant
$Z^\text{top}(M)$ is a \jw{cobordism invariant \cite{K1459,KW1458,FH160406527} for
invertible topological orders in the branch-independent bosonic systems.}

The above example \eq{Zabww} is exactly solvable when $M^5$ has no boundary,
since the partition function can be calculated exactly
\begin{align}
 Z = 2^{N_l+N_f} \ee^{\ii \pi \int_{M^5} \w_2 \w_3} 
\end{align}
where $N_l$ is the number of links (namely 1-simplices that can be paired with $a^{\Z_2}$) and 
$N_f$ is the number of faces (namely 2-simplices  that can be paired with $b^{\Z_2}$) in $M^5$.
Thus the summation $\sum_{a^{\Z_2}, b^{\Z_2}}$ gives a $2^{N_l+N_f}$ factor. 
After dropping the non-universal volume term $\ee^{-S^\text{eff}}=2^{N_l+N_f}$,
we see that the topological partition function is given by
\begin{align}
\label{topw2w3}
 Z^\text{top}(M^5) =\ee^{\ii \pi \int_{M^5} \w_2\w_3}.
\end{align}
The nontrivial topological invariant $\ee^{\ii \pi \int_{M^5} \w_2\w_3}$
suggests that the bosonic model \eq{Zabww} realizes a nontrivial topological
order.  The  topological order is invertible since the topological invariant is
a phase factor for any closed spacetime $M^5$.  We will refer such an
invertible topological order as the $\w_2\w_3$-topological order. 

\subsection{Generic Lattice Bosonic (GLB) System}

We also like to define a concept of \emph{generic lattice bosonic system}, as a
bosonic system on a spacetime simplicial complex, whose path integral may or may
not depend on the branch structures.  Indeed, when we define an generic
lattice bosonic system, the local branch structure of spacetime complex is
given, and the definition may depend on the choices of the local branch
structure.

Let us give an example of a generic lattice bosonic system, that happens to have 
``no degrees of freedom.'' 
 \jw{Here ``no degrees of freedom'' means that there is only one term in the path integral,
and also means that there is only one state $|0\rangle_j$ per single site $j$.}
By definition, the ground state of such a system is a
\emph{trivial tensor product state} $\otimes_j |0\rangle_j$ (or, simply, a \emph{product state}).  Since the  generic lattice bosonic system may have an
action that depend on the branch structure on the spacetime complex $M^5$, we
can choose the action amplitude to be
\begin{align}
\ee^{-S} &= 
\ee^{\ii \pi \int_{M^5} \w_2^{s}\hcupB{}{s}\w_3^{s}}
\ee^{\ii \pi \int_{\prt M^5} \nu(s, \w_2^{s},\w_3^{s})}
\nonumber\\
&=
\ee^{\ii \pi \int_{M^5} \w_2^{s}\hcupB{}{}\w_3^{s}}
=\ee^{\ii \pi \int_{M^5} \w_2^{s}\w_3^{s}}. 
\end{align}
This is possible since a branch simplicial complex fully
determines the cup product and the cocycles $\w_n^{s}$ that represents
Stiefel-Whitney cohomology classes \cite{GK150505856,WG170310937}.
The $\w_n^{s}$ depends on the branch structure and different choices of
branch structures $s$ can only change $\w_n^{s}$ by a coboundary.
Thus, when $M^5$ has boundaries, the action amplitude $\ee^{\ii \pi \int_{M^5}
\w_2^{s}\w_3^{s}}$ will depend on the branch structure.  So such an
action amplitude is only allowed by generic lattice bosonic systems.

Since there is no degrees of freedom, the path integral is only a summation of
one term and the topological partition function is given by the action
amplitude
\begin{align} 
\label{Zww}
Z^\text{top}(M^5) = \ee^{\ii \pi \int_{M^5} \w_2^{s}\w_3^{s}} .  
\end{align}
This implies that 
\frmbox{
In {\bf generic
lattice bosonic systems} (GLB),
a cobordism invariant $\ee^{\ii \pi
\int_{M^5} \w_2\w_3}$
can be realized by a system with no
degrees of freedom,
and corresponds to a trivial
 tensor product state with no SPT nor topological order.
Here \jw{``no degrees of freedom'' means that there is only one term in the path integral,
and also means that there is only one state $|0\rangle_j$ per single site $j$.
So the Hilbert space only contains a single 
tensor product state $\otimes_j |0\rangle_j$ for all sites.}
}
In contrast, \
\frmbox{In {\bf branch-independent bosonic systems} (BIB),
a cobordism invariant $\ee^{\ii \pi
\int_{M^5} \w_2\w_3}$ cannot be realized
by  a system with no
degrees of freedom, and thus
corresponds to a nontrivial invertible topological order.
}

In other words, we can smoothly deform the topological order characterized by
$\ee^{\ii \pi \int_{M^5} \w_2\w_3}$ into a trivial product state in the {parameter} space (or landscape) of
generic lattice bosonic systems, but such a smooth deformation path does not exist
in the {parameter} space (or landscape) of branch-independent bosonic systems.  
The smooth deformation is defined as the continuous and differentiable tuning of any
coupling strength of Lagrangian/Hamiltonian terms in the parameter space, that
do not cause any phase transitions. 

To construct the above
mentioned smooth deformation in the {parameter} space of generic lattice bosonic systems, let us
consider the following generic lattice bosonic system
\begin{align}
\label{ZabwwU}
 Z=\sum_{a^{\Z_2}, b^{\Z_2}} &
\ee^{\ii \pi \int_{M^5} 
(\w_2+\dd a^{\Z_2})
(\w_3+\dd b^{\Z_2})} 
\ee^{-U \int_{M^5}  |a^{\Z_2}|^2 + |b^{\Z_2}|^2} ,
\end{align}
where 
\begin{align}
 \int_{M^5}  |a^{\Z_2}|^2 \equiv \sum_{(ij)} |a^{\Z_2}_{ij}|^2, \ \ \ 
 \int_{M^5}  |b^{\Z_2}|^2 \equiv \sum_{(ijk)} |b^{\Z_2}_{ijk}|^2 .
\end{align}
Changing $U$ leads to the smooth deformation.  When $U=0$,  the path
integral \eq{ZabwwU} is \eq{Zabww}.  In the limit $U\to \infty$, the path
integral \eq{ZabwwU} becomes \eq{Zww}.  The path integral \eq{ZabwwU} is
still exactly solvable when $M^5$ has no boundaries
\begin{align}
 Z&=\sum_{a^{\Z_2}, b^{\Z_2}} 
\ee^{\ii \pi \int_{M^5} 
(\w_2+\dd a^{\Z_2})
(\w_3+\dd b^{\Z_2})} 
\ee^{-U \int_{M^5}  |a^{\Z_2}|^2 + |b^{\Z_2}|^2} 
\nonumber\\
&= (1+\ee^{-U})^{N_l+N_f}
\ee^{\ii \pi \int_{M^5} \w_2\w_3}
.
\end{align}
The system \jw{is gapped, namely, with a short-range correlation for any values of $U$.}  So the
deformation from $U=0$ to $U=+\infty$ is a smooth deformation that does
\emph{not} cause any phase transition.  This is why a $ \w_2\w_3$ topological
order and a product state belong to the same phase for generic lattice bosonic
systems.

From the above discussion, it is clear that invertible topological orders have
different classifications for branch-independent bosonic systems and for
generic lattice bosonic systems.  
However, the two kinds of invertible topological orders 
are still related.  
Let TP denotes a classification of ``Topological Phases'' following the notation of Freed-Hopkins \cite{FH160406527}.
Let
$\TP_d(\text{BIB})$ be the Abelian group that classifies the topological phases  
invertible topological orders for branch-independent bosonic (BIB) systems.  
Let
$\TP_d(\text{GLB})$ be the Abelian group that classifies invertible topological
orders for generic lattice bosonic (GLB) systems.  
Since invertible topological
orders for branch-independent bosonic systems can also be viewed as
invertible topological orders for generic lattice bosonic systems, we have a
group homomorphism
\begin{align}
 \TP_d(\text{BIB}) \xrightarrow{p_d} \TP_d(\text{GLB}).
\end{align}
Since invertible topological order characterized by Stiefel-Whitney classes all
become trivial for generic lattice bosonic systems, the map $p_d$ sends all the
Stiefel-Whitney class to zero in the topological invariant.  The map $p_d$ is
not injective.  The map $p_d$ may also not be surjective. So it may only tell
us a subset of invertible topological orders for generic lattice bosonic
systems.  In the rest of this article, we will mainly concentrate on
branch-independent bosonic systems and its $\w_2\w_3$ invertible topological
order.

\subsection{Lattice Bosonic Systems with time reversal symmetry}

The above discussion can be easily generalized to bosonic systems with time
reversal symmetry, defined via path integrals on a spacetime lattice \jw{with a real
action amplitude} $\ee^{-S}$.  In order words, we restrict ourselves to consider
only the real action amplitudes $\ee^{-S}$,  as a way to implement time
reversal symmetry, 

The simplest bosonic invertible topological order with time reversal symmetry
(or time-reversal SPT order) appears in 2d and is characterized by the
topological invariant $(-)^{\int_{M^2} \w_1^2}$ on a closed spacetime $M^2$.  A
branch-independent bosonic model that realizes the SPT order is given by
\begin{align}
\label{ZT1}
 Z =\sum_{g^{\Z_2}}
(-)^{\int_{M^2} (\w_1+\dd g^{\Z_2})^2}
\end{align}
where $\sum_{g^{\Z_2}}$ sums over all $\Z_2$-valued 0-cochains  $g^{\Z_2}$ and
the spacetime $M^2$ can have boundaries.  The branch-independence is ensured by the
invariance of the action amplitude $(-)^{\int_{M^2} (\w_1+\dd g^{\Z_2})^2}$
under the following transformation
\begin{align}
 \w_1 \to \w_1 +\dd v_0, \ \ \ \
g^{\Z_2} \to g^{\Z_2} + v_0 ,
\end{align}
even when $M^2$ has boundaries.

We also have a generic lattice bosonic model
given by
\begin{align}
\label{ZT2}
 Z =\sum_{g^{\Z_2}}
(-)^{\int_{M^2} (\w_1+\dd g^{\Z_2})^2}
\ee^{-U \int_{M^2} |g^{\Z_2}|^2}.
\end{align}
The model has the time-reversal symmetry since the action amplitude is real.
The model is exactly solvable when $M^2$ has no boundaries, and has the following partition function.
\begin{align}
  Z = (1+\ee^{-U})^{N_v} (-)^{\int_{M^2} \w_1^2}
\end{align}
where $N_v$ is the number of the vertices in the spacetime complex $M^2$.
Since the partition function only depends on the area of spacetime $M^2$, but
not depends on the shape of spacetime $M^2$, the model \eq{ZT2} is \jw{gapped (\ie
has short range correlations)} for any values of $U$.  There is no phase
transtion as we change $U$.

When $U=0$, the model \eq{ZT2} becomes \eq{ZT1} and realize the time-reversal
SPT order characterized by topological invariant $(-)^{\int_{M^2} \w_1^2}$.
When $U=\infty$, the model \eq{ZT2} becomes a model with no degrees of freedom
which must correspond to a trivial tensor product state.  We see that, for
generic lattice bosonic models with time reversal symmetry, the time-reversal
SPT order characterized by topological invariant $(-)^{\int_{M^2} \w_1^2}$ and
the  trivial tensor product state belong to the same phase.  This result
suggests that the bosonic invertible topological orders with time-reversal
symmetry (\ie with real action amplitudes) for generic lattice bosonic systems
are not classified by unoriented cobordism \jw{(\ie manifolds with orthogonal group O structures).}

\section{Boundaries of $\w_2\w_3$ invertible topological order}

\subsection{The branch independence and background gauge invariance}

In the last section, we studied a 5d branch-independent bosonic model on
spacetime complex $M^5$ defined by the path integral \eq{Zabww}.  That path
integral can be simplified by using the base branch structure to define the cup
product:
\begin{align}
\label{Zabww2}
 Z=\sum_{a^{\Z_2}, b^{\Z_2}}
\ee^{\ii \pi \int_{M^5} 
(\w_2^{s}+\dd a^{\Z_2})
(\w_3^{s}+\dd b^{\Z_2})} .
\end{align}
When $M^5$ is a closed manifold with no boundary, the above path integral gives
rise to a topological invariant (also known as a cobordism invariant) $\ee^{\ii
\pi \int_{M^5} \w_2\w_3}$.  In this section, we consider the case when
$M^5$ has a boundary.  We shall obtain the possible boundary theories for the
$\w_2\w_3$-topological order.  

One might guess that, when $M^5$ has a boundary, the partition function is
still given by $Z(M^5)=\ee^{\ii \pi \int_{M^5} \w_2^{s}\w_3^{s}}$.  If
this was true, we could conclude that the boundary is gapped, since the
partition function $Z(M^5)=\ee^{\ii \pi \int_{M^5} \w_2^{s}\w_3^{s}}$
does not depend on the metrics on the boundary $B^4 \equiv \prt M^5$. (A
gapless system must have a partition function that depends on the metrics (\ie
the shape and size) of the spacetime.)

Such a gapped boundary is possible for generic lattice bosonic systems, but it
is impossible for the branch-independent bosonic model \eq{Zabww2}.  This is
because the partition function in \eq{Zabww2} is independent of the branch
structure on $M^5$ even when $M^5$ has boundaries, while our guess
$Z(M^5)=\ee^{\ii \pi \int_{M^5} \w_2^{s}\w_3^{s}}$ depends on the
branch structure, since 
\begin{align}
\label{w2w3sZ}
 \w_2^{s} =  \w_2 +\dd v_1(s),\ \ \ \
 \w_3^{s} =  \w_3 +\dd v_2(s).
\end{align}
Thus $\ee^{\ii \pi \int_{M^5} \w_2^{s}\w_3^{s}}$ cannot be the partition
function of \eq{Zabww2} which must be independent of the branch structure.  

Due to \eq{w2w3sZ}, we see that we can encode the branch structure
independence, via the invariance under the following transformation
\begin{align}
\label{w2w3gauge}
 \w_2 \to  \w_2 +\dd v_1,\ \ \ \
 \w_3 \to  \w_3 +\dd v_2.
\end{align}
Thus the branch independence of \eq{Zabww2} can be rephrased as the
invariance of the following path integral
\begin{align}
\label{Zabww1}
 Z=\sum_{a^{\Z_2}, b^{\Z_2}}
\ee^{\ii \pi \int_{M^5} 
(\w_2+\dd a^{\Z_2})
(\w_3+\dd b^{\Z_2})} .
\end{align}
under the above transformation \eq{w2w3gauge}, even for spacetime $M^5$ with
boundaries.  We refer to \eq{w2w3gauge} as a ``background gauge
transformation,'' which is a change in the parameters in the Lagrangian 
rather than a change in the dynamical fields in the Lagrangian.
See the comparison between ``background gauge
transformation'' and ``dynamical gauge transformation''
in Appendix \ref{app:Gauge-Transformation}.

In the following, we will use the background gauge invariance under
\eq{w2w3gauge} to ensure the independence of branch structures.  For the
branch-independent bosonic model \eq{Zabww1}, the boundary must be
non-trivial. We may assume the boundary to be described by
\begin{align} \label{eq:5d-4d}
 Z(M^5) =\sum_{\phi} \ee^{\ii \pi \int_{M^5} \w_2\w_3 - \int_{\prt M^5} 
\cL_\text{bndry}(\phi,\w_2,\w_3)}.
\end{align}
The boundary Lagrangian $\cL_\text{bndry}$ is not invariant under the
background gauge transformation \eq{w2w3gauge}, which cancels the
non-invariance of $\w_2\w_3$ in $\ee^{\ii \pi \int_{M^5} \w_2\w_3}$ when $M^5$
has boundaries.  This cancellation of non-invariances of the bulk and boundary
theories is actually the idea of anomaly inflow \cite{CH8527}.

\subsection{4d $\Z_2$-gauge boundary of the $\w_2\w_3$ topological order}

In this section, we are going to explore the possibility that the boundary
Lagrangian $\cL_\text{bndry}$ describes a $\Z_2$ gauge theory, 
more precisely the dynamical Spin structure summed over in the path integral.  

\subsubsection{Effective boundary theory}
\label{eq:Z2EFT}

A 4d $\Z_2$ gauge theory can be described by $\Z_2$-valued 1-cochain $a^{\Z_2}$
and 2-cochain $b^{\Z_2}$ fields (for example, see \cite{Banks2010zn1011.5120,
W161201418, Putrov2016qdo1612.09298PWY}) with the following path integral
\begin{align}
\label{ZabZ2}
 Z &=\sum_{a^{\Z_2},b^{\Z_2}} 
\ee^{\ii \pi \int_{B^4} a^{\Z_2}\hcupB{}{s}\dd b^{\Z_2}}
\ee^{\ii \pi \int_{\prt B^4} \nu(s,a^{\Z_2},\dd b^{\Z_2})}
\\
&= 
\sum_{a^{\Z_2},b^{\Z_2}} 
\ee^{\ii \pi \int_{B^4} a^{\Z_2}\hcupB{}{}\dd b^{\Z_2}}
= 
\sum_{a^{\Z_2},b^{\Z_2}} 
\ee^{\ii \pi \int_{B^4} a^{\Z_2}\dd b^{\Z_2}}
,
\nonumber 
\end{align}
where $\sum_{a^{\Z_2},b^{\Z_2}}$ is a summation over $\Z_2$-valued 1-cochains
$a^{\Z_2}$ and 2-cochains $b^{\Z_2}$ on $B^4$.  But such a theory is invariant
under the background gauge transformation \eq{w2w3gauge}, and cannot cancel the
non-invariant of $\ee^{\ii \pi \int_{M^5} \w_2\w_3}$.

We can add coupling to $\w_2$ and $\w_3$ to fix this problem and obtain the
following boundary theory (with the bulk topological invariant included)
\begin{align}
\label{Z0Z2}
&\ \ \ \
 Z(M^5,B^4=\prt M^5)
\\
&=\sum_{a^{\Z_2},b^{\Z_2} \text{ on } B^4} \ee^{\ii \pi \int_{M^5} \w_2 \w_3 } \ee^{\ii \pi \int_{B^4} a^{\Z_2}\dd b^{\Z_2}
+ a^{\Z_2}\w_3 +\w_2 b^{\Z_2}} .
\nonumber 
\end{align}
Indeed, such a partition function is independent of the branch structure,
since it is invariant under the background gauge transformation \eq{w2w3gauge}.
To see this point, we note that the change in $\w_2$ and $\w_3$ can be absorbed
by $a^{\Z_2}$ and $b^{\Z_2}$.  In other words, the action amplitude $\ee^{\ii
\pi \int_{M^5} \w_2 \w_3 } \ee^{\ii \pi \int_{\prt M^5} a^{\Z_2}\dd b^{\Z_2}+
a^{\Z_2}\w_3 +\w_2 b^{\Z_2}} $ is invariant under the following generalized
transformation
\begin{align} 
\label{dynamical-gauge-a-b-w2-w3}
 a^{\Z_2} & \to a^{\Z_2} + v_1 , & 
 b^{\Z_2} & \to b^{\Z_2} + v_2 
\nonumber \\
\w_2  &\to \w_2 + \dd v_1,  & 
\w_3 & \to \w_3 + \dd v_2.
\end{align}
So \eq{Z0Z2} is a boundary theory of the branch-independent bosonic theory
\eq{Zabww1}.

In fact we can obtain \eq{Z0Z2} directly from \eq{Zabww1}
by assuming $M^5$ to have boundaries and not adding anything on the boundaries:
\begin{align}
\label{ZabwwB}
 Z &=\sum_{a^{\Z_2}, b^{\Z_2} \text{ on } M^5}
\ee^{\ii \pi \int_{M^5} 
(\w_2+\dd a^{\Z_2})
(\w_3+\dd b^{\Z_2})} 
\\
&=2^{N_l^b+N_f^b} 
\hskip -6mm
\sum_{a^{\Z_2},b^{\Z_2} \text{ on } B^4} 
\hskip -6mm
\ee^{\ii \pi \int_{M^5} \w_2 \w_3 } \ee^{\ii \pi \int_{B^4} a^{\Z_2}\dd b^{\Z_2}
+ a^{\Z_2}\w_3 +\w_2 b^{\Z_2}}
\nonumber 
\end{align}
where $N_2^b$ and $N_f^b$ are the number of links and faces in $M^5$
that are not on the boundary $\prt M^5$.

Both \eq{ZabZ2} and \eq{Z0Z2} describe some kinds of $\Z_2$ gauge theories.
However, the two $\Z_2$ gauge theory is very different.  \Eqn{ZabZ2} is
anomaly-free and can be realized by a branch-independent bosonic model in
4d.  In fact \eq{ZabZ2} itself is a branch-independent bosonic model in 4d
that realize the  anomaly-free $\Z_2$ gauge theory.  On the other hand,
\eq{Z0Z2} has an invertible gravitational anomaly.  It can only be realized as
a boundary of 5d invertible topological order.  In our example, \eq{Zabww1} is
a branch-independent bosonic model that realizes the 5d invertible
topological order, and \eq{Z0Z2} is a boundary theory of of the 5d model
\eq{Zabww1}.

Due to different  gravitational anomalies, the two $\Z_2$ gauge theories
\eq{ZabZ2} and \eq{Z0Z2} have different properties.  In the $\Z_2$ gauge
theory \eq{ZabZ2}, the $\Z_2$ gauge charge is a bosonic particle and the
$\Z_2$ gauge flux line behave like a bosonic string.  On the other hand, in the
$\Z_2$ gauge theory \eq{Z0Z2}, the $\Z_2$ gauge charge is a fermionic particle
and  the $\Z_2$ gauge flux line has certain fermionic nature.

To see the above result, we include the worldline of $\Z_2$ gauge charge and
worldsheet of $\Z_2$ gauge flux into the boundary theory \eq{Z0Z2}:
\begin{align}
  &\sum_{a^{\Z_2},b^{\Z_2}} \ee^{\ii \pi \int_{M^5} \w_2 \w_3 } 
\ee^{\ii \pi \int_{B^4} a^{\Z_2}\dd b^{\Z_2}
+ a^{\Z_2}\w_3 +\w_2 b^{\Z_2} }\cdot
\nonumber\\
&\ \ \ \ \ \ \ \ \
\ee^{\ii \pi \int_{B^4}  l_3 a^{\Z_2} + s_2 b^{\Z_2}}
\end{align}
where $l_3$ and $s_2$ are $\Z_2$-valued 3- and 2-cocycles which correspond to
the Poincar\'e dual of the worldline and the worldsheet.  But the above action
with the worldlines and worldsheets is not invariant under the transformation
\eq{dynamical-gauge-a-b-w2-w3} , and thus is not a correct boundary theory
for branch-independent bosonic systems.  We may fix this problem by
considering the following modified boundary theory
\begin{align}
 \sum_{{a^{\Z_2}},{b^{\Z_2}}} & \ee^{\ii \pi \int_{M^5} \w_2 \w_3 } 
\ee^{\ii \pi \int_{B^4} a^{\Z_2}\dd b^{\Z_2}
+ a^{\Z_2}\w_3 + b^{\Z_2}\w_2 } 
\nonumber \\
&
\ee^{\ii \pi \int_{B^4}  l_3 a^{\Z_2} + s_2 b^{\Z_2}}
\ee^{\ii \pi \int_{M^5}  l_3 \w_2  + s_2 \w_3 }.
\end{align}
In the above, we have assumed that $l_3$ and $s_2$ on the boundary $B^4$ can be
extended to the bulk $M^5$ as cocycles.  The invariance of the above path
integral under transformation \eq{dynamical-gauge-a-b-w2-w3} can be checked
directly.

But the above expression also has a problem: it depends on how we extend $l_3$
and $s_2$ on the boundary $B^4$ to the bulk $M^5$. To fix this problem, we
propose the path integral
\begin{align}
\label{ZfullZ2}
& Z  =\sum_{{a^{\Z_2}},{b^{\Z_2}}} \ee^{\ii \pi \int_{M^5} \w_2 \w_3 } 
\ee^{\ii \pi \int_{B^4} a^{\Z_2}\dd b^{\Z_2}
+ a^{\Z_2}\w_3 +\w_2 b^{\Z_2} }
\nonumber \\
&
\ee^{\ii \pi \int_{B^4}  l_3 a^{\Z_2} + s_2 b^{\Z_2}}
\ee^{\ii \pi \int_{M^5}  \Sq^2 l_3 + l_3 \w_2  + \Sq^2 {\Bs_{2}} s_2+s_2 \w_3 },
\end{align}
that is fully invariant under the transformations in
\eq{dynamical-gauge-a-b-w2-w3}.  Here $\Sq$ is the Steenrod square
\eq{Sqdef0} and $\Bs_{2}$ is the generalized Bockstein homomorphism \eq{BsDef}
that act on cocycles (see Appendix \ref{cochain} for details).

Let us first show that the term $\ee^{\ii \pi \int_{M^5}  \Sq^2 l_3 + l_3 \w_2
+ \Sq^2\Bs_{2} s_2+s_2 \w_3 }$ only depends on the fields on the boundary
$B^4=\prt M^5$, so that the above path integral is well-defined.  To do so, we
note that, according to Wu relation for a $\Z_2$-valued cocycle $x_{d-n}$ in the
top $d$-dimension on the complex $M^d$:
\begin{align} \label{eq:Wu-relation}
 \Sq^{n} x_{d-n} \se{2,\dd} u_{n} x_{d-n},
\end{align}
where $u_n$ is Wu class:
\begin{align}
u_0&\se{2,\dd}1, 
\ \ \ \ \
u_1\se{2,\dd}\w_1, 
\ \ \	 \ \
u_2\se{2,\dd}\w_1^2+\w_2, 
	\nonumber\\
u_3&\se{2,\dd}\w_1\w_2, 
\ \ \ \ \
u_4\se{2,\dd}\w_1^4+\w_2^2+\w_1\w_3+\w_4, 
 	\label{eq:Wu-class}\\
u_5&\se{2,\dd}\w_1^3\w_2+\w_1\w_2^2+\w_1^2\w_3+\w_1\w_4, 
	\nonumber\\
u_6&\se{2,\dd}\w_1^2\w_2^2+\w_1^3\w_3+\w_1\w_2\w_3+\w_3^2+\w_1^2\w_4+\w_2\w_4, 
	\nonumber\\
u_7&\se{2,\dd}\w_1^2\w_2\w_3+\w_1\w_3^2+\w_1\w_2\w_4, 
	\nonumber\\
u_8&\se{2,\dd}\w_1^8+\w_2^4+\w_1^2\w_3^2+\w_1^2\w_2\w_4+\w_1\w_3\w_4+\w_4^2
	\nonumber\\
	&\ \ \ \ 
	+\w_1^3\w_5 +\w_3\w_5+\w_1^2\w_6+\w_2\w_6+\w_1\w_7+\w_8. 
\nonumber 
\end{align}
From \eq{eq:Wu-relation} and \eq{eq:Wu-class}, we can show that $\Sq^2 l_3 +
l_3 \w_2$ is a coboundary on oriented $M^5$ with $\w_1\se{2}0$.  Thus $\ee^{\ii
\pi \int_{M^5}  \Sq^2 l_3 + l_3 \w_2}$ only depends on $l_3$ on the boundary
$B^4=\prt M^5$.  

The term $\ee^{\ii \pi \int_{M^5}  \Sq^2 l_3 + l_3 \w_2}$ makes $l_3$ on the
boundary to be a fermion worldline via a higher dimensional bosonization
\cite{BK160501640,W161201418,KT170108264,LW180901112,C191100017}.\footnote{\emph{Higher
dimensional bosonization} here especially in \cite{LW180901112} means to use
the purely bosonic commuting fields (\ie cochains) with only Steenrod algebras
(but without using Grassmann variables) to represent the fermionic systems with
the fermionic parity symmetry $\Z_2^f$.  The fermionic system here may be
regarded as a system with emergent fermions living on the boundary of bosonic
topological order.  See Appendix \ref{hDboson} for a summary.} (For details,
see Appendix \ref{fermsta}.) In other words, the anomalous $\Z_2$ gauge theory
on the boundary has a special property that the $\Z_2$ gauge charge is a
fermion.

{There is another way to understand why the $\Z_2$ gauge charge is a fermion.
The $\w_2\w_3$-topological order, as a bosonic state, can live on any
5-dimensional orientable manifold with a $\SO(5)$ connection for the tangent
bundle.  One way to obtain a gapped boundary of $\w_2\w_3$-topological order is
to trivialize $\w_2$ for the $\SO(5)$ connection on the boundary.  Such a
trivialization can be viewed as a group extension $\Spin(5) = \Z_2 \gext_{\w_2}
\SO(5)$ via
\begin{align} \label{eq:extended}
1 \to \Z_2 \to \Spin(5) \to \SO(5) \to 1.
\end{align} 
This is consistent with the fact that the
Spin manifold is necessary and sufficient condition 
for its second Stiefel-Whitney class $\w_2=0$ for its tangent bundle.
 Trivialize $\w_2$ on the boundary can be thought as promoting the
$\SO(5)$ connection in the bulk to a $\Spin(5)$ connection on the boundary, where
$\Spin(5)$ is a $\Z_2$ central extension of $\SO(5)$.  This implies that the
boundary is described by a twisted $\Z_2$ gauge theory, where the $\Z_2$
connection 1-cochain $a^{\Z_2}$ satisfies
\begin{align}
\label{eq:da=w2}
 \dd a^{\Z_2} \se{2} \w_2 ,
\end{align}
instead of $\dd a^{\Z_2} \se{2} 0$.  
The above relation can be obtained from \eq{Z0Z2} by integrating out
$b^{\Z_2}$ first.
In this case, the $\Z_2$ charge couples to
the $\Spin(5)$ connection on the boundary, instead of $\Z_2\times \SO(5)$
connection.  So the $\Z_2$ charge carries a half-integer spin representation of the spacetime $\Spin$ group if we interpret the extended 
normal $\Z_2$ as the fermion parity $\Z_2^f$ in \eq{eq:extended} as
\begin{align} \label{eq:extended-F}
1 \to \Z_2^f\to \Spin(5) \to \SO(5) \to 1.
\end{align} 
The odd $\Z_2$ gauge charge in \eq{eq:extended} is also the half-integer spin
representation of $\Spin$ group, which is also odd under the fermion parity
$\Z_2^f$ in \eq{eq:extended-F}.  Then according to the usual lattice belief in
terms of the spacetime-spin statistics relation, the half-integer spin
representation of this $\Z_2$ gauge charge is also a fermion.  The Spin
structure, which contains the emergent fermion parity $\Z_2^f$ on the boundary,
is also called the emergent dynamical Spin structure
\cite{WangWenWitten2018qoy1810.00844}.  }

However, a 4d $\Z_2$ gauge theory with a fermionic $\Z_2$ gauge charge may not
have gravitational anomaly, since such theory can be realized by a 4d
bosonic model
\begin{align}
 Z &= \sum_{a^{\Z_2}, b^{\Z_2} } \ee^{\ii \pi \int_{B^4} 
b^{\Z_2}(\w_2 +\dd a^{\Z_2})
} .
\end{align}
The action amplitude is
invariant under the following transformation, even when $B^4$ has boundaries 
\begin{align} 
 a^{\Z_2} \to a^{\Z_2} + v_1 , \ \ \
 b^{\Z_2} \to b^{\Z_2} , \ \ \
\w_2  \to \w_2 + \dd v_1 .
\end{align}
After including the worldline of $\Z_2$ charge and the worldsheet of $\Z_2$
flux, the above becomes
\begin{align}
\label{Z2ferm}
 Z  =\sum_{{a^{\Z_2}},{b^{\Z_2}} \text{ on } B^4} 
&
\ee^{\ii \pi \int_{B^4} a^{\Z_2}\dd b^{\Z_2}
 + b^{\Z_2}\w_2 }
 \\
&
\ee^{\ii \pi \int_{B^4}  l_3 a^{\Z_2} + s_2 b^{\Z_2}}
\ee^{\ii \pi \int_{M^5}  \Sq^2 l_3 + l_3 \w_2  },
\nonumber 
\end{align}
which is the path integral description of the anomaly-free $\Z_2$ gauge theory
with fermionic $\Z_2$ charge \cite{LW0316}

Therefore, the $\Z_2$-flux line must also have some special properties as a
realization of the anomaly in the $\Z_2$ gauge theory \eq{Z0Z2}.  Let us first
show that $ \Sq^2\Bs_{2} s_2+s_2 \w_3$ is also a coboundary on $M^5$.  Using
\begin{align}
 \Sq^1 x &\se{2} \Bs_{2} x ,
\nonumber\\
\Sq^1(\w_j) &\se{2,\dd} \w_1\w_j + (j-1) \w_{j+1},
\end{align}
we find that  
\begin{align} \label{eq:Bs2}
\Bs_{2} \w_2\se{2} \Sq^1 \w_2 \se{2,\dd} \w_1\w_2+\w_3.
\end{align}
Noticing $\w_1\se{2}0$, we find that
\begin{align} \label{eq:s2w3}
 s_2 \w_3 \se{2,\dd} s_2 \Bs_{2} \w_2 \se{2}  \w_2 \Bs_{2} s_2 +{\Bs_{2}(\w_2 s_2)}
\se{2,\dd} \w_2 \Bs_{2} s_2.
\end{align}
The first equality uses \eq{eq:Bs2}, and the second equality uses the chain
rule.  The third equality uses
$\beta_2(\w_2s_2)\se{2}\Sq^1(\w_2s_2)\se{2,\dd}\w_1\w_2s_2\se{2}0$, where we
have also used the Wu relation \eq{eq:Wu-relation} and $\w_1\se{2}0$.  Now,
according to \eq{eq:s2w3}, we have $ \w_2 \Bs_{2} s_2+s_2 \w_3\se{2,\dd} 0 $.
Then using \eq{eq:Wu-relation} and $\w_1\se{2} 0$, we have $\Sq^2 \Bs_{2} s_2+
s_2 \w_3   \se{2,\dd} 0$, and $\Sq^2\Bs_{2} s_2+\w_2 \Bs_{2} s_2  \se{2,\dd}
0$.  Thus $\ee^{\ii \pi \int_{M^5}  \Sq^2 \Bs_{2} s_2 + s_2 \w_3 } $ and
$\ee^{\ii \pi \int_{M^5}  \Sq^2 \Bs_{2} s_2 + \w_2\Bs_{2} s_2 } $ only differ
by a surface term on $B^4=\prt M^5$.  Also, $\ee^{\ii \pi \int_{M^5}  \Sq^2
\Bs_{2} s_2 + \w_2\Bs_{2} s_2 } $ itself is a surface term on $B^4=\prt M^5$.

The term $\ee^{\ii \pi \int_{M^5}  \Sq^2 \Bs_{2} s_2 + \w_2\Bs_{2} s_2 } $
makes $\Bs_{2} s_2$ on the boundary to be the Poincar\'e dual of a fermion
worldline via a higher dimensional bosonization
\cite{W161201418,KT170108264,LW180901112}. We note that if the 2d worldsheet
for the $\Z_2$ flux loop is \emph{orientable}, then its Poincar\'e dual $s_2$
is a $\Z$-valued 2-cocycle thus $s_2 \in Z^2(M^5;\Z)$.  In this case $\Bs_{2}
s_2 = \frac12 \dd s_2 =0$ because the cocycle condition imposes $\dd s_2 =0$.  

Therefore a nontrivial $\Bs_{2} s_2$ comes from an \emph{unorientable} 2d
worldsheet.\footnote{Although we require the \emph{oriented} and
\emph{orientable} special orthogonal SO symmetry for this 4d boundary and 5d
bulk theory, we do allow \emph{unorientable} worldsheets on 2d submanifolds.
Earlier we wrote  for \emph{oriented} worldsheet $s_2  \in Z^2(M^5;\Z)$ with
the topological term:
\begin{align} \label{eq:orientable-worldsheets}
\ee^{\ii \pi \int_{M^5}  \Sq^2 \Bs_{2} s_2 + \w_2\Bs_{2} s_2 }
=
\ee^{\ii \pi \int_{M^5}  \Sq^2 \Bs_{2} s_2 + \w_3 s_2}.
\end{align}
However, for \emph{unorientable} worldsheets $s_2  \in Z^2(M^5;\Z_2)$, we can use Steenrod square $\Sq^1$ to rewrite \eq{eq:orientable-worldsheets}
as
\begin{align}
 \ee^{\ii \pi \int_{M^5}  \Sq^2 \Sq^1 s_2 + \w_2\Sq^1s_2 }
 =
\ee^{\ii \pi \int_{M^5}  \Sq^2 \Sq^1 s_2 + \w_3 s_2}.
\end{align}
In Appendix \ref{sec:GeneralizedWu}, we prove a generalized Wu relation
\begin{align} \label{eq:GeneralizedWu}
\Sq^2\Sq^1x_{d-3}=(\w_1^3+\w_3)x_{d-3}
\end{align} 
on a manifold $M^d$ where $\w_i$ is the Stiefel-Whitney class of $M^d$.
}  
On an unorientable worldsheet, we have a worldline that marks
the reversal of the orientation, whose  Poincar\'e dual is $\Bs_{2} s_2$.  So
such an orientation-reversal worldline corresponds to a fermion worldline.

In other words, the anomalous $\Z_2$ gauge theory \eq{Z0Z2} on the boundary has
a special property that an unorientable worldsheet of the $\Z_2$-flux carries a
non-contractable fermionic worldline.  Such a fermionic worldline corresponds
to the orientation reversal loop on the unorientable  worldsheet.

\subsubsection{Trivialization picture}

We have used the trivialization picture to understand the half-integer spin and
the Fermi statistic of the $\Z_2$ charge.  Can we use the similar
trivialization picture to understand the above ``fermionic'' properties of
$\Z_2$-flux lines?

In the above we have associated the Fermi statistic of the $\Z_2$ charge
(described by $l_3$ in spacetime $B^4$) with the topological invariant
\begin{align}
\ee^{\ii \pi \int_{B^4}  l_3 a^{\Z_2} +\ii \pi \int_{M^5}  \Sq^2 l_3 + l_3 \w_2  }, \ \ \ B^4=\prt M^5,
\end{align}
expressed in one higher dimension $M^5$.  Similarly, we can associate the
``Fermi statistic'' of the $\Z_2$ flux line (described by $s_2$ in spacetime
$B^4$) with the topological invariant
\begin{align}
\ee^{\ii \pi \int_{B^4}  s_2 b^{\Z_2} +\ii \pi \int_{M^5}  \Sq^2 \Sq^1 s_2 
+ s_2 \w_3  }, \ \ \ B^4=\prt M^5.
\end{align}
To gain a better understanding of the ``Fermi'' statistics of $\Z_2$ charged
particle and $\Z_2$ flux line, We like to express topological invariants in the
same dimension rather than one-higher dimension.

We start with the path integral \eq{Z0Z2} describing the $\Z_2$ boundary of the
$\w_2\w_3$ invertible bosonic topological order.  Then we add the worldline for
$\Z_2$ charge and worksheet for $\Z_2$ flux line.  But here we assume the
worldline and worldsheet are boundaries.  Thus their Poincar\'e dual's, $l_3$
and $s_2$, are coboundaries
\begin{align}
l_3 = \dd \t l_2,\ \ \ \ s_2 = \dd \t s_1.
\end{align}
Adding the  worldline for $\Z_2$ charge and worksheet for $\Z_2$ flux line to
the boundary spacetime $B^4$, we obtain the following path integral
\begin{align}
Z = \sum_{{a^{\Z_2}},{b^{\Z_2}}} & \ee^{\ii \pi \int_{M^5} \w_2 \w_3 } 
\ee^{\ii \pi \int_{B^4} a^{\Z_2}\dd b^{\Z_2}
+ a^{\Z_2}\w_3 + b^{\Z_2}\w_2 }
\nonumber \\
&
\ee^{\ii \pi \int_{B^4}  (\dd \t l_2) a^{\Z_2} + (\dd \t s_1) b^{\Z_2}}
.
\end{align}
But the new term break the invariance under \eq{dynamical-gauge-a-b-w2-w3},
which to ensure the branch independence.
To fix this, we consider
\begin{align}
Z = \sum_{{a^{\Z_2}},{b^{\Z_2}}} & \ee^{\ii \pi \int_{M^5} \w_2 \w_3 } 
\ee^{\ii \pi \int_{B^4} a^{\Z_2}\dd b^{\Z_2}
+ a^{\Z_2}\w_3 + b^{\Z_2}\w_2 }
\nonumber \\
&
\ee^{\ii \pi \int_{B^4}  (\dd \t l_2) a^{\Z_2} 
+ \t l_2 \w_2
+ (\dd \t s_1) b^{\Z_2}
+\t s_1 \w_3
}
.
\end{align}
However, the fix causes another problem: $\t l_2$ and $\t l_2 +\bar l_2$
described the same worldline if $\bar l_2$ is a $\Z_2$-valued cocycle; $\t s_1$
and $\t s_1 +\bar s_1$ described the same worldsheet if $\bar s_1$ is a
$\Z_2$-valued cocycle.
Therefore, the path integral must be invariant
under the following transformations
\begin{align}
\label{lsbarls}
 \t l_2 &\to \t l_2 +\bar l_2, & \dd \bar l_2 &\se{2} 0; 
\nonumber\\
 \t s_1 &\to \t s_1 +\bar s_1, & \dd \bar s_1 &\se{2} 0.
\end{align}
To have such an invariance, we consider
\begin{align}
\label{Z2B4}
Z &= 
\hskip -0.5em
\sum_{{a^{\Z_2}},{b^{\Z_2}}}  
\hskip -0.5em
\ee^{\ii \pi \int_{M^5} \w_2 \w_3 } 
\ee^{\ii \pi \int_{B^4} a^{\Z_2}\dd b^{\Z_2}
+ a^{\Z_2}\w_3 + b^{\Z_2}\w_2 }
 \\
&
\ee^{\ii \pi \int_{B^4}  (\dd \t l_2) a^{\Z_2} 
+ \t l_2 \w_2 +\gSq^2 \t l_2 
+ (\dd \t s_1) b^{\Z_2}
+\t s_1 \w_3
 +\gSq^2 \gSq^1 \t s_1},
\nonumber
\end{align}
where $\gSq^n$ is the generalized Steenrod square introduced in
\Rf{LW180901112}, which acts on cochains and is defined in \eq{Sqdef}.
$\gSq^n$ is equal to Pontryagin square mod 2 when acting on $n$-cochains.
\jw{Thus the generalized Steenrod square $\gSq^n$ is in general \emph{not} the same 
as the convention Pontryagin square studied in \cite{Pontryagin1942, Whitehead1949}.}

Using \eq{Sqplus}, we find
\begin{align}
&
(\t l_2+\bar l_2) \w_2 +\gSq^2 (\t l_2 +\bar l_2)
\se{2,\dd}
(\t l_2+\bar l_2) \w_2 
+\gSq^2 \t l_2 + \gSq^2 \bar l_2
\nonumber\\
&\se{2,\dd}
(\t l_2+\bar l_2) \w_2 
+\gSq^2 \t l_2 + \w_2 \bar l_2
\se{2,\dd}
\t l_2 \w_2 +\gSq^2 \t l_2 ,
\end{align}
where we have used $ \gSq^2 \bar l_2 \se{2}  \Sq^2 \bar l_2 \se{2,\dd} 
(\w_2+\w_1^2)  \bar l_2$ and $\w_1 \se{2,\dd} 0$ for $B^4$.
This allows us to show the invariance of the path integral \eq{Z2B4} under
$\t l_2 \to \t l_2 +\bar l_2$.

Similarly, 
using \eq{Sqplus} and \eq{Sqd}, we find
\begin{align}
&\ \ \ \
(\t s_1+\bar s_1) \w_3 +\gSq^2 \gSq^1(\t s_1 +\bar s_1)
\nonumber\\
&
\se{2,\dd}
(\t s_1+\bar s_1) \w_3 
+\gSq^2\gSq^1 \t s_1 + \gSq^2\gSq^1 \bar s_1
\nonumber\\
&\se{2,\dd}
\t s_1\w_3 +\bar s_1\gSq^1 \w_2 
+\gSq^2 \gSq^1 \t s_1 + \w_2 \gSq^1 \bar s_1
\end{align}
where we have used $ \gSq^1 \w_2 \se{2,\dd}  \w_3$ when $\w_1 \se{2,\dd} 0$.
Using \eq{Sq1Bs2} and \eq{BsDef}, we have
\begin{align}
&\ \ \ \
\bar s_1\gSq^1 \w_2 + \w_2 \gSq^1 \bar s_1 \se{2}
\bar s_1\Bs_2 \w_2 + \w_2 \Bs_2 \bar s_1 
\nonumber\\
&
\se{2,\dd} (\Bs_2 \w_2) \bar s_1 + \w_2 \Bs_2 \bar s_1
\se{2,\dd} \Bs_2 (\w_2 \bar s_1) \se{\dd} 0
\end{align}
Therefore
\begin{align}
(\t s_1+\bar s_1) \w_3 +\gSq^2 \gSq^1(\t s_1 +\bar s_1)
\se{2,\dd}
\t s_1\w_3 
+\gSq^2 \gSq^1 \t s_1 .
\end{align}
This allows us to show the invariance of the path integral \eq{Z2B4} under $\t
s_1 \to \t s_1 +\bar s_1$.  Thus \eq{Z2B4} is a correct boundary theory for
$\w_2\w_3$ topological order.

Let us examine the $\t l_2$ terms in the theory.  The term $\ee^{\ii \pi
\int_{B^4}   \gSq^2 \t l_2 }$, quadratic in $\t l_2$, gives the $\Z_2$-charge
(described by $l_3=\dd \t l_2$) a Fermi statistics.  The accompanying linear
term $\ee^{\ii \pi \int_{B^4}   \w_2 \t l_2 }$ gives the $\Z_2$-charge a half
integer spin, which is associated with the statement that fermion is related to
the trivialization of $\w_2$ (for details, see Appendix  \ref{fermsta}).

Now let us examine the $\t s_1$ terms.  Due to the very similar structure, we
can say that the term $\ee^{\ii \pi \int_{B^4} \gSq^2 \gSq^1 \t s_1 }$ gives
the $\Z_2$-flux line (described by $s_2=\dd \t s_1$) an unusual statistics.  We
may also say that the accompanying linear term $\ee^{\ii \pi \int_{B^4}   \w_3
\t s_1 }$ is associated with the trivialization of $\w_3$.  So the unusual
statistics of the $\Z_2$-flux line is associated with the trivialization of
$\w_3$, just like the Fermi statistics of a particle is associated with the
trivialization of $\w_2$.

One way to characterize the unusual statistics of the $\Z_2$-flux line is to
notice that $\dd (\gSq^1 \t s_1) = \gSq^1 \dd \t s_1 = \gSq^1 s_2 
\se{2} \Bs_2 s_2 $ describes
the Poincar\'e dual of the orientation reversal line
of the worldsheet of the $\Z_2$-flux line.  
So the orientation reversal line of
the worldsheet behaves like a fermion worldline due to the term $\gSq^2 (\gSq^1 \t s_1)$.


From the relation $\bt_2 \w_2 \se{2,\dd} \w_3$, we see that the trivialization
of $\w_2$ on the boundary, also implies a trivialization of $\w_3$.
The worldsheet $s_2$ of the $\Z_2$-flux line
couples to a 2-cochain $b^{\Z_2}$ via
$\ee^{\ii \pi \int_{B^4} s_2 b^{\Z_2}}$ (see \eq{ZfullZ2}).
For our anomalous $\Z_2$ gauge theory,
the $\w_3$ is trivialized as a coboundary 
via the split into a lower-dimensional cochain
$b^{\Z_2}$, namely 
\begin{align} \label{eq:db=w3}
 \dd b^{\Z_2} \se{2} \w_3.
\end{align}
Such a relation can be obtained from \eq{Z0Z2} by integrating out
$a^{\Z_2}$ first.
{But \eq{eq:db=w3} is not independent from \eq{eq:da=w2}, because 
$\w_3 \se{\dd,2} \Sq^1 \w_2$.}

To summarize, the 4+1D invertible topological order has a boundary described
by\\ 
(1) a dynamical $\Z_2$ gauge theory with gravitational anomaly.\\
(2) In the continuum limit, the gauge charge
transforms as $\Z_2 \gext_{\w_2} \SO(\infty)=\Spin(\infty)$ under the $\Z_2$ gauge
transformation and spacetime rotation.\\ 
(3) Such a $\Z_2$ gauge charge is a fermion in the spin-statistics.  \\ 
(4) The \emph{orientation-reversal} worldline on the \emph{unorientable}
worldsheet of $\Z_2$-flux loop is a fermion worldline.  
But such a fermion worldline does not carry
the $\Z_2$ gauge charge.

\subsubsection{A physical consequence of the fermion-carrying $\Z_2$-flux
worldsheet}

In a usual anomaly-free $\Z_2$-gauge theory, if we proliferate the $\Z_2$-flux
worldsheets in spacetime, we will get new gapped state, which corresponds to a
confined phase $\Z_2$-gauge theory with no topological order.  Why proliferating
the $\Z_2$-flux worldsheet can give rise to a gapped state?  This is because
the path integral amplitude for the $\Z_2$-flux worldsheets is positive. If the
surface tension of the worldsheet is zero, the equal weight superposition of
the worldsheet leads to a short-range correlated state.

For the anomalous $\Z_2$-gauge theory on the boundary of $\w_2\w_3$ topological
order, the path integral amplitude for the $\Z_2$-flux worldsheets is no longer
positive.  It contains a $\pm$ sign from the braiding of the fermionic world
lines carried by the unorientable worldsheet.  In this case, we are not sure
that the proliferation the $\Z_2$-flux worldsheets can give rise to a gapped
$\Z_2$ confined state.  This result is consistent with our expectation that the
anomalous $\Z_2$ gauge theory at the boundary of the $\w_2\w_3$ topological
order cannot have a trivial confined phase.

If we only allow orientable worldsheet of the $\Z_2$-flux lines, then the path
integral amplitude for those  orientable worldsheet can be all positive.  We
can have a phase where the orientable worldsheet proliferate.  But the
proliferation of orientable worldsheet give rise to a $\U(1)$ gauge theory,
instead of $\Z_2$ confined phase.  Such a $\U(1)^f$-gauge boundary of the
$\w_2\w_3$ topological order will be discussed in the next subsection.

\subsection{4d $\U(1)^f$-gauge boundary of the $\w_2\w_3$ topological order}
\label{sec:4dU1}

The anomalous $\Z_2$-gauge theory is only one possible boundaries of the
$\w_2\w_3$ topological order.  In this section, we will discuss another
boundary --- a $\U(1)$-gauge theory with gravitational anomaly. 
To obtain such an anomalous
boundary $\U(1)$-gauge theory, we first rewrite the topological invariant 
$\w_2\w_3\se{2}\w_2 \Sq^1\w_2$ as
\begin{align}
\label{eq:w2Sq1w2}
 Z^\text{top}(M^5) &=\ee^{\ii \pi \int_{M^5} \w_2\Sq^1 \w_2}
\nonumber \\
&=\ee^{\ii \pi \int_{M^5} \w_2\Bs_{2} \w_2}
=\ee^{\ii \pi \int_{M^5} \w_2\frac{1}{2}\dd \w_2} ,
\end{align}
where we have used $\Sq^1 \w_2 \se{2,\dd} \Bs_2\w_2$ (see \eq{Sq1Bs2}) and
$\Bs_2\w_2 = \frac12 \dd \w_2$ (see \eq{BsDef}).  {In \eq{eq:w2Sq1w2},
both $(\w_2) (\Bs_{2} \w_2)$ and $(\w_2)(\frac{1}{2}\dd \w_2)$ pair between the
$\Z_2$-valued $\w_2$ and the $\Z$-valued $(\Bs_{2} \w_2)$ or $(\frac{1}{2}\dd
\w_2)$, which altogether can be well-defined in the $\Z_2$ value.}

We find that if $M^5$ has a $\Spin^c$ structure, then 
{$\Sq^1\w_2 = \Sq^1(c_1 \mod 2) \se{2,\dd} 0$},
thus
$\Bs_{2} \w_2 \se{\dd}0$ and
$Z^\text{top}(M^5)=1$ (See a proof in Appendix \ref{Cobordism-Constraint}'s Remark 3).  
Thus, the $\w_2\w_3$ is \emph{not} a cobordism invariant for the $\Spin^c$ structure.
{In this case, the $\SO(\infty)$ connection on $M^5$ can be
lifted into a 
$\U(1)^f \gext_{\frac12 \w_2} \SO(\infty)$ connection. 
The $\U(1)^f$ implies that 
$$
\U(1)^f \supset \Z_2^f
$$ contains the fermion parity as a normal subgroup.
Here $\U(1)^f
\gext_{\frac12 \w_2} \SO(\infty)$ is the $\U(1)^f$ extension of $\SO(\infty)$ characterized by
$\frac12 \w_2 \in H^2(\B\SO(\infty); \R/\Z )$.}
 
{Such a group extension to a total group 
$\Spin^c \equiv 
\Spin \times_{\Z_2} \U(1)^f
\equiv \frac{\Spin \times \U(1)^f}{{\Z_2^f}}$ via the short exact sequence 
\begin{align} \label{eq:extended-Spinc}
1 \to \U(1)^f \to \Spin^c(5) \to \SO(5) \to 1
\end{align} 
implies the $\w_2\w_3$ in SO 
is trivialized in
$\Spin^c$.
The $\Spin^c$ structure, which contains the emergent $\U(1)^f \supset \Z_2^f$ on the boundary,
is also called the emergent dynamical $\Spin^c$  structure \cite{WangWenWitten2018qoy1810.00844}.}

The above discussions also implies that the $\w_2\w_3$ topological order has
another boundary described by a $\U(1)^f$ gauge theory with gravitational
anomaly.
To write down such a $\U(1)^f$ gauge theory, we start with
a 5d branch-independent bosonic model that realize
the $\w_2\w_3$ invertible topological order:
\begin{align}
 Z = 
\sum_{a^{\R/\Z^f},b^{\Z}} 
&\ee^{\ii \pi \int_{M^5} (\w_2+2\dd a^{\R/\Z^f} )  
(\Bs_2 \w_2 + \dd b^\Z)
} . 
\end{align}
The brunching-independence is ensured by the invariance of the above path
integral under the following  dynamical gauge transformations ($\al_0$) and
background gauge transformations ($v_1$ and $v_2$):
\begin{align}
\label{gaugeU1}
 a^{\RZ^f} & \to a^{\RZ^f} +\dd \al_0 + \frac12 v_1,  \nonumber\\
 b^{\Z} & \to b^{\Z}+ v_2 ,
 \\
\w_2  &\to \w_2 - \dd v_1 - 2 v_2,  \nonumber \\
\Bs_2 \w_2 & \to \Bs_2 \w_2 - {\beta_2 \dd v_1} - \dd v_2  =  \Bs_2 \w_2 -  \dd v_2. \nonumber 
\end{align}
{The $v_1\in C^1(M^5;\Z)$ and $v_2\in C^2(M^5;\Z)$ are $\Z$-valued 1- and 2-cochains and
$\al_0\in C^0(B^4;\RZ)$ is a $\RZ$-valued function (\ie 0-cochain).}
{Note that $\beta_2 \dd v_1=0$ because $\beta_2 = \frac{1}{2} \dd$, while $\frac{1}{2}v_1\in C^1(M^5,\RZ)$, and $\dd\dd = \dd^2=0$ on $C^1(M^5,\RZ)$.}
{Here the gauge transformation of $a^{\RZ^f}$ is related to that of $\frac12a^{\Z_2}$ where the gauge transformation of $a^{\Z_2}$ is in \eq{dynamical-gauge-a-b-w2-w3}.}

When $M^5$ has boundaries and after integrating out $a^{\R/\Z^f},b^{\Z}$ in the
bulk, we obtain the following boundary theory
\begin{align}
 Z = 
\hskip -0.9em
\sum_{a^{\R/\Z^f},b^{\Z}} 
\hskip -0.6em
&\ee^{\ii \pi \int_{M^5} \w_2 \Bs_2 \w_2 } 
 \ee^{\ii 2\pi \int_{B^4} 
a^{\RZ^f} \dd b^{\Z} 
+ a^{\RZ^f} \Bs_{2} \w_2
+\frac12 \w_2 b^\Z 
}
\nonumber\\
 &\ee^{- \frac12 g\int_{B^4}  |b + \dd a^{\RZ^f} + \frac12 \w_2|^2}
\end{align}
where $a^{\RZ^f} \in C^1(B^4;\RZ)$ is a $\RZ$-valued
1-cochain\footnote{Note that the isomorphism $\RZ = \U(1)$, however we use
$\RZ$ to emphasize the group operation is addition as in $\RZ$, instead of
multiplication in $\U(1)$.  Also we denote $\RZ^f = \U(1)^f \supset \Z_2^f$ to
include the fermion parity normal subgroup.} and $b^{\Z} \in C^2(B^4;\Z)$ is a
$\Z$-valued 2-cochain.\footnote{First let us clarify why ${b^{\Z} \dd a^{\RZ^f}
}+  ({\Bs_{2}} \w_2) a^{\RZ^f} +  b^{\Z} (\frac12 \w_2)$ is well-defined in
$\R/\Z$-valued, for the cup product between a $\Z$-valued cohomology class
(e.g., here $b^{\Z}$, ${\Bs_{2}} \w_2$, etc.) and a $\RZ$-valued cohomology
class (e.g., here $\dd a^{\RZ^f}$, $a^{\RZ^f}$, $\frac12 \w_2$, etc.).  \\
{Generally, for $z\in \Z$ and the equivalence class $[{x}] \in \R/\Z$ where the
representative $x\in \R$, we have that the definition $z[{x}] \equiv [{zx}] \in
\R/\Z$ is well-defined since if $[x]=[y] \in \R/\Z$, then $x-y \in \Z$ and
$z(x-y) \in \Z$.  So $[zx]=[zy] \in \R/\Z$, thus we prove that $z[x] \in \R/\Z$
is well-defined.}
Thus we prove that ${b^{\Z} \dd a^{\RZ^f} }+  ({\Bs_{2}} \w_2) a^{\RZ^f} +
b^{\Z} (\frac12 \w_2)$ is well-defined in $\R/\Z$-valued.  }
{We add a gauge invariant term $ \ee^{- \frac12 g\int_{B^4}  |b + \dd
a^{\RZ^f} + \frac12 \w_2|^2} $, which
will produce the Maxwell term after we integrating out the $b^\Z$ field.}
We find that
a small $g$ leads to a semiclassical $\U(1)^f$ gauge theory.  Thus, the above
describes a $\U(1)^f$ gauge theory,\footnote{We can motivate better about the
4d action $\int_{B^4} {b^{\Z} \dd a^{\RZ^f} }$.  Schematically, without any
other source or operator insertions in the path integral, 
we have the equation of motion
$\dd {b^{\Z}} = 0$. 
Although the cocycle (closed) is not necessarily coboundary (exact), 
locally we can write
${b^{\Z}} = \dd {\rm v}^{\RZ^f}$,
since the $b^{\Z} \in C^2(B^4;\Z)$ has the integer quantized electric flux, then ${\rm v}^{\RZ^f} \in
C^1(B^4;{\RZ^f})$ is the dual gauge field (the 't Hooft magnetic gauge field).
Since the pure abelian U(1) gauge theory has the action $\dd {a}^{\RZ^f} \wedge
\star \dd a^{\RZ^f} = \star \dd {\rm v}^{\RZ^f} \wedge \dd {\rm v}^{\RZ^f}$
{with the U(1) gauge coupling suppressed}, we can also regard $b^{\Z}$ as the
integer quantized electric flux $\ointint \dd {\rm v}^{\RZ^f} =\ointint \star
\dd a^{\RZ^f}  \in \Z$  on a closed 2-cycle, while $\star b^{\Z}$  as the
integer quantized magnetic flux $\ointint \star \dd {\rm v}^{\RZ^f} = \ointint
\dd a^{\RZ^f} \in \Z$  on a closed 2-cycle.  Thus in this special case, we may
also treat $\int_{B^4} {b^{\Z} \dd a^{\RZ^f} }$ as $\int_{B^4}\dd {a}^{\RZ^f}
\wedge \star \dd {a}^{\RZ^f} =\int_{B^4}\star \dd {\rm v}^{\RZ^f} \wedge \dd
{\rm v}^{\RZ^f}$. 
{By the Maxwell equations, 
$\dd \dd a^{\RZ^f}=\dd \star \dd a^{\RZ^f}=0$, 
so $\dd a$ is a harmonic form (closed and co-closed) and the Hodge dual $\star \dd a^{\RZ^f}=\dd {\rm v}^{\RZ^f}$ is also a harmonic form. 
By the Hodge theorem, each cohomology class has a unique harmonic representative. 
So Hodge star is an isomorphism from the harmonic representatives of $H^k_{DR}(M)$ to the the harmonic representatives of $H^{n-k}_{DR}(M)$, here we can take $M=B^4$.
Poincar\'e duality says that $\langle\alpha,\beta\rangle=\int_M\alpha\wedge\beta$ is a perfect pairing between $H^k_{DR}(M)$ and $H^{n-k}_{DR}(M)$. So $\int_M\alpha\wedge \star \alpha>0$ for any nonzero $\alpha$ and we can define $\int_{M} |\alpha|^2$ as $\int_M\alpha\wedge \star \alpha$. 
The sum of $|\alpha|^2$ on a 2-simplex over the spacetime simplex is also the action $\int_{M} |\alpha|^2$.\\
For a cohomology class $\upalpha$, the integral $\int_M |\upalpha|^2$ is defined to be $\int_M |\alpha|^2$ for any cocycle representative $\alpha$ of the cohomology class $\upalpha$. In particular, we can choose the harmonic representative. 
For a harmonic form $\alpha$, 
we have $\int_M |\alpha|^2=\int_M \alpha\wedge\star\alpha$ since $\alpha\wedge\star\alpha=|\alpha|^2$.
}
}  with a gravitational anomaly of $\w_2 \w_3$.

In the presence of the 1d worldline of $\U(1)^f$-electric charge (\ie Wilson
line) and the worldline of $\U(1)^f$-magnetic monopole (\ie 't Hooft line) on
the boundary, the boundary theory becomes
\begin{align}
\label{ZU1}
 Z = 
\hskip -0.9em
\sum_{a^{\R/\Z^f},b^{\Z}} 
\hskip -0.6em
&\ee^{\ii \pi \int_{M^5} \w_2 \Bs_2 \w_2 } 
 \ee^{\ii 2\pi \int_{B^4} 
a^{\RZ^f} \dd b^{\Z} 
+ a^{\RZ^f} \Bs_{2} \w_2
+\frac12 \w_2 b^\Z 
}
\nonumber\\
 &\ee^{- \frac12 g\int_{B^4}  
|b + \dd a^{\RZ^f} + \frac12 \w_2 |^2
}
\nonumber\\
&
\ee^{\ii 2\pi \int_{B^4}  l_3^{e\; \Z} a^{\R/\Z^f} + \eta_2^{\R/\Z^f} b^{\Z}}
\nonumber\\
&
\ee^{\ii \pi \int_{M^5}  \Sq^2 l_3^{e\; \Z} + l_3^{e\; \Z} \w_2  + \Sq^2 {\dd} \eta_2^{\R/\Z^f}+\eta_2^{\R/\Z^f} {\dd} \w_2 }
\end{align}
where $l_3^{e\; \Z}$ is a $\Z$-valued 3-cocycle -- the Poincar\'e dual of the Wilson line 
$\ee^{\ii  \oint_{S^1} a^{\R/\Z^f}}$
(the worldline of the $\U(1)^f$ {electric} charge), and $\eta_2^{\R/\Z^f}$ is a $\R/\Z$-valued 2-cocycle, \ie it
satisfies
\begin{align} \label{eq:dual-tHooft}
 \dd \eta_2^{\R/\Z^f} = l^{m\; \Z}_3.
\end{align}
Here $l_3^m$ is a $\Z$-valued cocycle, which is the Poincar\'e dual of 
the 't Hooft line 
$\ee^{\ii  \oint_{S^1} v^{\R/\Z^f}}$ 
(the worldline of $\U(1)^f$ magnetic monopole)
written in terms of the dual gauge field.\footnote{Note that
$\ee^{\ii 2\pi \int_{B^4}  l_3^{e\; \Z} a^{\R/\Z^f} + \eta_2^{\R/\Z^f} b^{\Z}}$ in \eq{ZU1}
can be schematically rewritten 
as
\begin{align}
&\ee^{\ii 2\pi \int_{B^4}  l_3^{e\; \Z} a^{\R/\Z^f} + \eta_2^{\R/\Z^f} \dd {\rm v}^{\RZ^f}}
\nonumber\\
&=\ee^{\ii 2\pi \int_{B^4}  l_3^{e\; \Z} a^{\R/\Z^f} + (\dd \eta_2^{\R/\Z^f})  {\rm v}^{\RZ^f}} \nonumber\\
&=\ee^{\ii 2\pi \int_{B^4}  l_3^{e\; \Z} a^{\R/\Z^f} +  l_3^{m\; \Z}  {\rm v}^{\RZ^f}}
\end{align}
via $ b^{\Z} =  \dd {\rm v}^{\RZ^f} =  \star \dd a^{\RZ^f}$.  Hence we can
identify the Poincar\'e dual of the Wilson line ($a^{\R/\Z^f}$) as $l_3^{e\;
\Z}$, while the Poincar\'e dual of the 't Hooft line ($v^{\R/\Z^f}$) as
$l_3^{m\; \Z}$.  }  We note that  $\eta_2^{\R/\Z^f}$ (of $\R/\Z$ or U(1) value)
in \eq{ZU1} is related to $\frac{1}{2}s_2$ where $s_2$ (of $\Z_2$ or $\Z$
value) appears in \eq{ZfullZ2}.\footnote{If we map $\eta_2^{\R/\Z^f} \mapsto
\frac{1}{2}s_2$, both are $\R/\Z$-valued, then we can map between the
expressions in \eq{ZU1} and \eq{ZfullZ2}:
\begin{multline}
\Sq^2 {\dd} \eta_2^{\R/\Z^f}+\eta_2^{\R/\Z^f} {\dd} \w_2
\mapsto
\Sq^2 {\dd} \frac{1}{2}s_2 
+\frac{1}{2}s_2{\dd} \w_2 \\
=
\Sq^2 \Sq^1 s_2 
+s_2 \Sq^1 \w_2
=\Sq^2 (\beta_2 s_2)
+s_2  \w_3
\end{multline}
}

The theory \eq{ZU1} is invariant under the gauge transformation \eq{gaugeU1}.
Also, the term 
$$
\ee^{\ii \pi \int_{M^5}  \Sq^2 l_3^{e\; \Z} + l_3^{e\; \Z} \w_2  + \Sq^2 \dd
\eta_2^{\R/\Z^f}+\eta_2^{\R/\Z^f} \dd \w_2 }
$$ 
only depends on the fields on the boundary $B^4=\prt
M^5$.  Based on the relation \eq{eq:dual-tHooft}, this term can be also read schematically as
\begin{align}
\ee^{
\ii \pi \int_{M^5}  (\Sq^2 + \w_2)( l_3^{e\; \Z} +  l_3^{m\; \Z})
}.
\end{align}
This term makes the $\U(1)^f$ electric charge (Wilson line $\ee^{\ii
\oint_{S^1} a^{\R/\Z^f}}$ as Poincar\'e dual of $ l_3^{e\; \Z}$) and the
$\U(1)^f$ magnetic monopole ('t Hooft line $\ee^{\ii  \oint_{S^1} v^{\R/\Z^f}}$
as Poincar\'e dual of $l_3^{m\; \Z}$)
 to be fermions, via a higher dimensional bosonization
\cite{W161201418,KT170108264,LW180901112}.  Thus, we show that the boundary of
$\w_2\w_3$ topological order has the $\U(1)^f$ electric charge has the
fermionic statistics, the $\U(1)^f$ magnetic monopole has the fermionic
statistics, and their bound object $\U(1)^f$ dyon also has the fermionic
statistics.
This is known as the all fermion quantum electrodynamics (QED$_4$).

Before ending this section, we like to write down two bosonic theories with no
gravitational anomaly.  The first one is given by the following path integral
\begin{align}
 Z = 
\sum_{a^{\R/\Z^f},b^{\Z}} 
&
\ee^{\ii 2\pi \int_{B^4} a^{\RZ^f} \dd b^{\Z} 
-\frac12 g\int_{B^4} b \star b }
\\
&
\ee^{\ii 2\pi \int_{B^4}  l_3^{e\; \Z} a^{\R/\Z^f} + \eta_2^{\R/\Z^f} b^{\Z}}
\nonumber 
\end{align}
which is invariant under the following gauge transformation
\begin{align}
 a^{\RZ^f}  \to a^{\RZ^f} +\dd \al_0 , \ \ \ \ 
 b^{\Z}  \to b^{\Z}  .
\end{align}
When $g$ is small, the above bosonic model realizes a U(1) gauge theory at low
energies, where both electric charge described by $l_3^{e\; \Z}$ and magnetic
charge described by $\dd \eta_2^{m\; \RZ}$ are bosons.

The second bosonic model is given by
\begin{align}
 Z = & 
\hskip -0.9em
\sum_{a^{\R/\Z^f},b^{\Z}} 
\hskip -0.6em
 \ee^{\ii 2\pi \int_{B^4} 
a^{\RZ^f} \dd b^{\Z} 
+\frac12 \w_2 b^\Z }
\nonumber\\
&
\ee^{ - \frac12 g\int_{B^4}  (b +\dd a^{\R/\Z^f} + \frac12 \w_2 )
\star (b +\dd a^{\R/\Z^f} + \frac12 \w_2 )}
\nonumber\\
&
\ee^{\ii 2\pi \int_{B^4}  l_3^{e\; \Z} a^{\R/\Z^f} + \eta_2^{\R/\Z^f} b^{\Z}}
\ee^{\ii \pi \int_{M^5}  \Sq^2 l_3^{e\; \Z} + l_3^{e\; \Z} \w_2  }
\end{align}
which is invariant under the following gauge transformation
\begin{align}
 a^{\RZ^f} & \to a^{\RZ^f} +\dd \al_0 + \frac12 v_1, \ \ \ 
 b^{\Z}  \to b^{\Z} ,
 \nonumber \\
\w_2  &\to \w_2 - \dd v_1 
.
\end{align}
When $g$ is small, the above bosonic model realizes a U(1) gauge theory at
low energies, where the electric charge described by $l_3^{e\; \Z}$ is a
fermion, and  the magnetic charge described by $\dd \eta_2^{m\; \RZ}$ is a
boson.

\begin{widetext}
\onecolumngrid

\section{4d Boundary of SU(2) or Other Spin($n$) Internal Symmetric Theories}

So far we have formulated two kinds of 4d boundary theories of 5d $\w_2 \w_3$: 
the 4d $\Z_2$ gauge theories (where the local $\Z_2$ gauge field is more precisely the dynamical Spin structure summed over in the path integral)
and the 4d $\U(1)$ gauge theories (where the local $\U(1)$ gauge field is more precisely the dynamical $\U(1)$ gauge connection of Spin$^c$ structure summed over in the path integral).
We also have provided these boundary gauge theory constructions as the trivialization of $\w_2 \w_3$ of SO structure via its pullback $p$ to
the $p^*\w_2 \w_3 = 0$ in Spin or Spin$^c$ structures (Namely,
$1 \to \Z_2 \to \Spin \overset{p}{\to} \SO \to 1$
and
$1 \to \U(1)  \to \Spin^c  \overset{p}{\to}  \SO \to 1$
see further details in Appendix \ref{Cobordism}).
The $\Spin^c=\Spin \times_{\Z_2}  \U(1)$ structure constrains
that the fermions carry odd U(1) charge
while the bosons carry even U(1) charge. 
In this section, we consider the $\Spin^h=\Spin \times_{\Z_2}  \SU(2)$ structure
such that the fermions are in even dimensional representation (e.g., ${\bf 2}, {\bf 4}, \dots$; or isospin $\frac{1}{2},\frac{3}{2} ,\dots$) of SU(2)
while the bosons are in odd dimensional representation (e.g., ${\bf 1}, {\bf 3}, \dots$; or isospin $0,1,\dots$)  of SU(2). 
More generally, we can consider the $\Spin \times_{\Z_2}  \Spin(n)$ structure.
For example, for $n=10$, fermions are in the spinor representation of Spin(10) (e.g., ${\bf 16}, \dots$) 
while bosons are in other tensor representation of Spin(10) (e.g., ${\bf 10}, \dots$). 
We list down the cobordism invariants from $\TP_5({\Spin \times_{\Z_2}  \Spin(n)})$ in
Appendix \ref{Cobordism}.
In this section, we summarize and enumerate other 4d boundary theories of 5d $\w_2 \w_3$,
based on the  $\Spin \times_{\Z_2}  \Spin(n)$ structure construction, in particular $n=3$ and $10$.

 \begin{enumerate}

 \item Boundary with the SU(2) and  
 $\Spin^h=\Spin \times_{\Z_2}  \SU(2)=\Spin \times_{\Z_2}  \Spin(3)$ structures:

 \begin{enumerate}

 \item When the SU(2) is a global symmetry, 
 
$\bullet$ An odd number of the 
 fundamental {\bf 2}-dimensional representation (Rep) of SU(2) of the spacetime Weyl spinor $\psi$
cannot be gapped by quadratic mass term while preserving the Lorentz and SU(2) symmetries.
This is due to that the only quadratic mass term
$\epsilon^{\alpha \beta} \epsilon^{ij} \psi_{\alpha i} \psi_{\beta j}=0$ 
(where $\alpha, \beta$ are the Lorentz indices and ${i,j} \in \{1,2\}$ are SU(2) indices; we take both the singlet {\bf 1} out of
${\bf 2} \otimes {\bf 2} = {\bf 1} \oplus {\bf 3}$) vanish due to the fermi statistics.
This suggests a possible anomaly --- another hint is that the fermion spectrum
under the SU(2) gauge bundle over $S^4$ with an instanton number 1 background 
gives an odd number of fermion zero mode.
More generally, an odd number of ${\bf 4}  r +{\bf 2}$ Rep of SU(2) Weyl spinor
has the same anomaly known as Witten anomaly 
as a 't Hooft anomaly of the $\Spin^h$-symmetry
(See \Table{eq:SU2-rep}).
But these 4d theories live on the boundary of another
5d cobordism invariant, known as $\tilde{\eta} \PD(c_2(V_{\SU(2)}))$.\footnote{The 
$\tilde{\eta}$ is a mod 2 index of 1d Dirac operator
from $\TP_1(\Spin)=\Z_2$ or $\Omega_1^{\Spin}=\Z_2$.
A 1d manifold generator for the cobordism invariant $\tilde{\eta}$ is a 1d $S^1$ for fermions with periodic boundary condition,
so called the Ramond circle. A 4d manifold generator for the $c_2(V_{\SU(2)})$ is the nontrivial SU(2) bundle over the $S^4$,
such that the instanton number is 1.
The PD is Poincar\'e dual.
}
These 4d theories do not live on the boundary of the 5d $\w_2 \w_3$.
     
$\bullet$ An odd number of the 
  {\bf 4}-dimensional representation (Rep) of SU(2) of the spacetime Weyl spinor $\Psi$
also cannot be gapped by quadratic mass term while preserving the Lorentz and SU(2) symmetries.
The singlet of both Lorentz and SU(2) symmetry requires any quadratic mass term vanishes:
$\epsilon^{\alpha \beta} C^{IJ} \Psi_{\alpha I} \Psi_{\beta J}=0$.\footnote{Here 
${I,J} \in \{1,2,3,4\}$ are SU(2) indices which correspond 
to isospin-$\frac{3}{2}$  indices  $\{\frac{3}{2},\frac{1}{2},-\frac{1}{2},-\frac{3}{2}\}$.
We take the singlet {\bf 1} out of the tensor product of {\bf 4} of SU(2):
${\bf 4} \otimes {\bf 4} = {\bf 1} \oplus {\bf 3} \oplus {\bf 5} \oplus {\bf 7}$.
Based on the Clebsch-Gordan coefficients $\frac{1}{2} 
(|\frac{3}{2},\frac{3}{2}  \rangle |\frac{3}{2},-\frac{3}{2}  \rangle 
-|\frac{3}{2},\frac{1}{2}  \rangle |\frac{3}{2},-\frac{1}{2}  \rangle 
+ |\frac{3}{2}, -\frac{1}{2}  \rangle |\frac{3}{2},\frac{1}{2}  \rangle 
-|\frac{3}{2},-\frac{3}{2}  \rangle |\frac{3}{2},\frac{3}{2}  \rangle)= |0,0\rangle $, 
we have $C^{ij} \Psi_{\alpha i} \Psi_{\beta j} =\frac{1}{2} 
\big( (\Psi_{\alpha 1} \Psi_{\beta 4}-\Psi_{\alpha 4} \Psi_{\beta 1}) - (\Psi_{\alpha 2} \Psi_{\beta 3}-\Psi_{\alpha 3} \Psi_{\beta 2}) \big)$.
Since $\epsilon^{\alpha \beta}$ and $C^{ij}$ are both anti-symmetric, 
while all $\Psi$ are Grassmannian variables due to fermi statistics, thus 
$\epsilon^{\alpha \beta} C^{ij} \Psi_{\alpha i} \Psi_{\beta j}=0$.}
\Rf{WangWenWitten2018qoy1810.00844} shows that
on a 4d non-spin manifold, the complex projective space $\CP^2$,
with an appropriate large diffeomorphism by complex conjugation the $\CP^2$ coordinates $z_j \to \bar z_j$
and an appropriate SU(2) large gauge transformation,
we can construct a mapping torus
5d Dold manifold  $\CP^2 \rtimes S^1$
such that the large gauge-diffeomorphism is transformed along the fifth dimension.
Moreover, together with the SU(2) bundle, the whole theory is compatible with the $\Spin^h$ structure.
But the path integral gets a $(-1)$ sign under this large gauge-diffeomorphism transformation. 
This odd $(-1)$ non-invariance shows the new SU(2) anomaly.
More generally, an odd number of ${\bf 8}  r +{\bf 4}$ Rep of SU(2) Weyl spinor
has the same anomaly known as the new SU(2) anomaly 
as a 't Hooft anomaly of the $\Spin^h$-symmetry
(See \Table{eq:SU2-rep}).

\begin{table}[!h]
$\begin{array}{c |c c c c c c c c c  |c  c c c}
\hline
\text{SU(2) isospin} &0  & \frac{1}{2} &1  & \frac{3}{2} &2  & \frac{5}{2}&3 & \frac{7}{2} &\text{mod }4 & 2 r +\frac{1}{2} & 4 r +\frac{3}{2} & \text{mod }4 \\
\hline
\text{SU(2) Rep {\bf R} (dim)} & {\bf 1}  & {\bf 2} &{\bf 3}  & {\bf 4}& {\bf 5}  & {\bf 6} &{\bf 7} & {\bf 8} &\text{mod }8 & {\bf 4}  r +{\bf 2} & {\bf 8}  r +{\bf 4}  & \text{mod }8\\
\hline
\text{Witten SU(2) anomaly} & &\checkmark &  & &  &\checkmark& & & & \checkmark & &\\
\hline
\text{New SU(2) anomaly}    &  &  &  & \checkmark & &&& & &  & \checkmark & &\\
\hline
\end{array}
$
\caption{For a single
spacetime Weyl spinor ${\bf 2}_L$ in 4d,
it has a Witten SU(2) anomaly if the spacetime Weyl spinor is also the ${\bf 4}  r +{\bf 2}$-dimensional representation (Rep) of SU(2)  (or isospin $2 r +\frac{1}{2}$), for some nonnegative integer $r$. 
It has a new SU(2) anomaly
if the spacetime Weyl spinor is also the ${\bf 8}  r +{\bf 4}$-dimensional Rep of SU(2)  (or isospin $4 r +\frac{3}{2}$), for some nonnegative integer $r$. 
The checkmark $\checkmark$ means the fermion theory has the corresponding anomaly.
These SU(2) anomalies can be interpreted as either a 't Hooft anomaly of global symmetry (if the SU(2) is global symmetry not gauged),
or dynamical gauge anomaly (if the SU(2) is dynamically gauged).
}
\label{eq:SU2-rep}
\end{table}

 \item When the SU(2) is dynamically gauged, 

$\bullet$ Witten anomaly gives a dynamical gauge anomaly constraint such that
an odd number of ${\bf 4}  r +{\bf 2}$ Rep of SU(2) Weyl spinor coupled to dynamical SU(2) gauge fields are ill-defined.
It is not physically sensible to study its gauge dynamics.

$\bullet$ The gauge theories with new SU(2) anomalies are still well-defined theories with 
well-defined gauge dynamics on Spin manifolds, because $\w_2=0$ means no $\w_2 \w_3$ anomaly on the Spin manifolds.
However, their gauge dynamics become ill-defined in 4d on non-Spin manifolds.
 
\item Let us discuss further about the SU(2) theory with a ${\bf 4}$ Rep of SU(2) Weyl spinor.

$\bullet$ {\bf Explicit symmetry breaking}:
If we are allowed to break this SU(2) theory with the new SU(2) anomaly, for example by choosing the quadratic fermion mass term via the 
$\epsilon^{\alpha \beta} C'^{IJ} \Psi_{\alpha I} \Psi_{\beta J}$
 such that the pairing $C'^{IJ} \Psi_{\alpha I} \Psi_{\beta J}$ selects the
{\bf 3}-dimensional Rep of SU(2)
(which is also the vector Rep of SO(3), and the isospin-1 Rep of SU(2) and SO(3)),
then $C'^{IJ}$ is symmetric under $I \leftrightarrow J$, and such a mass term does not vanish under fermi statistics.\footnote{Here 
we take the {\bf 3} out of the tensor product of {\bf 4} of SU(2):
${\bf 4} \otimes {\bf 4} = {\bf 1} \oplus {\bf 3} \oplus {\bf 5} \oplus {\bf 7}$.
Based on the Clebsch-Gordan coefficients, we may choose 
$\big(\sqrt{\frac{9}{20}}
(|\frac{3}{2},\frac{3}{2}  \rangle |\frac{3}{2},-\frac{3}{2}  \rangle 
+
|\frac{3}{2},-\frac{3}{2}  \rangle |\frac{3}{2},\frac{3}{2}  \rangle)
-\sqrt{\frac{1}{20}}
(|\frac{3}{2},\frac{1}{2}  \rangle |\frac{3}{2},-\frac{1}{2}  \rangle 
+ |\frac{3}{2}, -\frac{1}{2}  \rangle |\frac{3}{2},\frac{1}{2}  \rangle) 
)= |1,0\rangle $, 
we have $C'^{IJ} \Psi_{\alpha I} \Psi_{\beta J} =
\big( \sqrt{\frac{9}{20}} 
(\Psi_{\alpha 1} \Psi_{\beta 4} + \Psi_{\alpha 4} \Psi_{\beta 1}) -
\sqrt{\frac{1}{20}}(\Psi_{\alpha 2} \Psi_{\beta 3}+\Psi_{\alpha 3} \Psi_{\beta 2}) \big)$.
Since $C'^{IJ}$ are symmetric, 
$\epsilon^{\alpha \beta} C'^{IJ} \Psi_{\alpha I} \Psi_{\beta J} \neq 0$.}
This mass term explicitly break the SU(2) down to U(1) symmetry.

$\bullet$ {\bf Spontaneous symmetry breaking}:
Instead we can consider the spontaneous symmetry breaking by introducing the Yukawa-Higgs term
$\epsilon^{\alpha \beta} (\vec{C'}^{IJ} \Psi_{\alpha I} \Psi_{\beta J}) \vec{\Phi}$
such that not merely $(\vec{C'}^{IJ} \Psi_{\alpha I} \Psi_{\beta J})$ is the {\bf 3} of SU(2)
but also the Higgs scalar $\vec{\Phi}$ is also the {\bf 3} of SU(2),
while we again select the SU(2) singlet ${\bf 1}$ out of their tensor product pairing
${\bf 3} \otimes {\bf 3} = {\bf 1} \oplus {\bf 3} \oplus {\bf 5}$.
Then the Higgs condensation $\langle \vec{\Phi} \rangle \neq 0$ only breaks the SU(2) \emph{spontaneously} down to U(1).

$\bullet$ {\bf SU(2) to all-fermion U(1) gauge theories}:  
 \Rf{WangWenWitten2018qoy1810.00844} shows that 
the spherical rotationally symmetric 't Hooft-Polyakov monopole (up to SU(2) gauge transformation)
in the presence of ${\bf 4}$ Rep of SU(2) Weyl spinor
traps fermionic zero modes in the spectrum.
\Rf{WangWenWitten2018qoy1810.00844} shows that the 't Hooft-Polyakov monopole becomes fermion.
This means under the breaking from SU(2) to U(1), such that
\begin{eqnarray}
&&\text{the isospin-$\frac{3}{2}$ Weyl fermion of SU(2)
 becomes the fermionic electric charge 1 of U(1).} \nonumber\\
&&\text{the fermionic 't Hooft-Polyakov monopole of SU(2)
becomes the fermionic magnetic monopole of U(1).}\nonumber
\end{eqnarray}
Together with the fermionic electric charge of U(1),
the dyon of U(1) is also a fermion.
 \Rf{WangWenWitten2018qoy1810.00844} suggests that
a possible renormalization group (RG) flow such that the
 high energy theory is an asymptotic free SU(2) gauge theory
 while the low energy theory is the all-fermion U(1) gauge theory.

The subtlety is that this 4d SU(2) gauge theory is ill-defined on a non-spin manifold and cannot be put on the boundary of 5d $\w_2\w_3$.
But once we break SU(2) down to U(1), the 4d  U(1) gauge theory can be put on a non-spin manifold on the boundary of 5d $\w_2\w_3$.

$\bullet$ {\bf All-fermion U(1) to $\Z_2$ gauge theories}:  
If we introduce another Higgs also the {\bf 3} of SU(2),
with a different vacuum expectation value,
we can further Higgs down the U(1) down to $\Z_2$, such that
\begin{eqnarray}
&&\text{the fermionic electric charge 1 of U(1)
becomes the fermionic electric charge 1 of $\Z_2$.} \nonumber\\
&&\text{the fermionic monopole's 1d 't Hooft line of U(1)
becomes the orientation-reversal 
1d fermionic worldline}\nonumber\\ 
&&\text{on an unorientable 2d worldsheet
 of  $\Z_2$ gauge theory.}\nonumber
\end{eqnarray}

\end{enumerate}

\item Boundary with the Spin(10) and  
 $\Spin \times_{\Z_2}  \Spin(10)$ structures:

 \begin{enumerate}

\item The standard $so(10)$ Grand Unification \cite{Georgi1974syUnityofAllElementaryParticleForces, Fritzsch1974nnMinkowskiUnifiedInteractionsofLeptonsandHadrons}
with Spin(10) internal symmetry group and with Weyl fermions in the {\bf 16} of Spin(10)
does not have the $\w_2 \w_3$ anomaly (more precisely, it is the $\w_2 \w_3 = \w_2' \w_3'$ mixed gauge-gravitational anomaly, which is also 
a nonperturbative global anomaly, with $\w_j =\w_j(TM)$ and $\w_j' =\w_j(V_{\SO(n=10)})$).
This is the only 5d cobordism invariant from  $\TP_5({\Spin \times_{\Z_2}  \Spin(10)})$, thus the only 4d global anomaly for  $\Spin \times_{\Z_2}  \Spin(10)$ structure. 
The absence of $\w_2 \w_3 = \w_2' \w_3'$ anomaly means that the standard $so(10)$ Grand Unification is free from all perturbative local and nonperturbative global anomalies
within $\Spin \times_{\Z_2}  \Spin(10)$ structure \cite{WangWen2018cai1809.11171, WangWenWitten2018qoy1810.00844, WW2019fxh1910.14668}.

\item However, it is possible to construct a modified $so(10)$ Grand Unification with Spin(10) internal symmetry group,
also with Weyl fermions in the {\bf 16} of Spin(10), but with additional discrete torsion class of Wess-Zumino-Witten like term written on the 4d boundary and 5d bulk coupled system \cite{Wang2106.16248,WangYou2111.10369GEQC,YouWang2202.13498}. Here we summarize the results in \cite{Wang2106.16248,WangYou2111.10369GEQC,YouWang2202.13498}:

$\bullet$ When Spin(10) internal symmetry group is treated as a global symmetry, this modified $so(10)$ Grand Unification can live on the boundary of 
5d $\w_2 \w_3 = \w_2' \w_3'$ invertible topological order. The $\w_2 \w_3 = \w_2' \w_3'$ anomaly is saturated by the discrete torsion class of Wess-Zumino-Witten like term alone.
The discrete torsion class of Wess-Zumino-Witten like term gives rise to various possible gauge theory realizations of low energy dynamics in 4d.
The various possible gauge theory realizations are the emergent gauge theories 
(similar to the emergent dynamical Spin structure of the $\Z_2$ gauge theory and emergent dynamical $\Spin^c$ structure of the all-fermion $\U(1)$ gauge theory
that we studied earlier).

$\bullet$ When Spin(10) internal symmetry group is dynamically gauged, the Spin(10) gauge field in the 5d bulk $\w_2 \w_3 = \w_2' \w_3'$
is also gauged. The 5d bulk is no longer a gapped invertible topological order; the 5d bulk becomes gapless and further long-range entangled.
Thus the Spin(10) gauge fields can live only on the 4d boundary, but also propagate into the 5d bulk.

\end{enumerate}

\end{enumerate}

\twocolumngrid

\end{widetext}
\newpage

%
%
%

\section{Acknowledgements} 
This work started in the summer of 2018  during X.-G.W.'s visit at IAS
Princeton, and is based on the unpublished notes of 2018.  We thank Edward
Witten for important discussions and previous collaborations
\cite{WangWenWitten2018qoy1810.00844} in 2018.  JW thanks Dan Freed and Ryan
Thorngren for discussions.  
ZW is supported by the Shuimu Tsinghua Scholar Program. 
JW is supported by NSF Grant DMS-1607871
``Analysis, Geometry and Mathematical Physics'' and Center for Mathematical
Sciences and Applications at Harvard University.  XGW is partially supported by
NSF DMR-2022428 and by the Simons Collaboration on Ultra-Quantum Matter, which
is a grant from the Simons Foundation (651446, XGW).

\newpage

\appendix

\section{Spacetime Complex with Branch Structure, Cochains, Higher Cup
Product} 
\label{cochain}

\subsection{Spacetime complex and branch structure}

In this article, we consider models defined on a spacetime lattice.  A
spacetime lattice is a triangulation of the $d$-dimensional spacetime,
which is denoted by $M^d$.  We will also call the triangulation $M^d$ as a
spacetime complex, which is formed by simplices -- the vertices, links,
triangles, \etc.  We will use $i,j,\cdots$ to label vertices of the spacetime
complex.  The links of the complex (the 1-simplices) will be labeled by
$(i,j),(j,k),\cdots$.  Similarly, the triangles of the complex  (the
2-simplices)  will be labeled by $(i,j,k),(j,k,l),\cdots$.

\begin{figure}[t]
\begin{center}
\includegraphics[scale=0.5]{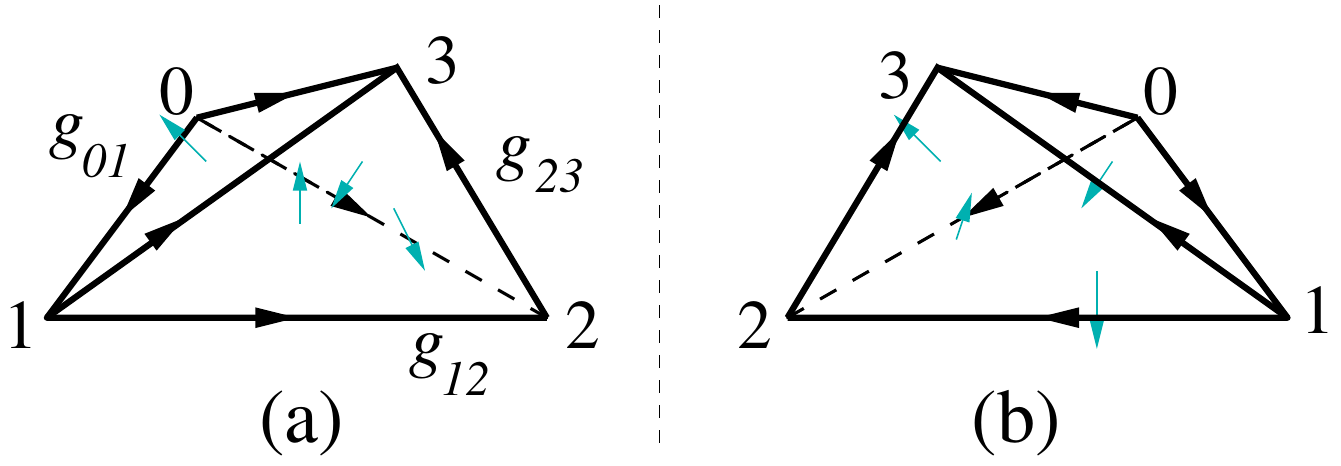} 
\end{center}
\caption{
Two branch simplices with opposite orientations.
(a) A branch simplex with positive orientation and (b) a branch simplex
with negative orientation.  }
\label{mir}
\end{figure}

In order to define a generic lattice theory on the spacetime complex
$M^d$ using local Lagrangian term on each simplex, it is important
to give the vertices of each simplex a local order.  A nice local scheme to
order  the vertices is given by a branch
structure.\cite{C0527,CGL1314,CGL1204} A branch structure is a choice of
orientation of each link in the $d$-dimensional complex so that there is no
oriented loop on any triangle (see Fig. \ref{mir}).

The branch structure induces a \emph{local order} of the vertices on each
simplex.  The first vertex of a simplex is the vertex with no incoming links,
and the second vertex is the vertex with only one incoming link, \etc.  So the
simplex in  Fig. \ref{mir}a has the following vertex ordering: $0,1,2,3$.  We
always use ordered vertices to label a simplex.  So  the simplex in  Fig.
\ref{mir}a and  \ref{mir}b are labeled as $[0,1,2,3]$.

The branch structure also gives the simplex (and its sub-simplices) a
canonical orientation.  Fig. \ref{mir} illustrates two $3$-simplices with
opposite canonical orientations compared with the 3-dimension space in which
they are embedded.  The blue arrows indicate the canonical orientations of the
$2$-simplices.  The black arrows indicate the canonical orientations of the
$1$-simplices.

\subsection{Chain, cochain, cycle, cocycle}

Given an Abelian group $(\M, +)$, 
$\M$ can also be viewed a $\Z$-module (\ie a vector space with
integer coefficient) that also allows scaling by an integer:  
\begin{align}
	 x+y & \in \M,\ \ \ \ x*y \in \M, \ \ \ \ mx = 
\underbrace{x+x +\cdots + x}_{m \text{ terms}} 
\in M,
	\nonumber\\
	x,y & \in \M,\ \ \ m \in \Z.
\end{align}
The direct sum of two modules
$\M_1\oplus \M_2$ (as vector spaces) is equal to the direct product of the two
modules (as sets):
\begin{align}
 \M_1\oplus \M_2 \stackrel{\text{as set}}{=} \M_1\times \M_2
\end{align}

An $n$-cochain $\al_n$ in $M^d$ is a formal
combination of $n$-simplexes in $M^d$ with coefficients in $\M$:
\begin{align}
 \al_n = 
\hskip -0.5em
\sum_{[i,j,\cdots,k]} 
\hskip -0.5em
\al_{n;i,j,\cdots,k} [i,j,\cdots,k],\ \ \
 \al_{n;i,j,\cdots,k} \in \M ,
\end{align}
where $\sum_{[i,j,\cdots,k]}$ sums over all simplexes in $M^d$.  The collection
of all such $n$-chains is denoted as $C_n(M^d;\M)$.  For example, a 2-chain can
be a 2-simplex: $[i,j,k]$, a sum of two 2-simplices: $[i,j,k]+[j,k,l]$, a more
general composition of 2-simplices: $[i,j,k]-2[j,k,l]$, \etc.

\begin{figure}[tb]
\begin{center}
\includegraphics[scale=0.5]{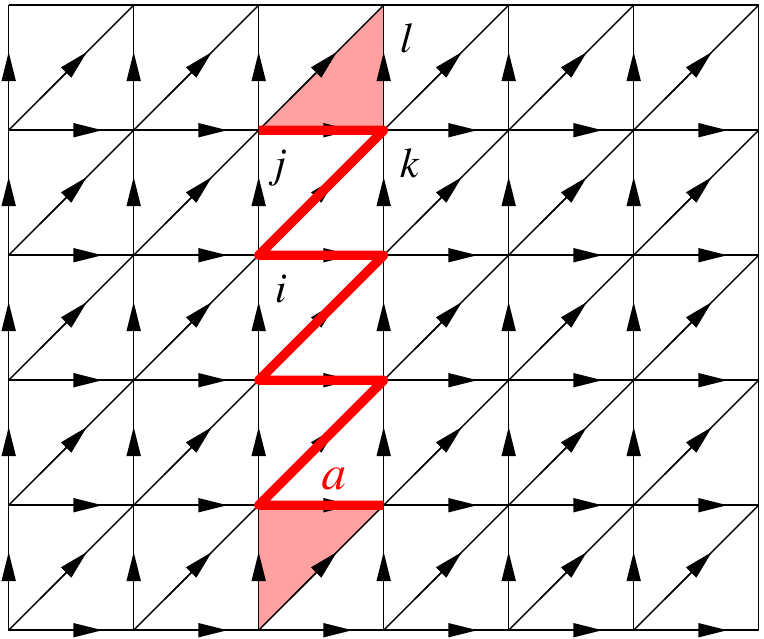} \end{center}
\caption{
A 1-cochain $a$ has a value $1$ on the red links: $ a_{ik}=a_{jk}= 1$ and a
value $0$ on other links: $ a_{ij}=a_{kl}=0 $.  $\dd a$ is non-zero on the
shaded triangles: $(\dd a)_{jkl} = a_{jk} + a_{kl} - a_{jl}$.  For such
1-cohain, we also have $a\smile a=0$.  So when viewed as a $\Z_2$-valued cochain,
$\Bs_2 a \neq a\smile a$ mod 2.
}
\label{dcochain}
\end{figure}

An $n$-cochain $f_n$ in $M^d$ is an assignment of values in $\M$ to each
$n$-simplex $M^d$. For example a value $f_{n;i,j,\cdots,k} \in \M$ is assigned
to $n$-simplex $(i,j,\cdots,k)$ (see Fig. \ref{dcochain}).  We also denote the
value $f_{n;i,j,\cdots,k}$ as $f_n([i,j,\cdots,k])$.  So \emph{a cochain $f_n$
can be viewed as a bosonic field, $f_n([i,j,\cdots,k])$, on the spacetime
lattice $M^d$}. 
To be more
precise $f_n$ is a linear map $f_n: n\text{-chain} \to \M$. We can denote the
linear map as $f_n(n\text{-chain})$, or
\begin{align}
 f_n(\al_n) = \sum_{[i,j,\cdots,k]} \al_{n;i,j,\cdots,k}f_n([i,j,\cdots,k]) \in \M.
\end{align}
where $\sum_{[i,j,\cdots,k]}$ sums over all
simplexes in $M^d$.

We will use $C^n(M^d;\M)$ to denote the set of all
$n$-cochains on $M^d$.  $C^n(M^d;\M)$ can also be
viewed as a set all $\M$-valued fields (or paths) on  $M^d$.  Note
that $C^n(M^d;\M)$ is an Abelian group under the $+$-operation.

The total spacetime lattice $M^d$ correspond to a $d$-chain.  We
will use the same $M^d$ to denote it:
\begin{align}
 M^d = \sum_{[i,j,\cdots,k]} s_{i,j,\cdots,k} [i,j,\cdots,k],
\end{align}
where $s_{i,j,\cdots,k}=\pm 1$, describing the relative orientation between
$M^d$ and $[i,j,\cdots,k]$.  Viewing $f_d$ as a linear map of $d$-chains, we
can define an ``integral'' over $M^d$:
\begin{align}
 \int_{M^d} f_d &\equiv f_d(M^d)
=\sum_{[i,j,\cdots,k]} f_{d;i,j,\cdots,k} s_{i,j,\cdots,k} 
\nonumber\\
&=\sum_{[i,j,\cdots,k]} f_d([i,j,\cdots,k]) s_{i,j,\cdots,k} 
.
\end{align}

\begin{figure}[tb]
\begin{center}
\includegraphics[scale=0.5]{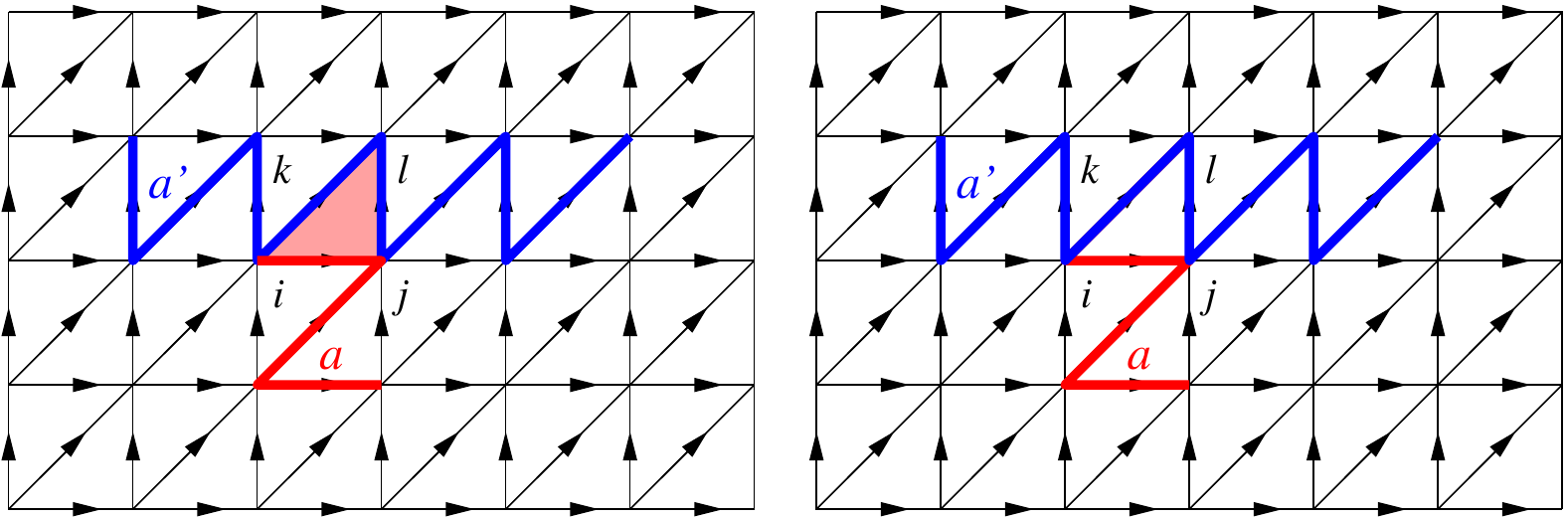} \end{center}
\caption{
A 1-cochain $a$ has a value $1$ on the red links, Another
1-cochain $a'$ has a value $1$ on the blue links.
On the left, $a\smile a'$ is non-zero on the shade triangles:
$(a\smile a')_{ijl}=a_{ij}a'_{jl}=1$.
On the right, $a'\smile a$ is zero on every triangle.
Thus $a\smile a'+a'\smile a$ is not a coboundary.
}
\label{cupcom}
\end{figure}

\subsection{Derivative operator on cochains}

We can define a derivative operator $\dd$ acting on an $n$-cochain $f_n$, which
give us an $(n+1)$-cochain (see Fig. \ref{dcochain}):
\begin{align} 
\label{eq:differential}
&\ \ \ \ (\dd f_n) ([i_0i_1i_2\cdots i_{n+1}])
\nonumber\\
&=\sum_{m=0}^{n+1} (-)^m 
f_n ([i_0i_1i_2\cdots\hat i_m\cdots i_{n+1}])
\end{align}
where $i_0i_1i_2\cdots \hat i_m \cdots i_{n+1}$ is the sequence
$i_0 i_1 i_2 \cdots i_{n+1}$ with $i_m$ removed, and
$i_0, i_1,i_2 \cdots i_{n+1}$ are the ordered vertices of the $(n+1)$-simplex
$(i_0 i_1 i_2 \cdots i_{n+1})$.

A cochain $f_n \in C^n(M^d;\M)$ is called a \emph{cocycle} if $\dd
f_n=0$.   The set of cocycles is denoted by $Z^n(M^d;\M)$.  A
cochain $f_n$ is called a \emph{coboundary} if there exist a cochain $f_{n-1}$
such that $\dd f_{n-1}=f_n$.  The set of coboundaries is denoted by
$B^n(M^d;\M)$.  Both $Z^n(M^d;\M)$ and
$B^n(M^d;\M)$ are Abelian groups as well.  Since $\dd^2=0$, a
coboundary is always a cocycle: $B^n(M^d;\M) \subset
Z^n(M^d;\M)$.  We may view two  cocycles differ by a coboundary as
equivalent.  The equivalence classes of cocycles, $[f_n]$, form the so called
cohomology group denoted by \begin{align} H^n(M^d;\M)=
Z^n(M^d;\M)/ B^n(M^d;\M), \end{align}
$H^n(M^d;\M)$, as a group quotient of $Z^n(M^d;\M)$ by
$B^n(M^d;\M)$, is also an Abelian group.

For the $\Z_N$-valued cochain $x_n$, we lift $\Z_N$ to $\Z$, via $\{0,1,\cdots,
N-1\} \subset \Z_N$ to $\{0,1,\cdots, N-1\} \subset \Z$, and define
\begin{align}
\label{BsDef}
 \Bs_N x_n \equiv \frac1N \dd x_n 
\end{align}
When $x_n$ is a cocycle, we have $\dd x_n \se{N} 0$. In this case, $\Bs_N x_n$
is a $\Z$-valued cocycle, and $\Bs_N$ is Bockstein homomorphism.

\subsection{Cup product and higher cup product}

From two cochains $f_m$  and $h_n$, we can construct a third cochain
$p_{m+n}$ via the cup product (see Fig. \ref{cupcom}):
\begin{align}
p_{m+n} &= f_m \smile h_n ,
\\
p_{m+n} ([0 \to {m+n}]) 
&= 
f_m ([0 \to m])\> 
h_n([m \to {m+n}]),
\nonumber 
\end{align}
where $i\to j$ is the consecutive sequence from $i$ to $j$: 
\begin{align}
i\to j\equiv i,i+1,\cdots,j-1,j. 
\end{align}
Note that the the order of vertices in a simplex $(0 \to m)$ and the notion of
consecutive sequence are determined by the branch structure.  Thus the cup
product (and the higher cup product below) on a simplicial complex can be
defined only after we assign a branch structure to the simplicial complex.
The value of the cup
product depends on the branch structure.

The cup product has the following property 
\begin{align}
\label{cupprop}
 \dd(h_n \smile f_m) &= (\dd h_n) \smile f_m + (-)^n h_n \smile (\dd f_m) 
\end{align}
for  cochains with global or local values.  
We see that $h_n \smile f_m $ is a
cocycle if both $f_m$ and $h_n$ are cocycles.  If both $f_m$ and $h_n$ are
cocycles, then $f_m \smile h_n$ is a coboundary if one of $f_m$ and $h_n$ is a
coboundary.  So the cup product is also an operation on cohomology groups
$\hcup{} : H^m(M^D;\M)\times H^n(M^D;\M) \to H^{m+n}(M^D;\M)$.  The cup product
of two \emph{cocycles} has the following property (see Fig. \ref{cupcom}) 
\begin{align}
 f_m \smile h_n &= (-)^{mn} h_n \smile f_m + \text{coboundary}
\end{align}

We can also define higher cup product $f_m \hcup{k} h_n$ which gives rise to a
$(m+n-k)$-cochain \cite{S4790}:
\begin{align}
\label{hcupdef}
&\ \ \ \
 (f_m \hcup{k} h_n)([0,1,\cdots,m+n-k]) 
\nonumber\\
&
 = 
\hskip -1em 
\sum_{0\leq i_0<\cdots< i_k \leq n+m-k} 
\hskip -3em  
(-)^p
f_m([0 \to i_0, i_1\to i_2, \cdots])\times
\nonumber\\
&
\ \ \ \ \ \ \ \ \ \
\ \ \ \ \ \ \ \ \ \
h_n([i_0\to i_1, i_2\to i_3, \cdots]),
\end{align} 
and $f_m \hcup{k} h_n =0$ for  $k<0$ or for $k>m \text{ or } n$.  Here $i\to j$
is the sequence $i,i+1,\cdots,j-1,j$, and $p$ is the number of permutations to
bring the sequence
\begin{align}
 0 \to i_0, i_1\to i_2, \cdots; i_0+1\to i_1-1, i_2+1\to i_3-1,\cdots
\end{align}
to the sequence
\begin{align}
 0 \to m+n-k.
\end{align}
For example
\begin{align}\label{eq:hcup1}
&
(f_m \hcup1 h_n) ([0\to m+n-1]) 
 = \sum_{i=0}^{m-1} (-)^{(m-i)(n+1)}\times
\nonumber\\
&
f_m([0 \to i, i+n\to m+n-1])
h_n([i\to i+n]).
\end{align} 
We can see that $\hcup0 =\smile$.  
Unlike cup product at $k=0$, the higher cup product $\hcup{k}$ of two
cocycles may not be a cocycle. For cochains $f_m, h_n$, we have
\begin{align}
\label{cupkrel}
& \dd( f_m \hcup{k} h_n)=
\dd f_m \hcup{k} h_n +
(-)^{m} f_m \hcup{k} \dd h_n+
\\
& \ \ \
(-)^{m+n-k} f_m \hcup{k-1} h_n +
(-)^{mn+m+n} h_n \hcup{k-1} f_m 
\nonumber 
\end{align}

Let $f_m$ and $h_n$ be cocycles and $c_l$ be a cochain, from \eq{cupkrel} we
can obtain
\begin{align}
\label{cupkrel1}
 & \dd (f_m \hcup{k} h_n) = (-)^{m+n-k} f_m \hcup{k-1} h_n 
\nonumber\\
&
\ \ \ \ \ \ \ \ \ \
 \ \ \ \ \ \ \
+ (-)^{mn+m+n}  h_n \hcup{k-1} f_m,
\nonumber\\
 & \dd (f_m \hcup{k} f_m) = [(-)^k+(-)^m] f_m \hcup{k-1} f_m,
\nonumber\\
& \dd (c_l\hcup{k-1} c_l + c_l\hcup{k} \dd c_l)
= \dd c_l\hcup{k} \dd c_l 
\nonumber\\
&\ \ \ -[(-)^k-(-)^l]
(c_l\hcup{k-2} c_l + c_l\hcup{k-1} \dd c_l) .
\end{align}

\subsection{Steenrod square and generalized Steenrod square}

From \eq{cupkrel1}, we see that, for $\Z_2$-valued cocycles $z_n$,
\begin{align}
\label{Sqdef0}
 \Sq^{n-k}(z_n) \equiv z_n\hcup{k} z_n
\end{align}
is always a cocycle.  Here $\Sq$ is called the Steenrod square.  More generally
$h_n \hcup{k} h_n$ is a cocycle if $n+k =$ odd and $h_n$ is a cocycle.
Usually, the Steenrod square is defined only for $\Z_2$-valued cocycles or
cohomology classes.  Here, we like to define a generalized
Steenrod square for $\M$-valued
cochains $c_n$:
\begin{align}
\label{Sqdef}
 \gSq^{n-k} c_n \equiv c_n\hcup{k} c_n +  c_n\hcup{k+1} \dd c_n .
\end{align}
From \eq{cupkrel1}, we see that
\begin{align}
\label{Sqd1}
 \dd \gSq^{k} c_n &= \dd(
c_n\hcup{n-k} c_n +  c_n\hcup{n-k+1} \dd c_n )
\\
&= \gSq^k \dd c_n +(-)^{n}
\begin{cases}
0, & k=\text{odd} \\ 
2  \gSq^{k+1} c_n  & k=\text{even} \\ 
\end{cases}
.
\nonumber 
\end{align}
In particular, when $c_n$ is a $\Z_2$-valued cochain, we have
\begin{align}
\label{Sqd}
  \dd \gSq^{k} c_n \se{2} \gSq^k \dd c_n.
\end{align}

Next, let us consider the action of $\gSq^k$ on the sum of two
 $\M$-valued cochains $c_n$ and $c_n'$:
\begin{align}
& \gSq^{k} (c_n+c_n')
 = \gSq^{k} c_n + \gSq^k c_n' +
\nonumber\\
&\ \ \
 c_n \hcup{n-k} c_n' + c_n' \hcup{n-k} c_n 
+ c_n \hcup{n-k+1} \dd c_n' + c_n' \hcup{n-k+1} \dd c_n 
\nonumber\\
&=\gSq^{k} c_n + \gSq^k c_n' 
+[1 + (-)^k]c_n \hcup{n-k} c_n'
\nonumber\\
&\ \ \
-(-)^{n-k} [ - (-)^{n-k} c_n' \hcup{n-k} c_n + (-)^n c_n \hcup{n-k} c_n']
\nonumber\\
&\ \ \
+ c_n \hcup{n-k+1} \dd c_n' + c_n' \hcup{n-k+1} \dd c_n
\end{align}
Notice that (see \eq{cupkrel})
\begin{align}
&\ \ \ \
- (-)^{n-k} c_n' \hcup{n-k} c_n + (-)^n c_n \hcup{n-k} c_n' 
\\
&= \dd(c_n'\hcup{n-k+1}c_n) 
- \dd c_n' \hcup{n-k+1} c_n 
-(-)^n c_n' \hcup{n-k+1} \dd c_n ,
\nonumber 
\end{align}
we see that
\begin{align}
& \gSq^{k} (c_n+c_n')
 = 
\gSq^{k} c_n + \gSq^k c_n' 
+[1 + (-)^k]c_n \hcup{n-k} c_n'
\nonumber\\
&
+(-)^{n-k} [ \dd c_n' \hcup{n-k+1} c_n +(-)^n c_n' \hcup{n-k+1} \dd c_n
]
\nonumber\\
&
-(-)^{n-k} 
\dd (c_n'\hcup{n-k+1}c_n) 
+c_n \hcup{n-k+1} \dd c_n'+ c_n' \hcup{n-k+1} \dd c_n
\nonumber\\
&=
\gSq^{k} c_n + \gSq^k c_n'  
+[1 + (-)^k]c_n \hcup{n-k} c_n'
\nonumber \\
&\  \ \
+[1+(-)^{k}]c_n' \hcup{n-k+1} \dd c_n 
-(-)^{n-k} \dd (c_n'\hcup{n-k+1}c_n)
\nonumber\\
&\ \ \
-[(-)^{n-k+1}\dd c_n' \hcup{n-k+1} c_n
- c_n \hcup{n-k+1} \dd c_n'] .
\end{align}
Notice that (see \eq{cupkrel})
\begin{align}
&\ \ \ \ 
 (-)^{n-k+1}\dd c_n' \hcup{n-k+1} c_n - c_n \hcup{n-k+1} \dd c_n'
\nonumber\\
&= \dd(\dd c_n' \hcup{n-k+2} c_n) +(-)^n \dd c_n' \hcup{n-k+2} \dd c_n  ,
\end{align}
we find
\begin{align}
& \gSq^{k} (c_n+c_n')
=
\gSq^{k} c_n + \gSq^k c_n'  
+[1 + (-)^k]c_n \hcup{n-k} c_n'
\nonumber \\
&\  \ \
+[1+(-)^{k}]c_n' \hcup{n-k+1} \dd c_n 
-(-)^{n-k} \dd (c_n'\hcup{n-k+1}c_n)
\nonumber\\
&\ \ \
-\dd (\dd c_n'\hcup{n-k+2} c_n )
-(-)^{n} \dd c_n'\hcup{n-k+2} \dd c_n 
\nonumber\\
&=
\gSq^{k} c_n + \gSq^k c_n'  
-(-)^{n} \dd c_n'\hcup{n-k+2} \dd c_n 
\nonumber \\
&\ \ \
+[1+(-)^{k}][c_n \hcup{n-k} c_n'+ c_n' \hcup{n-k+1} \dd c_n] 
\nonumber\\
&\ \ \
-(-)^{n-k} \dd (c_n'\hcup{n-k+1}c_n)
-\dd (\dd c_n'\hcup{n-k+2} c_n )
.
\label{Sqplus1}
\end{align}
We see that, if one of the $c_n$ and $c_n'$ is a cocycle,
\begin{align}
\label{Sqplus}
  \gSq^{k} (c_n+c_n') \se{2,\dd} \gSq^{k} c_n + \gSq^k c_n' .
\end{align}
We also see that
\begin{align}
\label{Sqgauge}
&\ \ \ \
 \gSq^{k} (c_n+\dd f_{n-1})
\\
& = \gSq^{k} c_n + \gSq^k \dd f_{n-1} +
[1+(-)^k] \dd f_{n-1}\hcup{n-k} c_n
\nonumber\\
&\ \ \
-(-)^{n-k} \dd (c_n\hcup{n-k+1}\dd f_{n-1})
-\dd (\dd c_n\hcup{n-k+2} \dd f_{n-1} )
\nonumber\\
& = \gSq^{k} c_n 
+ [1+(-)^k] [\dd f_{n-1}\hcup{n-k} c_n +(-)^n \gSq^{k+1}f_{n-1}]
\nonumber\\
&
+\dd [\gSq^k  f_{n-1}
-(-)^{n-k} c_n \hskip -0.5em \hcup{n-k+1} \hskip -0.5em \dd f_{n-1}
-\dd c_n \hskip -0.5em \hcup{n-k+2}  \hskip -0.5em \dd f_{n-1} ]
.
\nonumber 
\end{align}

Using \eq{Sqplus1}, we can also obtain the following result
if $\dd c_n =  \text{even}$
\begin{align}
\label{Sqplus2}
& \ \ \ \
 \gSq^k (c_n+2c_n')
\nonumber\\
& \se{4} \gSq^k c_n+2 \dd (c_n\hcup{n-k+1} c_n') +2 \dd c_n\hcup{n-k+1} c_n'
\nonumber\\
& \se{4} \gSq^k c_n+2 \dd (c_n\hcup{n-k+1} c_n') 
\end{align}

As another application, we note that, for a $\Q$-valued cochain $m_d$ and using
\eq{cupkrel},
\begin{align}
\label{Sq1Bs}
& \gSq^1(m_{d}) = m_{d}\hcup{d-1} m_{d} + m_{d}\hcup{d} \dd m_{d}
\nonumber\\
&=\frac12 (-)^{d} 
[\dd (m_{d}\hcup{d} m_{d}) 
-\dd m_{d} \hcup{d} m_{d}] 
+\frac12  m_{d} \hcup{d} \dd m_{d} 
\nonumber\\
&=
(-)^{d} \Bs_2 (m_{d}\hcup{d} m_{d}) -(-)^d \Bs_2 m_{d} \hcup{d} m_{d}
+  m_{d} \hcup{d} \Bs_2 m_{d}
\nonumber\\
&=
(-)^{d} \Bs_2  \gSq^0 m_{d} 
-2 (-)^d \Bs_2 m_{d} \hcup{d+1} \Bs_2 m_{d}
\nonumber\\
&=
(-)^{d} \Bs_2 \gSq^0 m_{d} 
-2 (-)^d \gSq^0 \Bs_2 m_{d} 
\end{align}
This way, we obtain a relation between Steenrod square and Bockstein
homomorphism, when $m_d$ is a $\Z_2$-valued cocycle 
\begin{align}
\label{Sq1Bs2}
  \gSq^1(m_{d}) \se{2} \Bs_2 m_{d} ,  
\end{align}
where we have used $\gSq^0 m_{d}= m_d$ when the value of the cochain is only
0,1.  

For a $k$-cochain $a_k$, $k=\text{odd}$, we find that
\begin{align}
&\ \ \ \
 \gSq^k a_k = a_ka_k+a_k\hcup{1}\dd a_k
\\
&=
\frac12 [\dd a_k \hcup{1} a_k -a_k \hcup{1}\dd a_k -\dd(a_k\hcup{1}a_k) ]
+a_k\hcup{1}\dd a_k
\nonumber\\
&= \frac12 [\dd a_k \hcup{2}\dd a_k-\dd (\dd a_k\hcup{2} a_k)] 
-\frac12 \dd(a_k\hcup{1}a_k) 
\nonumber\\
&= \frac14 \dd (\dd a_k \hcup{3}\dd a_k) 
-\frac12 \dd(a_k\hcup{1}a_k+\dd a_k\hcup{2} a_k)
\nonumber 
\end{align}
Thus $\gSq^k a_k$ is always a $\Q$-valued coboundary, when $k$ is odd.

\subsection{Branch structure dependence}
\label{branchdep}

Note that the concepts of chain and cochain do not depend on the branch
structure.  Although the definition of the derivative operator $\dd$ formally
depends on the branch structure, in fact, \jw{it is independent of the branch
structure as a map between cochains.}

However, the cup product and higher cup product do depend on the branch structure,
as maps from two cochains to one cochain.  To stress this dependence, we write
a higher cup product as $\hcupB{k}{B}$, where $B$ denotes the branch
structure.  In this section, we like to study this branch structure
dependence.  First we need to find a quantitative way to describe a change of
branch structures.

Let us compare two branch structures $B_0$ and $B$.  We can use a
$\Z_2$-valued 1-cochain $s$ to describe the difference between $B_0$ and $B$:
$s_{ij}=1$ if the arrow on the link $(ij)$ is different for $B_0$ and $B$,
and $s_{ij}=0$ otherwise.  However, not every 1-cochain $s$ corresponds
to the difference between two branch structures. We find that $s$
describes the difference between two branch structures if and only if (iff)
on every triangle $(ijk)$, $i<j<k$, $s$ has a form
\begin{align}\label{eq:diff-branch}
& s_{ij}=1,\ \ s_{jk} =0,\ \ s_{ik} =0; 
\nonumber\\
\text{or }\ \ & s_{ij}=0,\ \ s_{jk} =1,\ \ s_{ik} =0; 
\nonumber\\
\text{or }\ \ & s_{ij}=0,\ \ s_{jk} =1,\ \ s_{ik} =1; 
\nonumber\\
\text{or }\ \ & s_{ij}=1,\ \ s_{jk} =0,\ \ s_{ik} =1; 
\nonumber\\
\text{or }\ \ & s_{ij}=1,\ \ s_{jk} =1,\ \ s_{ik} =1. 
\end{align}
where the
order $i<j<k$ is determined by the base branch structure $B_0$.  Thus, after
we fixed a base branch structure $B_0$, all other branch structure can be
described by $s$.  We may write higher cup product as $\hcupB{k}{s}$.
The higher cup product for the base branch structure $B_0$ is written as
$\hcup{k}$, which correspond to $s=0$.

We believe that, for cocycles $f,g$, $f\hcupB{}{s}g -f\hcupB{}{}g $ is a
coboundary.  Thus
\begin{align}
 f\hcupB{}{s}g + \dd \nu(s,f,g) =f\smile g .
\end{align}
If $f$ and $g$ are 1-cocycles, then we find that
\begin{align}\label{eq:diff-cup}
&f\smile g-f\hcupB{}{s} g\nonumber\\
=& \dd(s\hcup{1} f\hcup{1} g)+2(s\hcup{1} f)\smile g+2f\smile (s\hcup{1} g)\nonumber\\
&-2(s\hcup{1} f)\smile (s\hcup{1} g)-2(s\hcup{1} g)\hcup{1}(s\smile f)\nonumber\\
&-2(s\hcup{1} f)\hcup{1}(g\smile s)+2(s\hcup{1} g)\hcup{1}(s\smile (s\hcup{1} f))\nonumber\\
&+2(s\hcup{1} f)\hcup{1}((s\hcup{1} g)\smile s)
\end{align}
holds on a triangle $(ijk)$ for all the 5 choices of $s$ in \eq{eq:diff-branch}. 
We prove it as follows. The value of the right hand side of \eq{eq:diff-cup} on a triangle $(ijk)$ with the branch structure $B_0$ is
\begin{align}\label{eq:RHS-diff-cup}
&s_{ij}f_{ij}g_{ij}+s_{jk}f_{jk}g_{jk}-s_{ik}f_{ik}g_{ik}\nonumber\\
&+2s_{ij}f_{ij}g_{jk}+2f_{ij}s_{jk}g_{jk}-2s_{ij}f_{ij}s_{jk}g_{jk}\nonumber\\
&+2s_{ik}g_{ik}s_{ij}f_{jk}+2s_{ik}f_{ik}g_{ij}s_{jk}\nonumber\\
&-2(g_{ik}f_{jk}+g_{ij}f_{ik})s_{ij}s_{jk}s_{ik}
\end{align}
where we have used \eq{eq:hcup1}.
\begin{itemize}
\item
If $s_{ij}=1$, $s_{jk} =0$, $s_{ik} =0$, then the value of the left hand side of \eq{eq:diff-cup} on the triangle $(ijk)$ is
$f_{ij}g_{jk}-f_{ji}g_{ik}$, while the value of \eq{eq:RHS-diff-cup} is $f_{ij}g_{ij}+2f_{ij}g_{jk}=f_{ij}g_{jk}-f_{ji}g_{ik}$ since $g_{ij}+g_{jk}-g_{ik}=0$ where we have used the cocycle condition for $g$.

\item
If $s_{ij}=0$, $s_{jk} =1$, $s_{ik} =0$, then the value of the left hand side of \eq{eq:diff-cup} on the triangle $(ijk)$ is
$f_{ij}g_{jk}-f_{ik}g_{kj}$, while the value of \eq{eq:RHS-diff-cup} is $f_{jk}g_{jk}+2f_{ij}g_{jk}=f_{ij}g_{jk}-f_{ik}g_{kj}$ since $f_{ij}+f_{jk}-f_{ik}=0$ where we have used the cocycle condition for $f$.

\item
If $s_{ij}=0$, $s_{jk} =1$, $s_{ik} =1$, then the value of the left hand side of \eq{eq:diff-cup} on the triangle $(ijk)$ is
$f_{ij}g_{jk}-f_{ki}g_{ij}$, while the value of \eq{eq:RHS-diff-cup} is 
\begin{align}
&f_{jk}g_{jk}-f_{ik}g_{ik}+2f_{ij}g_{jk}+2f_{ik}g_{ij}\nonumber\\
=&f_{ij}g_{jk}+f_{ik}g_{ij}+(f_{ij}+f_{jk})g_{jk}+f_{ik}(g_{ij}-g_{ik})\nonumber\\
=&f_{ij}g_{jk}-f_{ki}g_{ij}+f_{ik}g_{jk}-f_{ik}g_{jk}\nonumber\\
=&f_{ij}g_{jk}-f_{ki}g_{ij}
\end{align}
since $f_{ij}+f_{jk}-f_{ik}=0$ and $g_{ij}+g_{jk}-g_{ik}=0$ where we have used the cocycle condition for $f$ and $g$.

\item
If $s_{ij}=1$, $s_{jk} =0$, $s_{ik} =1$, then the value of the left hand side of \eq{eq:diff-cup} on the triangle $(ijk)$ is
$f_{ij}g_{jk}-f_{jk}g_{ki}$, while the value of \eq{eq:RHS-diff-cup} is 
\begin{align}
&f_{ij}g_{ij}-f_{ik}g_{ik}+2f_{ij}g_{jk}+2f_{jk}g_{ik}\nonumber\\
=&f_{ij}g_{jk}+f_{jk}g_{ik}+f_{ij}(g_{ij}+g_{jk})+g_{ik}(f_{jk}-f_{ik})\nonumber\\
=&f_{ij}g_{jk}-f_{jk}g_{ki}+f_{ij}g_{ik}-g_{ik}f_{ij}\nonumber\\
=&f_{ij}g_{jk}-f_{jk}g_{ki}
\end{align}
since $f_{ij}+f_{jk}-f_{ik}=0$ and $g_{ij}+g_{jk}-g_{ik}=0$ where we have used the cocycle condition for $f$ and $g$.

\item
If $s_{ij}=1$, $s_{jk} =1$, $s_{ik} =1$, then the value of the left hand side of \eq{eq:diff-cup} on the triangle $(ijk)$ is
$f_{ij}g_{jk}-f_{kj}g_{ji}$, while the value of \eq{eq:RHS-diff-cup} is 
\begin{align}
&f_{ij}g_{ij}+f_{jk}g_{jk}-f_{ik}g_{ik}+2f_{ij}g_{jk}\nonumber\\
&+2f_{jk}g_{ik}+2f_{ik}g_{ij}-2(g_{ik}f_{jk}+g_{ij}f_{ik})\nonumber\\
=&f_{ij}g_{jk}+f_{ij}(g_{ij}+g_{jk})+f_{jk}g_{jk}-f_{ik}g_{ik}\nonumber\\
=&f_{ij}g_{jk}+(f_{ij}-f_{ik})g_{ik}+f_{jk}g_{jk}\nonumber\\
=&f_{ij}g_{jk}+f_{jk}(g_{jk}-g_{ik})\nonumber\\
=&f_{ij}g_{jk}-f_{jk}g_{ij}\nonumber\\
=&f_{ij}g_{jk}-f_{kj}g_{ji}
\end{align}
since $f_{ij}+f_{jk}-f_{ik}=0$ and $g_{ij}+g_{jk}-g_{ik}=0$ where we have used the cocycle condition for $f$ and $g$.

\end{itemize}
Thus we have proved that \eq{eq:diff-cup} holds on a triangle $(ijk)$ for all the 5 choices of $s$ in \eq{eq:diff-branch}. 
So $f\hcupB{}{}g -f\hcupB{}{s}g $ is a coboundary modulo 2 if $f$ and $g$ are 1-cocycles.

\subsection{Poincar\'e dual and pseudo-inverse of Poincar\'e dual}
\label{sec:Poincaredual}

The Poincar\'e dual of a cochain $ f \in C^n(K;\Z_2)$ is defined to be the cap
product $[K]\frown  f \in C_{m-n}(K;\Z_2)$ where $[K]$ is the fundamental class
of $K$ (the sum modulo 2 of all $m$-simplices of $K$).  The cap product
$\sigma\frown f $ for an $m$-simplex $\sigma=[v_0,\dots,v_n,\dots, v_m]$ and $ f
\in C^n(K;\Z_2)$ is an $(m-n)$-chain, which is defined as:
\begin{align}
\sigma\frown f := f ([v_0,\dots,v_n])[v_n,\dots,v_m].
\end{align}
So the Poincar\'e dual $\PD( f )=[K]\frown  f $ is
\begin{align}\label{eq:Poincare-dual}
\PD( f )=\sum_{[v_0,\dots,v_n,\dots, v_m]} f ([v_0,\dots,v_n])[v_n,\dots,v_m]
\end{align}
where $\sum_{[v_0,\dots,v_n,\dots, v_m]}$ is the sum of all $m$-simplices of
$K$. 

Since the Poincar\'e dual is an isomorphism between cohomology and homology, it
has an pseudo-inverse (defined on cycles and cocycles and up to a boundary or
coboundary).  The pseudo-inverse Poincar\'e dual of a {\em cycle} $\psi\in
C_{m-n}(K;\Z_2)$ is a {\em cocycle} $\overline{\PD}(\psi)\in C^{n}(K;\Z_2)$, which is
defined via its values on all the $n$-simplices $[v_0,\dots,v_n]$:
first assume that no summand of $\psi$ is of the
form $[v,\dots]$ where $v$ is any one of the first $n$ vertices according to the order given by the branch structure.
We can first determine the value of $\overline{\PD}(\psi)$ on the ``minimal'' subset of all the $n$-simplices $[v_0,\dots,v_n]$:
\begin{align}\label{eq:inv-Poincare-dual}
 \overline{\PD}(\psi)([v_0,\dots,v_n])
&= 
\psi_{v_n,\dots,v_m} 
\\
\text{for } \psi &= \sum_{[v_n,\dots,v_m]}
\psi_{v_n,\dots,v_m} [v_n,\dots,v_m]
\nonumber 
\end{align}
where 
$\psi_{v_n,\dots,v_m} = 0,1$.
Here by minimal we mean:
since $\overline{\PD}(\psi)$ has to be a cocycle, its value on any boundary is zero, if we have determined the values of $\overline{\PD}(\psi)$ on the $n$-simplices consist of vertices that are prior according to the order given by the branch structure, then
 we can determine the values of $\overline{\PD}(\psi)$ on other $n$-simplices $[v_0,\dots,v_n]$ and $\overline{\PD}(\psi)$ is defined up to a coboundary.

Note that the
summand of the Poincar\'e dual of any cochain $ f \in C^n(K;\Z_2)$ 
can not be of the form $[v,\dots]$ where $v$ is any one of the first $n$ vertices according to the order given by the branch structure. If $\psi\in C_{m-n}(K;\Z_2)$ is a cycle,
we can modify $\psi$ by a boundary such that no summand of $\psi$ is of the
form $[v,\dots]$ where $v$ is any one of the first $n$ vertices according to the order given by the branch structure. So the definition of the pseudo-inverse Poincar\'e dual is complete.

For example, let $K$ be the surface of a tetrahedron and $|K|=S^2$. Given the
branch structure on $K$ so that the 4 vertices of $K$ are ordered as $v_0$,
$v_1$, $v_2$, and $v_3$, see Figure \ref{figure:tetrahedron}. If
$\psi=[v_0,v_1]+[v_0,v_3]+[v_1,v_2]+[v_2,v_3]$, then modify $\psi$ by a
boundary $[v_0,v_1]+[v_0,v_3]+[v_1,v_3]$ such that no summand of $\psi$ is of
the form $[v_0,\dots]$, we get $\psi'=[v_1,v_2]+[v_1,v_3]+[v_2,v_3]$. By
\eq{eq:inv-Poincare-dual}, $\overline{\PD}(\psi')$ takes value 1 on $[v_0,v_1]$,
so the sum of the values of $\overline{\PD}(\psi')$ on $[v_0,v_2]$ and $[v_1,v_2]$ is 1. 
The values of $\overline{\PD}(\psi')$ on other 1-simplices can also be determined and $\overline{\PD}(\psi')$ is determined up to a coboundary.
By \eq{eq:Poincare-dual},
$\PD(\overline{\PD}(\psi'))=[v_1,v_2]+[v_1,v_3]+[v_2,v_3]$. So $\PD(\overline{\PD}(\psi'))$ and $\psi'$ are equal.

Since $v_0, \dots, v_n,\dots,v_m$ are ordered according to the branch
structure, the Poincar\'e dual of a cochain and the pseudo-inverse Poincar\'e
dual of a chain depends on the branch structure, \ie the same cochain can have
different Poincar\'e duals and the same chain can have different pseudo-inverse
Poincar\'e duals for different branch structures.

\section{Comparison with Standard Mathematical Conventions and Stiefel-Whitney
Class} \label{sec:Conventions}

In the main text of this article, we use the Stiefel-Whitney \emph{cocycle} $\w_n$ and
the Steenrod algebra for cochains, for example, summarized in \Sec{cochain}. In
this section, instead, we make the comparison with the standard mathematical
conventions and Stiefel-Whitney \emph{characteristic class} $w_n$.  {Some of the math
notations/conventions can be found also in the summary of
\cite{WanWang2018bns1812.11967}.  

{Let us define mathematically carefully about Stiefel-Whitney class.} The
Stiefel-Whitney classes of a real vector bundle $\xi:\R^n\to
\mathcal{E}(\xi)\to\mathcal{B}(\xi)$ (here $\mathcal{E}(\xi)$ is the total
space of $\xi$ and $\mathcal{B}(\xi)$ is the {base} of $\xi$) are the
cohomology classes $w_j(\xi)\in H^j( \mathcal{B}(\xi);\Z_2)$ ($j=0,1,2,\dots$)
satisfying the following axioms:\\ A1: $w_0(\xi)=1\in
H^0(\mathcal{B}(\xi);\Z_2)$, $w_j(\xi)=0$, {for} $\forall j>n$.\\ A2: {\bf
Naturality} --- For $f: \mathcal{B}(\xi)\to \mathcal{B}(\eta)$ covered by a
bundle map (so that $\xi=f^*\eta$), $w_j(\xi)=f^*w_j(\eta)$.\\ A3:\;{\bf
Whitney sum formula} --- If $\xi$ and $\eta$ are vector bundles over the same
base $\mathcal{B}$, then $w_k(\xi\oplus \eta)=\sum_{j=0}^kw_j(\xi)\smile
w_{k-j}(\eta)$.\\ A4: For a canonical line bundle {$\gamma_1^1$} over $\RP^1$,
$w_1(\gamma_1^1)\ne0$ (\ie The $\gamma_1^1$ is the M\"obius strip.  The
$\gamma_n^1$ is the canonical line bundle over $\RP^n$).\\

The Steenrod square is $\Sq^{n-k} c_n \equiv c_n\hcup{k} c_n$ for a
$\Z_2$-valued $n$-cohomology class $c_n \in H^n(-,\Z_2)$.  The first Steenrod
square $\Sq^1: H^n(-, \Z_2) \mapsto H^{n+1}(-,\Z_2)$ is the Bockstein
homomorphism associated with the group extension $\Z_2 \to \Z_4 \to \Z_2$.  The
$\beta_2: H^n(-,\Z_2) \mapsto H^{n+1}(-,\Z)$ is the Bockstein homomorphism
associated with the group extension $\Z \to \Z \to \Z_2$.  Poincar\'e dual
means $\PD(B)=B\frown [M]$ where $\frown$ is the cap product, the PD maps a
cohomology class $B$ to a homology class, and $[M]$ is the fundamental class of
the manifold.  So PD is the cap product between a cohomology class and the
fundamental class of the manifold.  The cup product $\smile$ is a product
between a cochain and another cochain.  We shall make the cup product $\smile$
implicit whenever the product is clear written between cochains. }

\section{Emergence of Half-Integer Spin and Fermi Statistics} \label{fermsta}

In the following, we like to explain more carefully why $\Z_2$ gauge charge
current $l_3$ is a fermion current, or why the $\Z_2$ gauge charge is a
fermion.  We like to show that for a twisted $\Z_2$-gauge theory satisfying $
\dd a^{\Z_2} \se{2} \w_2$, the corresponding $\Z_2$ gauge charge is a fermion.
This is because $\dd a^{\Z_2} \se{2} \w_2$ implies that under a combined
$\Z_2$-gauge and $\SO(\infty)$ spacetime rotation transformation, the $\Z_2$
gauge charge transforms as $\Z_2 \gext_{\w_2} \SO(\infty)$.  In other words,
the $\Z_2$ gauge charge couple to a $\Z_2 \gext_{\w_2} \SO(\infty)$ connection
in spacetime.  Let us use
\begin{align}
 (a_{ij}^{\Z_2}, \ga_{ij}),\ \ \ a_{ij}^{\Z_2} \in \Z_2, \ \ \ga_{ij}\in \SO(\infty)
\end{align}
on link $(ij)$ to describe a $\Z_2 \gext_{\w_2} \SO(\infty)$.  Here we use a pair
\begin{align}
 (a^{\Z_2}, \ga),\ \ \ a^{\Z_2} \in \Z_2, \ \ \ga\in \SO(\infty)
\end{align}
to label an element in $\Z_2 \gext_{\w_2} \SO(\infty)$.  The group multiplication in $\Z_2
\gext_{\w_2} \SO(\infty)$ is given by
\begin{align}
 (a_1^{\Z_2}, \ga_1) \ (a_2^{\Z_2}, \ga_2)
&=\big(a_1^{\Z_2}+a_2^{\Z_2}+\w_2(\ga_1,\ga_2) , \ga_1\ga_2 \big)
\end{align}
where $\w_2(\ga_1,\ga_2) \in H^2(B\SO(\infty);\Z_2)$.

For a nearly flat connection $(a_{ij}^{\Z_2}, \ga_{ij})$ on a triangle $(ijk)$,
we have
\begin{align}
 (a_{ij}^{\Z_2}, \ga_{ij}) \ (a_{jk}^{\Z_2}, \ga_{jk})
&=\big(a_{ij}^{\Z_2}+a_{jk}^{\Z_2}+\w_2(\ga_{ij},\ga_{jk}) , \ga_{ij}\ga_{jk} \big)
\nonumber\\
&\approx (a_{ik}^{\Z_2} , \ga_{ik}).
\end{align}
We see that
\begin{align}
 \w_2(\ga_{ij},\ga_{jk}) \se{2}  a_{ik}^{\Z_2} - a_{ij}^{\Z_2} - a_{jk}^{\Z_2} 
\end{align}
which is $ \w_2 \se{2} \dd a^{\Z_2}$.  This way we show that the twisted
$\Z_2$-gauge theory has a $\Z_2$ gauge charge that transforms as $\Z_2 \gext_{\w_2} \SO(\infty) =
\Spin(\infty)$ simply denoted as Spin. In other words, the $\Z_2$ gauge charge carries a half-integer spin, and is a
fermion using the spin-statistics theorem.

We may also compute the statistics of the $\Z_2$ gauge charge directly (which
is phrased as the high dimensional bosonization in
\Rf{W161201418,KT170108264}).  Let us assume the worldline of the $\Z_2$
gauge charge is a \emph{boundary}.  In this case, the Poincar\'e dual of the worldline
is a \emph{coboundary}
\begin{align}
 l_3 \se{2} \dd b_c^{\Z_2}.
\end{align}
Now we can rewrite
\begin{align}
&\ \ \ \
 \ee^{\ii \pi \int_{M^5}  \Sq^2 l_3 + l_3 \w_2  }
= \ee^{\ii \pi \int_{M^5}  \Sq^2 \dd b_c^{\Z_2}  + \dd b_c^{\Z_2} \w_2  }
\nonumber\\
&= \ee^{\ii \pi \int_{M^5}  \gSq^2 \dd b_c^{\Z_2}  + \dd b_c^{\Z_2} \w_2  }
= \ee^{\ii \pi \int_{M^5}  \dd \gSq^2 b_c^{\Z_2}  + \dd b_c^{\Z_2} \w_2  }
\nonumber\\
&= \ee^{\ii \pi \int_{B^4}  \gSq^2 b_c^{\Z_2}  + b_c^{\Z_2} \w_2  }.
\end{align}
Here $\gSq$ is a generalized Steenrod square that acts on a cochain $c_n$
\begin{align}
\label{gSqdef}
 \gSq^{n-k} c_n \equiv c_n\hcup{k} c_n +  c_n\hcup{k+1} \dd c_n ,
\end{align}
where $\hcup{k}$ is the higher cup product.  It has properties
\begin{align}
\label{Sqd2}
&
  \dd \gSq^{k} c_n  \se{2} \gSq^k \dd c_n ,
\nonumber\\
&
  \gSq^{k} (c_n+c_n')   \se{2}
\gSq^{k} c_n + \gSq^k c_n'  
+ \dd c_n'\hcup{n-k+2} \dd c_n 
\nonumber\\
&\ \ \
+ \dd (c_n'\hcup{n-k+1}c_n)
+ \dd (\dd c_n'\hcup{n-k+2} c_n )
\end{align}
Now the term in the path integral that contain $l_3$
becomes
\begin{align}
\label{l3phase}
\ee^{\ii \pi \int_{B^4}  l_3 a^{\Z_2} + \gSq^2 b_c^{\Z_2}  + b_c^{\Z_2} \w_2  },
\end{align}
where $ b_c^{\Z_2}$ is given by $l_3$ via $\dd b_c^{\Z_2} \se{2} l_3$.  
The above phase factor has a gauge invariance
\begin{align}
 \w_2 \to \w_2+\dd u_1,\ \ \ \
a^{\Z_2} \to a^{\Z_2} + u_1.
\end{align}

We note that $b_c^{\Z_2}$ is determined up to a cocycle
$b_0^{\Z_2}$.
Since
\begin{align}
&\ \ \ \
\ee^{\ii \pi \int_{B^4}  l_3 a^{\Z_2} + \gSq^2 (b_c^{\Z_2}+b_0^{\Z_2})  + 
(b_c^{\Z_2}+b_0^{\Z_2}) \w_2  }
\nonumber\\
&=
\ee^{\ii \pi \int_{B^4}  l_3 a^{\Z_2} + \gSq^2 b_c^{\Z_2}+ \Sq^2 b_0^{\Z_2}  + 
(b_c^{\Z_2}+b_0^{\Z_2}) \w_2  }
\nonumber\\
&=
\ee^{\ii \pi \int_{B^4}  l_3 a^{\Z_2} + \gSq^2 b_c^{\Z_2}  + b_c^{\Z_2} \w_2  }
\end{align}
therefore the phase factor \eq{l3phase} does not depend on this $b_0^{\Z_2}$
ambiguity.  In the above, we have used $\Sq^2 b_0^{\Z_2}
+ b_0^{\Z_2} \w_2 \se{2,\dd} 0$ since $\w_1\se{2,\dd}0$.

The linear $l_3$-term in the phase factor
\begin{align}
\ee^{\ii \pi \int_{B^4}  l_3 a^{\Z_2} + b_c^{\Z_2} \w_2  },
\end{align} 
describes the coupling to the background $\Z_2 \gext_{\w_2} \SO(\infty)$-connection,
which indicates that the $\Z_2$ gauge charge carry half-integer spin.  The quadratic
$l_3$-term 
\begin{align}
\ee^{\ii \pi \int_{B^4}  \gSq^2 b_c^{\Z_2}  }
=\ee^{\ii \pi \int_{B^4}   (b_c^{\Z_2})^2 + l_3 \hcup{1} b_c^{\Z_2}  }
\end{align} 
describes the Fermi statistics of the $\Z_2$ gauge charge.  The absence of
$b_0^{\Z_2}$ cocycle ambiguity requires the linear $l_3$-term and the quadratic
$l_3$-term to appear together as a combination \eq{l3phase}.  Similarly, the
WZW-like phase factor $\ee^{\ii \pi \int_{M^5}  \Sq^2 l_3 + \w_2 l_3 }$ will
not depend on how we extend from $B^4$ to $M^5$ only when the linear $l_3$-term
$\w_2 l_3$ and the quadratic $l_3$-term $\Sq^2 l_3$  to appear together.  This
corresponds to the spin statistical theorem.

To summarize, adding a phase factor $\ee^{\ii \pi \int_{M^5}  \Sq^2 l_3 + \w_2
l_3 }$ will make the current $l_3$ on the boundary $B^4=\prt M^5$ to become a
fermion current where the fermions carry a half-integer spin.  Similarly,
adding a phase factor  $\ee^{\ii \pi \int_{M^5}  \Sq^2 \Bs_{2} s_2 + \w_2\Bs_{2} s_2 }$
will make $\Bs_{2} s_2$  on the boundary to become a fermion current as well.

\section{Cobordism Group Data and Anomaly Classification}
\label{Cobordism}

Let us systematically enumerate the pertinent cobordism group $\TP_d(G)$
with some spacetime-internal $G$ symmetry in \Table{table:cobordism}.

\begin{table}[!h]
\centering
\begin{tabular}{c c c c c c c c c}
\hline
$d$ & 1 &  2 & 3& 4 & 5 & 6 & \dots  \\
\hline
$\TP_d(\SO)$ & $0$ & $0$ & $\Z$  & $0$  & $\Z_2$  & $0$  & \dots  \\
\hline
$\TP_d(\Spin)$ & $\Z_2$ & $\Z_2$ & $\Z$  & $0$  & $0$  & $0$  & \dots  \\
\hline
$\TP_d(\Spin \times \U(1))$ & $\Z \times \Z_2$ & $\Z_2$ & $\Z^2$  & $0$  & $\Z^2$  & $0$  & \dots  \\
\hline
$\TP_d(\Spin \times_{\Z_2}  \U(1))$ & $\Z$    & $0$ & $\Z^2$  & $0$  & $\Z^2$  & $0$  & \dots  \\
\hline
$\TP_d(\Spin \times \SU(2))$ & $\Z_2$   & $\Z_2$ & $\Z^2$  & $0$  & $\Z_2$  & $\Z_2$ & \dots    \\
\hline
$\TP_d(\Spin \times_{\Z_2}  \SU(2))$  & 0 & $0$ & $\Z^2$  & $0$  & $\Z_2^2$  & $\Z_2^2$  & \dots  \\
\hline
$\TP_d(\Spin \times  \SO(3))$  & $\Z_2$  & $\Z_2^2$ & $\Z^2$  & $0$  & $0$  & 0  & \dots  \\
\hline
$\begin{array}{l}
\TP_d(\Spin \times  \Spin(n \geq 7)) \\ 
\TP_d(\Spin \times  \Spin(10))
\end{array}$ 
& $\Z_2$ & $\Z_2$ & $\Z^2$  & $0$  & 0  & 0 & \dots  \\
\hline
$\begin{array}{l}
\TP_d(\Spin \times_{\Z_2}  \Spin(n \geq 7 )) \\ 
\TP_d(\Spin \times_{\Z_2}  \Spin(10))
\end{array}$ 
& 0 & $0$ & $\Z^2$  & $0$  & $\Z_2$  & $\Z_2$  & \dots  \\
\hline
$\begin{array}{l}
\TP_d(\Spin \times  \SO(n \geq 7 )) \\ 
\TP_d(\Spin \times  \SO(10))
\end{array}$ 
& $\Z_2$ & $\Z_2^2$ & $\Z^2$  & $\Z_2$  & $0$  & $0$  & \dots  \\
\hline
\end{tabular}
\caption{The cobordism group $\TP_d(G)$ classifies the invertible topological phases or invertible topological  field theories 
(including both the $G$-SPT state and the invertible topological order with $G$-symmetry)
of spacetime-internal symmetry $G$ in the $d$-dimensional spacetime.
See the cobordism computations in 
\cite{WangWen2018cai1809.11171, WanWang2018bns1812.11967, WW2019fxh1910.14668}.
}
\label{table:cobordism}
\end{table}

In particular, we focus on $\TP_5(G)$ which
classifies the invertible topological phases of spacetime-internal symmetry $G$ in the $5$-dimensional spacetime.
We would like to comment why the invertible topological order characterized by $w_2 w_3$ is present or absent 
in the given $G$ symmetry.

Here we denote the
$\w_j= \w_j(TM) =\w_j(a^{\SO})$ as the $j^\text{th}$ Stiefel-Whitney class of the
spacetime tangent bundle ($TM$) of the spacetime manifold $M$,
while $\w_j'=\w_j(V_{\SO(n))})$ is the $j^\text{th}$ Stiefel-Whitney class of the
associated vector bundle of the principal gauge bundle of ${\SO(n)}=\Spin(n)/\Z_2$.

Below let us explain why the cobordism invariant $\w_2 \w_3$ 
vanishes in some $G$ symmetry ($e.g.$, Spin and $\Spin^c \equiv \Spin \times_{\Z_2}  \U(1)$),
but why the $\w_2 \w_3$ persists in other symmetry ($e.g.$, SO and  $\Spin^h=\Spin \times_{\Z_2}  \SU(2)$).

\subsection{Spacetime and gauge bundle constraint}
\label{Cobordism-Constraint}

\begin{enumerate}

\item 
There is no particular constraint on $\w_2$ or  $\w_3$
for the $\SO$ structure and SO manifold, thus the cobordism invariant $\w_2 \w_3$ derived in the cobordism group of an $\SO$ structure still survives.

\item 
The constraint for the $\Spin$ structure and Spin manifold is $\w_2=0$, thus $\w_3= \Sq^1 \w_2=0$, and $\w_2 \w_3=0$.
We had also used the symmetry extension to trivialize the  $\w_2 \w_3$ term in an $\SO$ structure
via a pullback to 0 in a $\Spin$ structure under the group extension $1 \to \Z_2 \to \Spin \to \SO \to 1$.

\item 
The constraint for the $\Spin^c  \equiv \Spin \times_{\Z_2} \U(1)$ structure is $\w_2=c_1 \mod 2$ also $\w_3=0$, so $\w_2 \w_3=0$.
To derive this, we use the $\w_3= \Sq^1\w_2 =\Sq^1(c_1 \mod 2)=0$, since $\Sq^1\rho=0$ where $\rho$ is the mod 2 map.
The $\Sq^1=\rho\beta_2$ where $\beta_2$ is the Bockstein associated with the short exact sequence $1 \to \Z \to \Z \to \Z_2 \to 1$,
which induces the fiber sequence in their classifying spaces as $\dots \to \B^2 \Z \to \B^2 \Z_2  \to \B^3 \Z \to \dots$.
The $\beta_2$ sends the $\rho c_1 =c_1 \mod 2 \in \Z_2$ in $H^2(M;\Z_2)$
to  $\beta_2 \rho c_1 = \beta_2 (c_1 \mod 2 ) \in \Z$ in $H^3(M;\Z)$.
Moreover, 
the group of homotopy classes of the maps from $M$ to the higher classifying space $\B^nG$ is the cohomology group $H^n(M;G)$,
which implies the long exact sequence of cohomology groups
 $\dots \to H^2(M; \Z) \overset{\rho}{\to} H^2(M; \Z_2)  \overset{\beta_2}{\to} H^3(M; \Z) \to \dots$.
This implies that the $\text{im} \rho = \ker \beta_2$, thus $\beta_2\rho=0$.
We derive the $\w_3= \Sq^1\w_2 =\Sq^1(c_1 \mod 2) =\Sq^1(\rho c_1)=\rho \beta_2 (\rho c_1) =\rho (\beta_2 \rho) c_1 =0$, thus $\w_2 \w_3=0$.

The constraint for the $ \Spin \times \U(1)$ structure still requires a Spin manifold ($\w_2=0$) 
with a tensor product structure of spacetime tangent bundle and the principal U(1) gauge bundle,
thus $\w_2 \w_3=0$.

\item
The constraint for the $\Spin^h \equiv \Spin \times_{\Z_2} \SU(2)$ structure includes $\w_2=\w_2' $, 
where we denote $\w_j' =\w_j(V_{\SO(3))})$.
Thus $\Sq^1 \w_2=\Sq^1 \w_2' \Rightarrow \w_3=\w_3'$, so $\w_2\w_3=\w_2' \w_3'$ can be non-zero.

The constraint for the $ \Spin \times \SU(2)$ or $ \Spin \times \SO(3)$ structure still requires a Spin manifold ($\w_2=0$) 
with a tensor product structure of spacetime tangent bundle and the principal SU(2) or SO(3) gauge bundle,
thus $\w_2 \w_3=0$.

\item Now we discuss
$\Spin \times  \Spin(n \geq 7 )$,
$\Spin \times_{\Z_2}  \Spin(n \geq 7 )$
and $\Spin \times  \SO(n \geq 7 )$,
especially when $n=10$ or $18$ suitable for Grand Unified Theories 
\cite{Fritzsch1974nnMinkowskiUnifiedInteractionsofLeptonsandHadrons, WilczekZee1981iz1982Spinors, FujimotoSO1819811982}.

The constraint for the $\Spin \times  \Spin(n \geq 7 )$ and $\Spin \times  \SO(n \geq 7 )$ structure still requires a Spin manifold ($\w_2=0$) 
with a tensor product structure of spacetime tangent bundle and the principal SU(2) or SO(3) gauge bundle,
thus $\w_2 \w_3=0$.

The constraint for the $\Spin \times_{\Z_2}  \Spin(n \geq 7 )$ 
structure includes $\w_2=\w_2' $, 
where we denote $\w_j' =\w_j(V_{\SO(n))})$.
Thus $\Sq^1 \w_2=\Sq^1 \w_2' \Rightarrow \w_3=\w_3'$, so $\w_2\w_3=\w_2' \w_3'$ can be non-zero.

\end{enumerate}

\subsection{Cobordism invariants}

To summarize their 5d cobordism invariants:

\begin{enumerate}

\item $\TP_5(\SO) = \Z_2$
is generated by the cobordism invariant $\w_2 \w_3$.
The manifold generator is a non-Spin manifold 
such as a Wu manifold $$\SU(3)/\SO(3)$$ or 
a Dold manifold  $$\CP^2 \rtimes S^1$$ 
(which identifies the complex conjugation of coordinates in $z \in \CP^2$
with the antipodal inversion of $x \in S^1$, so $(z,x) \sim (\bar{z}, -x)$).

\item $\TP_5(\Spin) = 0$ trivializes the 
cobordism invariant $\w_2 \w_3$ to none via a pullback from $\SO$ to $\Spin$.

\item $\TP_5(\Spin \times  \U(1))=\Z^2$ classes are generated by two 
5d cobordism invariants $a c_1^2$ and $\mu (\PD(c_1))$.
The 5d $a c_1$ corresponds to the perturbative local anomaly captured by Feynman diagram of $\U(1)$-$\U(1)$-$\U(1)$ fields acting on the three vertices of the triangle diagram.
The 5d $\mu (\PD(c_1))$ corresponds to the perturbative local anomaly captured by Feynman diagram of $\U(1)$-$\text{gravity}$-$\text{gravity}$ fields acting on the three vertices of the triangle diagram.

Here the $a$ is the ${\U(1)}$ 1-form gauge connection.
Here the first Chern class $c_1 = c_1(V_{\U(1)})$ is written as the associated vector bundle of $\U(1)$.
The $\mu (\PD(c_1))$ is the 3d Rokhlin invariant of $\PD(c_1)$, where $\PD(c_1)$ is the
submanifold of a Spin 5-manifold which represents the Poincar\'e dual of $c_1$.
{In general, the Poincar\'e dual means
$\PD(B)=B\frown [M]$ where $\frown$ is the cap product, PD maps a cohomology class $B$ to a homology class,
and $[M]$ is the fundamental class of the manifold.}
  
The
$\TP_5(\Spin \times  \U(1))=\Z^2$
are also descended from the two 6d topological invariants
of the bordism group $\Omega_6^{\Spin}(\B \U(1))$: $c_1^3$
and 
$\frac{1}{8} (\sigma (\PD(c_1)) - \rF \cdot \rF)$ from 
the free part of the bordism group $\Omega_6^{\Spin}(\B \U(1))$. 
The $\PD(c_1)$ 
is the submanifold of a Spin 6-manifold which represents the Poincar\'e dual of $c_1=c_1(V_{\U(1)})$.
The $\sigma(\PD(c_1))$ is the signature of the 4-manifold $\PD(c_1)$. 
The $\rF $ 
is a 2d characteristic surface of the 4-manifold $\PD(c_1)$, 
where $\rF$ represents $\PD(B)$ where $B \in H^2(\PD(c_1);\Z)$. 
The $\rF \cdot \rF$ is the intersection form of the 
4-manifold $\PD(c_1)$. 
{The intersection form  $\rF \cdot \rF=\langle B\smile B,[\PD(c_1)]\rangle$ 
is computed via the pairing between a cohomology class with a homology class,
where $[\PD(c_1)]$ is the fundamental class of $\PD(c_1)$.}
By Rokhlin's theorem, $\sigma(\PD(c_1)) - \rF \cdot \rF$ is a multiple of 8
and $\frac{1}{8} (\sigma(\PD(c_1)) - \rF \cdot \rF) ={ {\rm Arf}(\PD(c_1), \rF) \mod 2}$.
The ${\rm Arf}(\PD(c_1), \rF)$ 
is the Arf invariant of a quadratic form $\tilde q: H_1(F;\Z_2) \to \Z_2$, it is $\Z_2$-valued, the LHS is $\Z$-valued and equals to the RHS modulo 2.
The 
$\rF \cdot h=h \cdot h \mod 2$ is true for all $h \in H_2(\PD(c_1);\Z)$.

\item $\TP_5(\Spin \times_{\Z_2}  \U(1))=\TP_5(\Spin^c) = \Z^2$ classes are generated by
two 5d cobordism invariants $\frac{1}{2} a c_1^2$ and {$\frac{1}{48} c_1 \CS_3(TM)$.}
The 5d $\frac{1}{2} a c_1^2$ corresponds to the perturbative local anomaly captured by Feynman diagram of $\U(1)$-$\U(1)$-$\U(1)$ fields acting on the three vertices of the triangle diagram.
The 5d $\frac{1}{48} c_1 \CS_3(TM)$ corresponds to the perturbative local anomaly captured by Feynman diagram of $\U(1)$-$\text{gravity}$-$\text{gravity}$ fields acting on the three vertices of the triangle diagram.

The
$\TP_5(\Spin^c)=\Z^2$
are also descended from the two 6d topological invariants
of the bordism group $\Omega_6^{\Spin^c}$: $\frac{1}{2}c_1^3$
and 
$ \frac{1}{16} \sigma(\PD(c_1))$ from 
the free part of the bordism group $\Omega_6^{\Spin^c}$. 
The $\PD(c_1)$ is a Spin submanifold of the $\Spin^c$
6-manifold which represents the Poincar\'e dual of $c_1$.

\item $\TP_5(\Spin \times  \SU(2))=\Z_2$ class is generated by a 
5d cobordism invariant $\tilde{\eta} \PD(c_2(V_{\SU(2)}))$,
where the $\tilde{\eta}$ is a mod 2 index of 1d Dirac operator
from $\TP_1(\Spin)=\Z_2$ or $\Omega_1^{\Spin}=\Z_2$.
A 1d manifold generator for the cobordism invariant $\tilde{\eta}$ is a 1d $S^1$ for fermions with periodic boundary condition,
so called the Ramond circle. A 4d manifold generator for the $c_2(V_{\SU(2)})$ is the nontrivial SU(2) bundle over the $S^4$,
such that the instanton number is 1.
So the 5d manifold generator for the cobordism invariant $\tilde{\eta} \PD(c_2(V_{\SU(2)}))$
is the $$S^1 \times S^4$$ with the fermion periodic boundary condition on $S^1$ and the SU(2) bundle over $S^4$
with an instanton number 1. The 4d boundary for a 5d $\tilde{\eta} \PD(c_2(V_{\SU(2)}))$ captures the Witten SU(2) anomaly \cite{Witten1982fp}.

\item $\TP_5(\Spin \times_{\Z_2}  \SU(2))=\TP_5(\Spin^h) = \Z_2^2$ classes are generated by
two 5d cobordism invariants. One is the similar cobordism invariant as that of the $\TP_5(\Spin \times  \SU(2))=\Z_2$
whose 4d boundary has the Witten SU(2) anomaly \cite{Witten1982fp}.
The other is the $\w_2 \w_3=\w_2' \w_3'$ with
the $\w_j' =\w_j(V_{\SO(3))})$. 

\item $\TP_5(\Spin \times_{\Z_2}  \Spin(n))=\Z_2$ has a $\Z_2$ class generated 
by $\w_2 \w_3=\w_2' \w_3'$ with
the $\w_j' =\w_j(V_{\SO(n))})$. 
  
\end{enumerate}

\subsubsection{Anomalies in SU(2) vs SO(3): cobordism vs homotopy group}

We note that the 4d nonperturbative global anomalies
of an internal SU(2) symmetric theory on non-spin manifolds are classified by
$$\TP_5(\Spin \times_{\Z_2}  \SU(2))= \Z_2^2,$$
whose generators are the Witten SU(2) anomaly and the new SU(2) anomaly $\w_2 \w_3 =\w_2' \w_3'$;
while on spin manifolds are classified by
$$\TP_5(\Spin \times  \SU(2))=\Z_2,$$
whose generator is the Witten SU(2) anomaly.
The global anomalies
of an internal SO(3) symmetric theory on non-spin manifolds are classified by
$$\TP_5(\SO \times  \SO(3))= \Z_2^2,$$
whose generators are the $\w_2 \w_3$ gravitational anomaly from the $\w_j(TM)$ of spacetime tangent bundle $TM$ 
and the $\w_2' \w_3'$ gauge anomaly from the $\w_j'(V_{\SO(3)})$ of internal gauge bundle;
while on spin manifolds are classified by
$$\TP_5(\Spin \times  \SO(3))=0.$$

We shall compare the cobordism group classification of global anomalies
with the traditional homotopy group analysis \cite{Witten1982fp} of global anomalies.
We will find that the homotopy group analysis is insufficient,
such that the homotopy group sometimes leads to incomplete or misleading results.

For example, in \Table{table:homotopy},
we learn that the homotopy group $\pi_4(\SU(2))=\Z_2$ only gives the Witten SU(2) anomaly  \cite{Witten1982fp}
but misses the new SU(2) anomaly \cite{WangWenWitten2018qoy1810.00844}.
We also learn that the homotopy group $\pi_4(\SO(3))=\Z_2$ gives a possible global anomaly
but in fact the topological invariant in the homotopy theory
does not correspond to any 5d cobordism invariant on Spin manifolds (with $\Spin \times  \SO(3)$ structures).
Thus there is no corresponding Witten SU(2) anomaly in an internal SO(3) symmetric theory.
 
\begin{table}[!h]
\centering
\begin{tabular}{c c c c c c c c c}
\hline
$d$ & 1 &  2 & 3& 4 & 5 & 6 & \dots  \\
\hline
$\pi_d(\SU(2))$ & $0$ & $0$ & $\Z$  & $\Z_2$  & $\Z_2$  & $\Z_{12}$   & \dots  \\
$\pi_d(\SO(3))$ & $\Z_2$ & $0$ & $\Z$  & $\Z_2$   & $\Z_2$  & $\Z_{12}$   & \dots  \\
\hline
\end{tabular}
\caption{The homotopy group $\pi_d(G)$ 
sometimes leads to incomplete or misleading results of global anomalies.
The lesson is that we should use 
{the $d$th cobordism group $\TP_d(G)$
to classify perturbative local and nonperturbative global anomalies in the $(d-1)$d spacetime,
such as the cobordism group data in \Table{table:cobordism}.
See the cobordism computations in 
\cite{WanWang2018bns1812.11967, WW2019fxh1910.14668}.}
}
\label{table:homotopy}
\end{table}

\subsection{Trivialization via group extension}

 \begin{enumerate}

 \item
Trivialization via the pullback $p^*$ in
$$1 \to \Z_2 \to \Spin(5) \overset{p}{\to} \SO(5) \to 1,$$ 
following \eq{eq:extended} and the symmetry extension method \cite{WW170506728},
gauging the normal subgroup $\Z_2$
provides a boundary $\Z_2$ gauge theory construction of the 5d bulk $\w_2 \w_3$.

$\bullet$ The 5d cobordism invariant $\w_2 \w_3$ for the 5d spacetime
with the $\SO$ symmetry becomes trivialized to $0$ for the
spacetime
with the $\Spin$ symmetry, because the Spin structure requires 
$\w_2=0$ and $\w_3= \Sq^1\w_2=0$
on the Spin manifold.

$\bullet$ The $\w_2 \w_3$ of 
$\TP_5(\SO)$ is trivialized as 
$p^*(\w_2 \w_3)=0$ in $\TP_5(\Spin)$.

$\bullet$ So there is \emph{no} $\Spin$ symmetric theory 
with some internal global 
$\Z_2^f$ fermionic parity symmetry in 4d with the boundary anomaly
 of  5d $\w_2 \w_3$.
 
$\bullet$ This means the 4d boundary anomaly of 5d $\w_2 \w_3$ of SO
is also vanished in Spin. However, dynamically gauging the normal $\Z_2$ subgroup
provides a boundary $\Z_2$ gauge theory that preserves the SO symmetry but
with the 't Hooft anomaly of $\w_2 \w_3$.

 \item
Trivialization via the pullback $p^*$ in
$$1 \to \U(1)  \to \Spin^c(5)  \overset{p}{\to}  \SO(5) \to 1,$$ 
following the symmetry extension method \cite{WW170506728},
gauging the normal subgroup $\U(1)$
provides a boundary $\U(1)$ gauge theory construction of the 5d bulk $\w_2 \w_3$.

$\bullet$ The 5d cobordism invariant $\w_2 \w_3$ for the 5d spacetime
with the $\SO$ symmetry becomes trivialized to $0$ for the
spacetime
with the $\Spin^c$ symmetry, because the $\Spin^c$ structure requires 
$\w_2=c_1 \mod 2$ and $\w_3= \Sq^1\w_2=\Sq^1(c_1 \mod 2)=0$
on the Spin$^c$ manifold.

$\bullet$ The $\w_2 \w_3$ of 
$\TP_5(\SO)$ is trivialized as 
$p^*(\w_2 \w_3)=0$ in $\TP_5(\Spin^c)$.

$\bullet$ So there is \emph{no} $\Spin^c$ symmetric theory 
with some internal global $\U(1)$ symmetry in 4d with the boundary anomaly
 of  5d $\w_2 \w_3$.

$\bullet$ This means the 4d boundary anomaly of 5d $\w_2 \w_3$ of SO
is also vanished in $\Spin^c$. 
However, dynamically gauging the normal $\U(1)$ subgroup
provides a boundary $\U(1)$ gauge theory that preserves the SO symmetry but
with the 't Hooft anomaly of $\w_2 \w_3$.

 \item
The group extension
$$1 \to \SU(2) \to \Spin^h(5)  \overset{p}{\to} \SO(5) \to 1$$
however does \emph{not} provide the trivialization of $\w_2 \w_3$.

$\bullet$ The 5d cobordism invariant $\w_2 \w_3$ for the 5d spacetime
with the $\SO$ symmetry becomes $\w_2 \w_3=\w_2' \w_3'$ for the
spacetime
with the $\Spin^h$ symmetry, because the $\Spin^h$ structure requires 
$\w_2=\w_2' $ and $\w_3= \Sq^1\w_2= \Sq^1\w_2' =\w_3'$
on the Spin$^h$ manifold.
This also means the 4d gravitational anomaly on the boundary of 5d $\w_2 \w_3$ term becomes the 4d mixed gauge-gravitational anomaly
on the boundary of 5d $\w_2' \w_3'=\w_2 \w_3$ term.

$\bullet$ The $\w_2 \w_3$ of 
$\TP_5(\SO)$ is \emph{not} trivialized but becomes 
$p^*(\w_2 \w_3)=\w_2 \w_3=\w_2' \w_3'$ in $\TP_5(\Spin^h)$.

$\bullet$ So there indeed exists certain $\Spin^h$ symmetric theory with some 
internal global $\SU(2)$ symmetry in 4d with the boundary 't Hooft anomaly
 of  5d $\w_2' \w_3'= \w_2 \w_3$. 
 In fact, the Weyl fermion 
 as a 2-component spacetime spinor and a 4-component internal $\SU(2)$ spinor,
 in the representation of $(\mathbf{2}_L, \mathbf{4})$ of $
 \Spin \times_{\Z_2} \SU(2) \equiv \Spin^h$
 has this precise so-called new SU(2) anomaly \cite{WangWenWitten2018qoy1810.00844}
 of 5d $\w_2' \w_3'= \w_2 \w_3$.

$\bullet$ This means the 4d boundary anomaly of 5d $\w_2 \w_3$ of SO
does not need to vanish in $\Spin^h$. 
However, we can ask whether it is sensible to 
dynamically gauge the normal $\SU(2)$ subgroup
in this $\Spin^h$ symmetric Weyl fermion 
theory with the new SU(2) anomaly \cite{WangWenWitten2018qoy1810.00844}. \\
--- If we only restrict to the Spin manifold with $\w_2=\w_2'=0$ 
thus also $\w_3=\w_3'=0$, then, yes, we can obtain a well-defined SU(2)
gauge theory on a Spin manifold (such as a flat Euclidean or Minkowski spacetime)
and we can study its dynamics \cite{WangWenWitten2018qoy1810.00844}.\\
--- If we construct this SU(2) gauge theory on a generic 
non-Spin manifold with a $\Spin^h$ structure,
then we have $\w_2=\w_2' \neq 0$ 
thus also $\w_3=\w_3'\neq 0$. Then, no, we obtain an ill-defined SU(2)
gauge theory by summing over the SU(2) bundle with the SU(2) connections 
on a generic non-Spin manifold. 
We cannot study the dynamics
of an ill-defined SU(2) gauge theory 
with dynamical gauge-gravitational anomaly uncanceled
\cite{WangWenWitten2018qoy1810.00844}.

 \item
The group extension
$$1 \to \Spin(n \geq 3) \to \Spin(5) \times_{\Z_2}  \Spin(n \geq 3)   \overset{p}{\to} \SO(5) \to 1$$
however also does \emph{not} provide the trivialization of $\w_2 \w_3$, but modifies the $\w_2 \w_3$ 
to $\w_2 \w_3=\w_2' \w_3'$.
The situation for $n \geq 3$ is similar to our previous remark on 
$\Spin(3)=\SU(2)$.

$\bullet$ In comparison to the $\Spin(2)=\U(1)$ case, there \emph{exists} an all fermion QED$_4$ as a U(1) gauge theory
definable on a generic non-Spin manifold
with a pure 4d gravitational anomaly as a 't Hooft anomaly of the spacetime diffeomorphism SO symmetry
from the 5d $\w_2 \w_3$.

$\bullet$ But for $\Spin(n \geq 3)$,
we do \emph{not} have a $\Spin(n \geq 3)$ gauge theory
--- such a 4d gauge theory
is \emph{not} definable on a generic non-Spin manifold
--- with a pure 4d gravitational anomaly as a 't Hooft anomaly of the spacetime diffeomorphism SO symmetry
from the 5d $\w_2 \w_3$.

$\bullet$  For $\Spin(n \geq 3)$,
we do have a $\Spin(n)$-symmetric theory
 definable on a generic non-Spin manifold
with a 4d mixed gauge-gravitational anomaly as a 't Hooft anomaly of the gauge-diffeomorphism symmetry
from the 5d $\w_2 \w_3=\w_2' \w_3'$.\\
--- Dynamically gauging the $\Spin(n)$ in 4d alone makes sense only on a Spin manifold, which results in a 4d $\Spin(n)$ gauge theory
with a well-defined dynamics on a 4d Spin manifold.\\
--- Dynamically gauging the $\Spin(n)$ in 4d alone on a non-Spin manifold is ill-defined.
But dynamically gauging the $\Spin(n)$ 
on a non-Spin manifold can result in a well-defined 4d-5d coupled fully gauged system.
This 4d-5d coupled system for $\Spin(n=10)$ is studied in \cite{Wang2106.16248,WangYou2111.10369GEQC,YouWang2202.13498}.

\end{enumerate}

\section{Oriented Bordism Groups and Manifold Generators}

In Thom's famous 1954 article \cite{Thom1954}, he showed that the oriented bordism ring is isomorphic to stable homotopy groups of the Thom spectrum $M\SO$: $\Omega_*^{\SO}=\pi_*(M\SO)$.
All of the homotopy groups are a direct sum $\Z^r\oplus\Z_2^s$. Bordism classes of oriented manifolds are completely determined by their Pontryagin and Stiefel-Whitney numbers.
The mod 2 cohomology of $M\SO$ is the same as the mod 2 cohomology of $\B\SO$, a polynomial ring on the Stiefel-Whitney classes $\w_2, \w_3, \dots$ whose Poincar\'e series is
\[\prod_{i\ge2}\frac{1}{1-t^i}.\]
Rationally, the oriented bordism ring is a polynomial algebra $\Q[x_4, x_8, x_{12}, \dots ]$ on generators in degrees that are a multiple of 4. This tells us the rank $r$ of each group. The
Poincar\'e series for the free part of $\Omega_*^{\SO}$ is thus
\[p_{\text{free}}(t)=\prod_{i\ge1}\frac{1}{1-t^{4i}}.\]
2-locally, the Thom spectrum $M\SO$ is a wedge sum of
suspensions of Eilenberg-Mac Lane spectra $H\Z_2$ and $H\Z$.
This allows us to write
\begin{multline}
H^*(M\SO;\Z_2)\cong\\
\bigoplus_{\text{free summands}}H^*(H\Z;\Z_2)\oplus\bigoplus_{\text{torsion summands}}H^*(H\Z_2;\Z_2).
\end{multline} 
Let the Poincar\'e series for $H^*(H\Z_2;\Z_2)$ and $H^*(H\Z;\Z_2)$ be $p_{H\Z_2}(t)$ and $p_{H\Z}(t)$ respectively, then by \cite{Serre1953}, we have
\[p_{H\Z_2}(t)=\prod_{k=2^i-1}\frac{1}{1-t^k}\]
and 
\[p_{H\Z}(t)=\frac{1}{1+t}p_{H\Z_2}(t).\]

Since the Poincar\'e series $p_{\text{tors}}(t)$ of the
torsion part in $\Omega_*^{\SO}$ satisfies
\[p_{\text{tors}}(t)\cdot p_{H\Z_2}(t)+p_{\text{free}}(t)\cdot p_{H\Z}(t)=\prod_{k\ge2}\frac{1}{1-t^k},\]
we can solve
\[p_{\text{tors}}(t)=[(1-t)\prod_{k\ge2,k\ne2^i-1}(\frac{1}{1-t^k})]-[\frac{1}{1+t}\prod_{k\ge1}(\frac{1}{1-t^{4k}})].\]
In particular, we have Table \ref{table:oriented} \cite[page 203]{Milnor-Stasheff}.
\begin{table}
\begin{tabular}{c|c|c} 
\hline
$d$  & $\Omega_d^{\SO}$  & manifold generators \\
\hline
$0$ &  $\Z$  \\
$1$ & $0$  \\
$2$  &  $0$ \\
$3$ &  $0$ \\
$4$ & $\Z$ & $\CP^2$ \\
$5$ & $\Z_2$ & 
$\begin{array}{c} 
\text{$Y^5$, Wu $\SU(3)/\SO(3)$, or} \\
\text{Dold $S^1\times_{\tau}\CP^2$ manifolds}
\end{array}$ 
\\
$6$ & $0$ \\
$7$ & $0$ \\
$8$ & $\Z^2$ & $\CP^4,\CP^2\times\CP^2$\\
$9$ & $\Z_2^2$ & $Y^9,Y^5\times\CP^2$ \\
$10$ &  $\Z_2$ & $Y^5\times Y^5$ \\
$11$ &  $\Z_2$ & $Y^{11}$\\
\hline
 \end{tabular} 
\caption{Oriented bordism groups and manifold generators. As manifold $Y^5$ (respectively $Y^9,Y^{11}$) we may take the nonsingular hypersurface of degree $(1,1)$ in the product $\RP^2\times\RP^4$ (respectively $\RP^2\times\RP^8$ or $\RP^4\times\RP^8$) of real projective spaces. These manifolds are called real Milnor manifolds. 
The 5d Wu manifold is $\SU(3)/\SO(3)$. The 5d Dold manifold is $S^1\times_{\tau}\CP^2$ where the involution $\tau$ sends $(x,[y])$ to $(-x,[\bar{y}])$.}
\label{table:oriented}
\end{table}

In 5d, the bordism group $\Omega_5^{\SO}=\Z_2$ and the bordism invariant is $\w_2\w_3$ since the only non-vanishing Stiefel-Whitney number of oriented 5-manifolds is $\w_2\w_3$.

\section{Combinatorial Formula for Stiefel-Whitney Classes}
\label{app:SW}

In 1940, Whitney obtained an explicit combinatorial formula for the
Stiefel-Whitney classes \cite{Whitney1940}.  The formula is as follows.  Let
$K$ be an $m$-dimensional combinatorial manifold and $K'$ the first barycentric
subdivision of $K$.  Let $C_n$ be the sum modulo 2 of all $(m-n)$-dimensional
simplices of $K'$.  Then the chain $C_n$ is a cycle modulo 2 and represents the
homology class $W_n$ Poincar\'e dual to the $n$-th Stiefel-Whitney class of
$K$.

In \cite{Goldstein-Turner}, the authors obtained a formula for the
Stiefel-Whitney homology classes in the original triangulation without passing
to the first barycentric subdivision. Their formula is as follows.  A branch
structure on a triangulation is an orientation of the links with no closed
loops which in turn provides an order to the vertices of simplices.  Given a
branch structure on $K$ so that any representation of a simplex in $K$ is
written with its vertices in increasing order. Let $s$ be an $(m-n)$-simplex in
$K$, say $s=[v_0,v_1,\dots,v_{m-n}]$. Let $t$ be another simplex which has $s$
as a face; \ie $s\subset t$ ($s$ may be equal to $t$).\\ 
Let $B_{-1}=$ set of
vertices of $t$ less than $v_0$,\\ $B_0=$ set of vertices of $t$ strictly
between $v_0$ and $v_1$,\\ $B_k=$ set of vertices of $t$ strictly between $v_k$
and $v_{k+1}$,\\ $B_{m-n}=$ set of vertices of $t$ greater than $v_{m-n}$.\\ We
say that $s$ is regular in $t$, if $B_k$ is empty for every odd $k$. Let
$\partial_{m-n}(t)$ denote the mod 2 chain which consists of all
$(m-n)$-simplices $s$ in $t$ so that $s$ is regular in $t$.  Then
$C_n'=\sum_{\dim t\ge m-n}\partial_{m-n}(t)$ is a chain which represents the
homology class $W_n$ Poincar\'e dual to the $n$-th Stiefel-Whitney class of
$K$.

\begin{figure}
\includegraphics{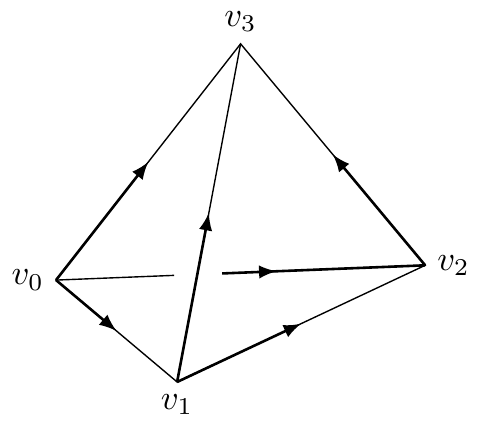}
\caption{The surface of a branch tetrahedron.}
\label{figure:tetrahedron}
\end{figure}

\begin{figure}
\includegraphics{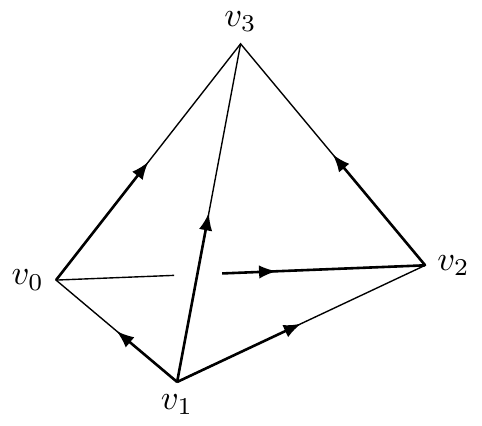}
\caption{The surface of a tetrahedron with another branch structure in which the order between $v_0$ and $v_1$ is reversed.}
\label{figure:flip}
\end{figure}

For example, let $K$ be the surface of a tetrahedron and $|K|=S^2$. Then
$C_1'=\sum_{\dim t\ge1}\partial_1(t)$. Given the branch structure on $K$ so
that the 4 vertices of $K$ are ordered as $v_0$, $v_1$, $v_2$, and $v_3$, see
Figure \ref{figure:tetrahedron}. For $\dim t\ge1$, $t$ can be chosen as
$[v_0,v_1]$, $[v_0,v_2]$, $[v_1,v_2]$, $[v_0,v_3]$, $[v_1,v_3]$, and
$[v_2,v_3]$ if $\dim t=1$ and $[v_0,v_1,v_2]$, $[v_0,v_1,v_3]$,
$[v_0,v_2,v_3]$, and $[v_1,v_2,v_3]$ if $\dim t=2$.  If $\dim t=1$ and $s\in
\partial_1(t)$, then $s=t$, if $\dim t=2$ and $s\in \partial_1(t)$, then $s$ is
the 1-simplex whose two vertices are the smallest and the greatest vertices of
$t$.  Therefore, $C_1'=[v_0,v_1]+[v_0,v_3]+[v_1,v_2]+[v_2,v_3]$.  If another
branch structure is given on $K$ so that the order between $v_0$ and $v_1$
is reversed while other orders remain the same, see Figure \ref{figure:flip},
then $C_1'$ changes to $[v_0,v_1]+[v_0,v_2]+[v_1,v_3]+[v_2,v_3]$ and the
difference with the original $C_1'$ is
$[v_0,v_3]+[v_0,v_2]+[v_1,v_2]+[v_1,v_3]$ which is a boundary. So $C_n'$
depends on the branch structure and different choices of branch
structures can only change $C_n'$ by a boundary.

The Poincar\'e dual (see Appendix \ref{sec:Poincaredual}) also depends on the branch structure. 
Thus $\w_n$ depends on the branch structure and different choices of
branch structures can only change $\w_n$ by a coboundary.


\section{Compute $\w_2\w_3$ on Real Milnor, Wu, and Dold Manifolds}

The 5d real Milnor manifold \cite{Milnor1965} $Y^5=H(2,4)$ is the submanifold of $\RP^2\times \RP^4$ given by
\begin{multline}
H(2,4)=\{([x_0,x_1,x_2],[y_0,\dots,y_4])\in\RP^2\times\RP^4:\\
\sum_{i=0}^2x_iy_i=0\}.
\end{multline}
In fact, $H(2,4)$ is the submanifold of $\RP^2\times \RP^4$ Poincar\'e dual to $(a+b)$ where $a$ and $b$ are the generators of $H^*(\RP^2;\Z_2)$ and $H^*(\RP^4;\Z_2)$ respectively. Note that $a^3=0$ and $b^5=0$.
The total Stiefel-Whitney class $\w(H(2,4))$ of $H(2,4)$ is given by the restriction to $H(2,4)$ of the expression
\[\frac{(1+a)^3(1+b)^5}{(1+a+b)}.\]
By direct computation, we find that $\w_2=a^2+ab$ and $\w_3=ab^2+a^2b$. So $\w_2\w_3=a^2b^3$ and the Stiefel-Whitney number $\langle \w_2\w_3,[Y^5]\rangle=\langle (a+b)\w_2\w_3,[\RP^2\times\RP^4]\rangle=1$.

The Wu manifold $W:= \SU(3)/\SO(3)$ has cohomology ring $H^*(W;\Z_2) = \Z_2[z_2,z_3]/(z_2^2,z_3^2)$ with the total Stiefel-Whitney class $\w (W)= 1 + z_2 + z_3$, $\Sq(z_2) = z_2 + z_3 $, and $\Sq(z_3) = z_3 + z_2z_3$ where $\Sq:=\Sq^0+\Sq^1+\Sq^2+\cdots$ is the total Steenrod square. So the Stiefel-Whitney number $\langle \w_2\w_3,[W]\rangle=1$.

The 5d Dold manifold \cite{Dold1956} $P(1,2)$ is the quotient $S^1\times_{\tau}\CP^2$ where the involution $\tau$ sends $(x,[y])$ to $(-x,[\bar{y}])$.
The ring structure of $H^*(P(1,2);\Z_2)$ is 
\[H^*(P(1,2);\Z_2)=[\Z_2[c]/(c^2=0)]\otimes[\Z_2[d]/(d^3=0)],\]
and the total Stiefel-Whitney class of $P(1,2)$ is
\[\w(P(1,2))=(1+c)(1+c+d)^3,\]
where $c\in H^1(P(1,2);\Z_2)$ and $d\in H^2(P(1,2);\Z_2)$.
The Steenrod squares act by 
\[\Sq^0=\text{id},\;\Sq^1(c)=0,\;\Sq^1(d)=cd,\;\Sq^2(d)=d^2,\]
and all other Steenrod squares act trivially on $c$ and $d$.
By direct computation, we find that $\w_2=d$ and $\w_3=cd$. 
So the Stiefel-Whitney number $\langle \w_2\w_3,[P(1,2)]\rangle=1.$

\section{Generalized Wu Relation}
\label{sec:GeneralizedWu}
The classical Wu relation \eq{eq:Wu-relation} expresses the action of a single Steenrod square $\Sq^{n}$ on a $\Z_2$-valued cocycle $x_{d-n}$ in the top $d$-dimension on a manifold $M^d$ as the cup product $u_{n}x_{d-n}$ where $u_n$ is the Wu class \eqref{eq:Wu-class}.
In this section, we generalize this Wu relation to other elements in the mod 2 Steenrod algebra $\A_2$. 

By Adem relation, $\Sq^1\Sq^1=0$ and $\Sq^1\Sq^2=\Sq^3$. So the simplest element in $\A_2$ which is not a single Steenrod square is $\Sq^2\Sq^1$.
We claim that $\Sq^2\Sq^1x_{d-3}=(\w_1^3+\w_3)x_{d-3}$ on a manifold $M^d$ where $\w_i$ is the Stiefel-Whitney class of $M^d$.
In fact, 
\begin{align} \label{eq:generalizedWu}
&\Sq^2\Sq^1x_{d-3}\nonumber\\
=&(\w_1^2 + \w_2) (\Sq^1 x_{d-3})\nonumber\\
=&\Sq^1(\w_1^2x_{d-3})+(\Sq^1\w_2)x_{d-3}+\Sq^1(\w_2x_{d-3})\nonumber\\
=&\w_1^3x_{d-3}+(\w_1\w_2+\w_3)x_{d-3}+\w_1\w_2x_{d-3}\nonumber\\
=&(\w_1^3+\w_3)x_{d-3}.
\end{align}
In the first equality, we used the Wu relation \eq{eq:Wu-relation} for $\Sq^2$.
In the second equality, we used the product formula for Steenrod square $\Sq^k(x \smile y)=\sum_{i+j=k}\Sq^i x \smile \Sq^j y$
and $\Sq^1(\w_1^2)=0$.
In the third equality, we used the Wu relation \eq{eq:Wu-relation} for $\Sq^1$ and $\Sq^1\w_2=\w_1\w_2+\w_3$.
This \eq{eq:generalizedWu} is a new generalized Wu relation, which is mentioned in \eq{eq:GeneralizedWu}.

\section{Pullback Construction of Branch-Independent Bosonic Models}
\label{pullback}

In this section, we are going to present an general systematic construction of
branch-independent bosonic models.  We will first construct a model with a
finite $G$ symmetry, realizing a $G$-SPT order.  The degrees of freedom in our
model are described by $g_i \in G$ on each vertex-$i$.
The model on space time $M^d$ is defined by the path integral
\begin{align}
 Z(M^d) = \sum_{g_i} \ee^{-S(g_i)}.
\end{align}
We can rewrite that model as
\begin{align}
 Z(M^d) = \sum_{g_i} \ee^{-S(a_{ij}^G)},\ \ \ \
a_{ij}^G = g_i g_j^{-1}.
\end{align}
We can add a background flat $G$-gauge field $A_{ij}^G\in G$
\begin{align}
 A_{ij}^GA_{jk}^G=A_{ik}^G
\end{align}
to describe the symmetry twist, and consider the following gauged model
\begin{align}
\label{ZaA}
 Z(M^d, A^G) = \sum_{g_i} \ee^{-S(a_{ij}^G)},\ \ \ \
a_{ij}^G = g_i A_{ij}^G g_j^{-1}.
\end{align}
Note that $a_{ij}^G$ satisfy a flat condition
\begin{align}
\label{aflat}
 a_{ij}^Ga_{jk}^G=a_{ik}^G.
\end{align}
When $A_{ij}^G=1$,
the partition function in
\eq{ZaA} automatically has the
$G$ symmetry
\begin{align}
 g_i \to g_i h, \ \ \ h\in G,
\end{align}
even for spacetime $M^d$ with boundaries.  This implies that the $G$ symmetry
is anomaly-free.

We can choose a proper $S(a_{ij}^G)$ so that the bosonic model \eq{ZaA}
realized a bosonic SPT order.  To do so, let us consider the classifying space
of the group $G$, and choose an one-vertex triangulation of the classifying
space.  We denote the resulting simplicial complex as $\B G$. For the details
about the one-vertex triangulation, see \Rf{ZW180809394}.  $\B G$ has the
properties that $\pi_1(\B G)=G$ and $\pi_{n\neq 1}(\B G)=0$.  Since $\B G$ has
only one vertex, the links in $\B G$ are labeled by $\bar a_{ij}^G \in G$, that
satisfy the condition
\begin{align}
\label{baflat}
 \bar a_{ij}^G \bar a_{jk}^G = \bar a_{ik}^G
\end{align}
for every triangles $(ijk)$ in $\B G$.

$a_{ij}^G$ on the links of spacetime simplicial complex $M^d$ defines a
homomorphism of simplicial complex 
\begin{align}
M^d \xrightarrow{\phi(a_{ij}^G)} \B G .
\end{align}
$\phi(a_{ij}^G)$ maps the vertices in $M^d$ to the only vertex in $\B G$.
$\phi(a_{ij}^G)$ maps the link in $M^d$ with value $a_{ij}^G$ to the link in
$\B G$ labeled by $\bar a_{ik}^G = a_{ik}^G$.  The two flat conditions \eq{aflat}
and \eq{baflat} ensure that $\phi(a_{ij}^G)$ maps the triangles in $M^d$ to
triangles in $\B G$, \etc. Thus $\phi(a_{ij}^G)$ is a homomorphism of simplicial
complex, and 
$a_{ij}^G$ is the pullback of
$\bar a_{ij}^G$ by $\phi(a_{ij}^G)$:
\begin{align}
 a_{ij}^G = \phi^*(a_{ij}^G) \ \bar a_{ij}^G .
\end{align}

Let $\bar \om_d$ be a $\RZ$-valued cocycle in $\B G$:
\begin{align}
 \bar \om_d \in H^d(\B G; \RZ).
\end{align}
Now we can construct a bosonic model using $\bar \om_d$:
\begin{align}
\label{ZaAom}
 Z(M^d, A^G) = \sum_{g_i} \ee^{\ii 2\pi \int_{M^d}
\phi^*(a_{ij}^G) \bar \om_d
},\ \ \ \
a_{ij}^G = g_i A_{ij}^G g_j^{-1}.
\end{align}
We refer to this construction as pullback construction.  Clearly, the resulting
bosonic model is branch-independent, since during the whole construction, the
branch structure is not even specified.  We also note that the action amplitude
$ \ee^{\ii 2\pi \int_{M^d} \phi^*(a_{ij}^G) \bar \om_d }$ does not depend on
the Stiefel-Whitney cocycle $\w_n$.

The branch-independent bosonic model \eq{ZaAom} realizes the SPT order labeled
by $\bar \om_d \in H^d(\B G; \RZ)$, and described by the SPT invariant
\cite{HW1339,W1447}
\begin{align}
 Z^\text{top}(M^d, A^G) =
\ee^{\ii 2\pi \int_{M^d}
\phi^*(A_{ij}^G) \bar \om_d
} .
\end{align}
This is consistent with the group
cohomology classification of SPT order \cite{CGL1314}.

The above construction of branch-independent bosonic models can be
generalized to compact continuous group $G$ with less rigor, where the
branch-independent bosonic model is given by the following path integral
\begin{align}
\label{ZaAomL}
 Z(M^d, A^G) = \sum_{\phi} \ee^{\ii 2\pi \int_{M^d} \phi^* \bar \om_d }
\ee^{-\int_{M^d} \cL(\phi)}.
\end{align}
Here we give the classifying space of $G$ a triangulation and denote the
resulting simplicial complex as $\B G$.  The trianglation temporarily breaks
the $G$ symmetry.  $\phi$ is a homomorphism from spacetime simplicial complex
$M^d$ to classifying space simplicial complex  $\B G$:
\begin{align}
M^d \xrightarrow{\phi} \B G .
\end{align}
The path integral is a sum over the homomorphisms $\phi$.  This is similar to
the definition of non-linear $\si$-model on a continuum spacetime, where the
path integral is a sum over the continuous maps from spacetime manifold to
target space.  The term $\ee^{-\int_{M^d} \cL(\phi)}$ is chosen such that the
model \eq{ZaAomL} describes a disordered state.  In the limit of fine
triangulation the model has a $G$ symmetry, and the disordered state is $G$
symmetric.

Also, $\bar \om_d$ in the path integral \eq{ZaAomL} is a $\RZ$-valued cocycle
in $\B G$:
\begin{align}
 \bar \om_d \in H^d(\B G; \RZ).
\end{align}
If $H^{n}(\B G;\Z)=\Z\oplus \Z_n\oplus \cdots$, then $H^{n}(\B G;\RZ)=\RZ\oplus \Z_n\oplus
\cdots$ according to the universal coefficient theorem 
\begin{align}
\label{ucf}
 H^d(X;\M)
&\simeq  \M \otimes_{\Z} H^d(X;\Z)
\oplus
\Tor( \M, H^{d+1}(X;\Z) )  .
\end{align}
We see that the torsion part of $H^{d}(\B G;\RZ)$ and $H^{d+1}(\B G;\Z)$
coincide:
\begin{align}
 \Tor H^{d}(\B G;\RZ)
=\Tor H^{d+1}(\B G;\Z).
\end{align}

The cocycle that generate $\RZ$ part of $H^{n}(\B G;\RZ)$ does not have a
quantized coefficient and can be continuously changed to zero.  So the  $\RZ$
part of $H^{d}(\B G;\RZ)$ does not characterize a topological phase.  Only
$\bar \om_d \in  \Tor H^{d}(\B G;\RZ) =  \Tor (\RZ, H^{d+1}(\B G;\Z))$ gives
rise to distinct topological phase via the model \eq{ZaAomL}, with is a $G$-SPT
phase.

When $G$ is continuous, some $G$-SPT order can belong to $\Z$ class which is
not a torsion. To construct branch-independent bosonic model to realize this
kind of SPT order, we need to generalize the above model \eq{ZaAomL} to the
following form
\begin{align}
\label{ZaAomLN}
 Z(M^d, A^G) 
&= \sum_{\phi} 
\ee^{-\int_{M^d} \cL(\phi)}
\ee^{\ii 2\pi [\int_{M^d} \phi^* \bar \om_d 
+ \int_{N^{d+1}} \phi^*_N \bar \nu_{d+1} ]},
\nonumber\\
M^d &= \prt N^{d+1}.
\end{align}
The term $\ee^{\ii 2\pi \int_{N^{d+1}} \phi^*_N \bar \nu_{d+1}}$, living in one
higher dimension, is a  Wess-Zumino-Witten like term, and $\phi_N$ is a
homomorphism of simplicial complex
\begin{align}
N^{d+1} \xrightarrow{\phi_N} \B G 
\end{align}
such that at the boundary $M^d = \prt N^{d+1}$, $\phi_N = \phi$.
$\bar\nu_{d+1}$ is a $\R$-valued cocycle that satisfies a quantization
condition
\begin{align}
 \int_{N^{d+1}}  \bar \nu_{d+1} \in \Z,
\end{align}
for all closed $N^{d+1}$ in $ \B G$.  In other words, the $\R$-valued cocycle
$\bar\nu_{d+1}$ represents a cohomology class in the free part of $H^{d+1} (\B
G; \Z)$:
\begin{align}
\bar \nu_{d+1} \in \text{Free}(H^{d+1} (\B G; \Z) ) .
\end{align}

In the disordered phase, \eq{ZaAomLN} realizes a $G$-SPT order characterized by
$(\bar \om_{d}, \bar \nu_{d+1})$ in $\text{Tor}(H^{d} (\B G; \RZ) )
=\text{Tor}(H^{d+1} (\B G; \Z) )$ and $\text{Free}(H^{d+1} (\B G; \Z) )$.  In
other words, the $G$-SPT order characterized is characterized by the elements
in $H^{d+1} (\B G; \Z)$, which agree with the group cohomology theory of SPT
order for symmetries described by compact groups.

When $G=\SO(\infty)$, the term $\ee^{\ii 2\pi \int_{N^{d+1}} \phi^*_N \bar
\nu_{d+1}}$ gives rise to the $\SO(\infty)$ Chern-Simons term on $M^d =\prt
N^{d+1}$, whose generator is the Pontryagin class (for $d+1 = 0$ mod 4).  The
pullback of different maps $\phi$
\begin{align}
M^d \xrightarrow{\phi} \B \SO(\infty) 
\end{align}
give rise to different $\SO(\infty)$ bundle over $M^d$.  If we restrict
$\sum_{\phi}$ in \eq{ZaAomLN} to a subset of maps $\phi$, such that the
$\SO(\infty)$ bundle over $M^d$ is the same as the stabilized tangent bundle of
$M^d$, the model \eq{ZaAomLN} may realizes a bosonic invertible topological
order.  The $\Z$-class of  invertible topological orders are described by
gravitational Chern-Simons term, which is also $\SO(\infty)$ Chern-Simons term.
We see that the  model \eq{ZaAomLN} can only realize gravitational Chern-Simons
terms generated by Pontryagin classes, which have no framing anomaly. Thus in
3d, the  model \eq{ZaAomLN} can only realize  invertible topological orders
generated by E$_8^3$ topological order.  The  E$_8$ topological orders is
characterized by a gravitational Chern-Simons term that corresponds to
$\frac13$ of the first  Pontryagin class, which has a framing anomaly.

\section{{Background vs Dynamical Gauge Transformations 
}}
\label{app:Gauge-Transformation}

In \Sec{eq:Z2EFT}, we describe the invariance or non-invariance of path integral in terms of the change of
coboundaries and branch structures. 
Here we fill in some additional terminology more accessible for quantum field theorists:
in terms of \emph{background gauge transformations} vs \emph{dynamical gauge transformations}.

For 4d $\Z_2$ gauge theory \eq{ZabZ2} described by $\Z_2$-valued 1-cochain
and 2-cochain dynamical fields $a^{\Z_2}$ and $b^{\Z_2}$ 
$$
 Z=\sum_{a^{\Z_2},b^{\Z_2}} \ee^{\ii \pi \int_{B^4} a^{\Z_2}\dd b^{\Z_2}},
$$
where $\sum_{a^{\Z_2},b^{\Z_2}}$ is a summation over $\Z_2$-valued 1- and
2-cochains. 
The above theory has 
\emph{dynamical gauge transformations} for dynamical fields:
\begin{align} \label{eq:dynamical-gauge-a-b}
 a^{\Z_2} \to a^{\Z_2}+\dd u_0, \ \ \ 
 b^{\Z_2} \to b^{\Z_2}+\dd u_1,
\end{align}
where $u_0\in C^0(B^4;\Z_2)$ and $u_1\in C^1(B^4;\Z_2)$ are $\Z_2$-valued 0- and 1-cochain fields.  

Now let us discuss another interpretation of \eq{Z0Z2}.
{The Stiefel-Whitney classes 
$\w_2 \in H^2(M^5;\Z_2)$ and $\w_3 \in H^3(M^5;\Z_2)$ 
are special cohomology classes satisfying 
the extra axioms A1-A4 listed earlier, with the base manifold $M^5$ for the real vector bundle.
Since $M^5$ is orientable, we have $\w_1\se{2}0$ and $\w_3\se{2}\Sq^1\w_2$.

When $M^5$ has a boundary, the
partition function \eq{Z0Z2} depends on the choice of the coboundaries
in $\w_2$ and $\w_3$. \ie under the following \emph{background gauge transformation} for non-dynamical fields:
\begin{eqnarray} \label{topw2w3-background-gauge}
&&\w_2 \to \w_2 +\dd v_1,\\ \nn
&&\w_3 \to \w_3 +{\Sq^1\dd v_1 + \dd v_2}
\to \w_3 +{\dd v_2.}
\end{eqnarray}
Although the Stiefel-Whitney classes have the relation $\Sq^1\w_2 = \w_3$
so that the transformation $\dd v_1$ can be related to $\Sq^1\dd v_1$,
but they can differ by a coboundary $\dd v_2$ which thus absorbs $\Sq^1\dd v_1$.

An anomalous $\Z_2$-gauge theory {(that has 't Hooft anomaly of spacetime SO diffeomorphism)}
on the boundary $B^4=\prt M^5$ of
the topological state is described by 
\eq{Z0Z2}  
which has not only the \emph{dynamical gauge transformations} 
(involving $u_0$ and $u_1$ in \eq{eq:dynamical-gauge-a-b})
but also additional {\emph{background gauge transformations}}
(involving $v_1$ in \eq{topw2w3-background-gauge}):
\begin{align} 
\label{eq:dynamical-gauge-a-b-w2-w3}
 a^{\Z_2} & \to a^{\Z_2} +\dd u_0+ v_1 , & 
 b^{\Z_2} & \to b^{\Z_2} +\dd u_1 +  v_2,
\nonumber \\
\w_2  &\to \w_2 + \dd v_1,  & 
\w_3 & \to \w_3 + \dd v_2.
\end{align}
It turns out that the \emph{background gauge transformations at the lattice
scale} of the simplicial complex are important to ensure the anomaly inflow or
anomaly cancellation between the bulk and boundary for the 't Hooft anomaly of
global symmetries.  In contrast, the  \emph{dynamical gauge transformations at
the lattice scale} of the simplicial complex are not so crucial or fundamental
--- the  \emph{dynamical gauge invariance} at the lattice scale, even if we
break them locally, the \emph{dynamical gauge invariance} can re-emerge at a
larger length scale.  So only the \emph{emergent dynamical gauge invariance} is
crucial.

\section{$\mathbb{Z}_2$ topological order with emergent fermion and higher
dimensional bosonization}
\label{hDboson}

In this section, we review and summarize the higher dimensional bosonization
following \Rf{LW180901112}.  In $d+1$-dimensional spacetime, a bosonic model
that realizes a $\Z_2$ topological order is described by the following path
integral
\begin{align}
 Z(M^{d+1}) =
\sum_{\dd a^{\Z_2} \se{2} 0} 1,
\end{align}
where $\sum_{\dd a^{\Z_2} \se{2} 0}$ sums over all $\Z_2$ valued 1-cocycles
$a^{\Z_2}$.  The low energy effective theory of the $\Z_2$ topological order is
a $\Z_2$ gauge theory where the point-like $\Z_2$ charge is a boson.  Such a
$\Z_2$ topological order  has another realization in terms of $\Z_2$-valued
$d-1$-cocyles $l^{\Z_2}$
\begin{align}
 Z(M^{d+1}) =
\sum_{\dd l^{\Z_2}_{d-1} \se{2} 0} 1 .
\end{align}

There is a twisted $\Z_2$ topological order \cite{LW0316}, whose low energy
effective theory is a twisted $\Z_2$ gauge theory where the point-like $\Z_2$
charge is a fermion.  Such a twisted $\Z_2$ topological order is realized by
the following bosonic model
\begin{align}
 Z(M^{d+1}) =
\sum_{\dd l^{\Z_2}_{d-1} \se{2} 0} \ee^{\ii \pi \int_{M^{d+1}} 
\gSq^2 l^{\Z_2}_{d-1} } .
\end{align}
The above path integral does not contain a $\Z_2$ charge. 
To include a $\Z_2$ charge, we note that the worldline of a particle can be described by its
Poincar\'e dual $f_d$, which is a $\Z_2$-valued $d$-coboundary.  The path
integral including such a worldline is given by
\begin{align}
\label{ZlF}
 Z(M^{d+1}) =
\sum_{\dd l^{\Z_2}_{d-1} \se{2} f_d} \ee^{\ii \pi \int_{M^{d+1}} 
\gSq^2 l^{\Z_2}_{d-1} } .
\end{align}
The term $\ee^{\ii \pi \int_{M^{d+1}} \gSq^2 l^{\Z_2}_{d-1} }$ gives the
worldline (described by $f_{d-1}$) a Fermi statistics.

The $(d+1)$d twisted $\Z_2$ topological order has a $d$-dimensional boundary,
formed by condensing the $\Z_2$ flux.  Such a boundary contains only point-like
topological excitations, which are fermions, coming from the bulk fermionic
$\Z_2$ charge.  Such a boundary is the canonical boundary of the path integral
\eq{ZlF}, which is described by the following path integral
\begin{align}
\label{ZlFB}
 Z(M^{d+1}) =
\sum_{\dd l^{\Z_2}_{d-1} \se{2} 0} \ee^{\ii \pi \int_{M^{d+1}} 
\gSq^2 l^{\Z_2}_{d-1} } , \ \ \ \ B^d =\prt M^{d+1}.
\end{align}
The cocycle $l^{\Z_2}_{d-1}$ on the boundary $B^d$ can be viewed as the
Poincar\'e dual of the worldline of boundary fermions.

Now, let us try to view the above path integral \eq{ZlFB} as a path integral
on the boundary $B^d$ for the field  $l^{\Z_2}_{d-1}$,  and view the term
$\ee^{\ii \pi \int_{M^{d+1}} \gSq^2 l^{\Z_2}_{d-1} }$ as a Wess-Zumino-Witten
like term defined in one higher dimension.  But such a viewpoint is not
quite correct, since the $\ee^{\ii \pi \int_{M^{d+1}} \gSq^2 l^{\Z_2}_{d-1} }$ not
only depends on $l^{\Z_2}_{d-1}$ on the boundary $B^d=\prt M^{d+1}$, but also
depends on $M^{d+1}$, \ie how we extend $B^d$.  However, when $M^{d+1}$ is
oriented and spin, $\w_1\se{2,\dd} 0$ and $\w_2\se{2,\dd} 0$, $\gSq^2
l^{\Z_2}_{d-1} \se{2,\dd} 0$.  In this case, $\ee^{\ii \pi \int_{M^{d+1}}
\gSq^2 l^{\Z_2}_{d-1} }$ only depends on $l^{\Z_2}_{d-1}$ on the boundary
$B^d=\prt M^{d+1}$, and can indeed by viewed as Wess-Zumino-Witten like term.

We can modify the path integral \eq{ZlFB} to relax the requirement for
$M^{d+1}$ to be spin:
\begin{align}
\label{ZlFB1}
 Z(M^{d+1}) &=
\hskip -0.5em
\sum_{\dd l^{\Z_2}_{d-1} \se{2} 0} 
\hskip -0.5em
\ee^{\ii \pi \big( \int_{B^d}  l^{\Z_2}_{d-1} A^{\Z_2^f} 
+ \int_{M^{d+1} } 
\gSq^2 l^{\Z_2}_{d-1} +\w_2l^{\Z_2}_{d-1} \big)} 
\nonumber\\
 B^d &= \prt M^{d+1},\ \ \ \ \w_2 \se{2} \dd A^{\Z_2^f} \text{ on } B^d.
\end{align}
In the above, $M^{d+1}$ is orientable but may not be spin, and $B^d$ is spin
such that $\w_2 \se{2} \dd A^{\Z_2^f} $ on $B^d$.  In this case, $ \ee^{\ii \pi
\int_{M^{d+1}} \gSq^2 l^{\Z_2}_{d-1} +\w_2l^{\Z_2}_{d-1} }$ is a
Wess-Zumino-Witten like term, that only depend on $l^{\Z_2}_{d-1}$ and $\w_2$
on $B^d$.  Also, \eq{ZlFB1} is invariant under the ``gauge'' transformation
\begin{align}
 \w_2 \to \w_2 +\dd v_1^{\Z_2}, \ \ \ \
 A^{\Z_2^f} \to A^{\Z_2^f} + v^{\Z_2}_1
,
\end{align}
so that it is branch independent.

$ A^{\Z_2^f}$ is a $\Z_2$-valued 1-cochain, which corresponds to the spin
structure on $B^d$.  The relation $\w_2 \se{2} \dd A^{\Z_2^f} $ tells us that
the worldline $l^{\Z_2}_{d-1}$ couples to the $SO(n)$ tangent bundle of $B^d$
in such a way that the worldline corresponds to an half-integer spin.

\bibliography{BSM-GEQC.bib,../../bib/all,../../bib/publst,../../bib/allnew,../../bib/publstnew} 

\end{document}